\documentclass[10pt,a4paper]{article}
\usepackage[a4paper,text={160mm,247mm},centering]{geometry}
\usepackage{anyfontsize}

\newif\ifpublic\publictrue
\usepackage[normalem]{ulem}
\usepackage{style} 
\usepackage[style=alphabetic,maxnames=5]{biblatex} 
\usepackage{import} 
\usepackage{blindtext} 
\usepackage{macros} 
\usepackage{tcolorbox}
\usepackage{stix}
\ifpublic\else
\usepackage[notcite]{showkeys}
\fi
\usepackage{bbm}
\usepackage{multicol}
\usepackage{tikz}
\usepackage{empheq}
\usepackage{fancybox}
\usetikzlibrary{matrix,arrows,positioning,shapes,fadings,decorations.pathmorphing,
decorations.pathreplacing,automata,shadows,decorations.markings,patterns}
\usepackage[thicklines]{cancel}
\usetikzlibrary{calc,shapes.callouts,shapes.arrows}
\usepackage{enumitem}
\usepackage{tikz-cd}
\usepackage[font=small,labelfont=bf,width=0.85\textwidth]{caption}

\DeclareMathSymbol{\mlq}{\mathord}{operators}{``}

\usepackage{titlesec}

\titleformat{\section}
  {\normalfont\Large\bfseries}
  {\S\thesection}{1em}{}

\title{\texorpdfstring{
Gauging The Diamond:\\
Integrable Coset Models from Twistor Space
}{
Gauging The Diamond: Integrable Coset Models from Twistor Space
}}
\author{Lewis T. Cole, Ryan Cullinan, Ben Hoare, Joaquin Liniado and Dan Thompson}

\let\barefrac=\frac
\renewcommand{\frac}[2]{\mathinner{\barefrac{#1}{#2}}}

\let\baresqrt=\sqrt
\makeatletter
\renewcommand{\sqrt}{\@ifnextchar[\@sqrt@space@a\@sqrt@space@b}
\def\@sqrt@space@a[#1]#2{\mathinner{\mathchoice{\mkern-3mu}{\mkern-3mu}{}{}\baresqrt[#1]{#2}}}
\def\@sqrt@space@b#1{\mathinner{\mathchoice{\mkern-3mu}{\mkern-3mu}{}{}\baresqrt{#1}}}
\makeatother

\makeatletter
\let\per@dot@old=\.
\def\.{\ifmmode\def\per@dot@sel{\mkern3mu}\else\def\per@dot@sel{\per@dot@old}\fi\per@dot@sel}
\makeatother

\let\barefootnote=\footnote
\renewcommand{\footnote}[1]{\barefootnote{#1\vspace{3pt}}}

\providecommand{\hypersetup}[1]{}
\providecommand{\texorpdfstring}[2]{#1}
\hypersetup{plainpages=false}
\hypersetup{pdfpagemode=UseNone}
\hypersetup{bookmarksnumbered=true}
\hypersetup{pdfstartview=FitH}
\hypersetup{colorlinks=false}
\hypersetup{citebordercolor={1 1 1}}
\hypersetup{urlbordercolor={1 1 1}}
\hypersetup{linkbordercolor={1 1 1}}

\makeatletter
\let\@keywords\@empty
\let\@subject\@empty
\providecommand{\keywords}[1]{\gdef\@keywords{#1}}
\providecommand{\subject}[1]{\gdef\@subject{#1}}
\def\thetitle{\@title}
\def\theauthor{\@author}
\def\thesubject{\@subject}
\def\thedate{\@date}
\def\thekeywords{\@keywords}
\makeatother
\AtBeginDocument{
\hypersetup{pdftitle={\thetitle}}
\hypersetup{pdfauthor={\theauthor}}
\hypersetup{pdfsubject={\thesubject}}
\hypersetup{pdfkeywords={\thekeywords}}}

\begin{document}

\pdfbookmark[1]{Title Page}{title}
\thispagestyle{empty}

\vspace*{2cm}
\begin{center}
\begingroup\Large\bfseries\thetitle\par\endgroup

\vspace{1cm}



\setcounter{page}{1}

\begingroup
Lewis T. Cole$^{\textrm{a,}}$\footnote{\texttt{l.t.cole@pm.me}},
Ryan A. Cullinan$^{\textrm{b,}}$\footnote{\texttt{ryan.a.cullinan@durham.ac.uk}},
Ben Hoare$^{\textrm{b,}}$\footnote{\texttt{ben.hoare@durham.ac.uk}},
Joaquin Liniado$^{\textrm{c,}}$\footnote{\texttt{jliniado@iflp.unlp.edu.ar}}
and Daniel C. Thompson$^{\textrm{a,d,}}$\footnote{\texttt{d.c.thompson@swansea.ac.uk}}\par\endgroup
\vspace{1cm}

\textit{\small
$^{\textrm{a}}$\,Department of Physics, Swansea University,
Swansea SA2 8PP, UK
\\
$^{\textrm{b}}$\,Department of Mathematical Sciences, Durham University, Durham DH1 3LE, UK
\\$^{\textrm{c}}$\,Instituto de F\'{i}sica La Plata (CONICET and Universidad Nacional de La Plata) \\ CC 67 (1900) La Plata, Argentina
\\
$^{\textrm{d}}$\,Theoretische Natuurkunde, Vrije Universiteit Brussel, and the International Solvay Institutes, \\ Pleinlaan 2, B-1050 Brussels, Belgium
}
\vspace{1cm}
 
\textbf{Abstract}
\vspace{5mm}

\begin{minipage}{12.5cm}\small
Recent work has shown that certain integrable  and conformal field theories in two dimensions can be given a higher-dimensional origin from holomorphic Chern-Simons in six dimensions.  Along with anti-self-dual Yang-Mills and four-dimensional Chern-Simons, this gives rise to a diamond correspondence of theories.
In this work we extend this framework to incorporate models realised through gaugings. As well as describing a higher-dimensional origin of coset CFTs, by choosing the details of the reduction from higher dimensions, we obtain rich classes of two-dimensional integrable models including homogeneous sine-Gordon models and generalisations that are new to the literature.  
\end{minipage}


\end{center}
\newpage 

\tableofcontents
 
\newpage

\section{Introduction}

Quantum field theories (QFTs) in two dimensions have both direct applications in condensed matter systems and as the worldsheet theories of strings, and can provide a tractable sandpit for the study of quantum field theory more generally. Special examples are provided by conformal field theories (CFTs) and integrable field theories (IFTs), for which powerful infinite-dimensional symmetries enable us to exactly determine certain key properties and observables.  

One longstanding goal has been to provide a constructive origin of these integrable systems from some putative parent theory, perhaps in higher dimensions. For instance, Ward suggested \cite{Ward:1985gz} that all integrable equations may arise as reductions of the 4d anti-self-dual Yang-Mills (ASDYM) equation.
Given a choice of complex structure on $\bR^{4}$, the ASDYM equation are
\begin{align}
\label{eq:F20}
F^{2,0} &= 0 = F_{z w} \ , \\
\label{eq:F02}
F^{0,2} &=  0 = F_{\bar{z}\bar{w}} \ , \\
\label{eq:F11}
\varpi \wedge F^{1,1} &= 0 = F_{z\bar{z}} + F_{w\bar{w}} \ ,
\end{align}
where $\varpi$ is the K{\"a}hler form.  There are (at the very least) two senses in which ASDYM can be viewed as an integrable theory in its own right.  First is that the ASDYM equations can be exactly solved by the ADHM construction \cite{ADHM}.  Second is that these equations admit a zero curvature formulation in terms of a Lax pair of differential operators:
$$L= \nabla_z -\zeta \nabla_{\bar{w}} \, , \quad M= \nabla_w  + \zeta \nabla_{\bar{z}}  \, ,  \qquad [L,M] = 0 \quad \forall \, \zeta \iff F = -\star F\, .  $$  
Accordingly, in this work, we will denote four-dimensional QFTs whose equations of motion can be recast as the anti-self duality of some connection as IFT$_{4}$.

A prominent example in this class of theories is the 4d Wess-Zumino-Witten model (WZW$_{4}$) \cite{Donaldson:1985zz, Nair:1990aa, Nair:1991ab, Losev:1995cr}, which arises as a partial gauge fixing of the ASDYM equation. Up to a gauge transformation, one may parameterise a generic connection that solves equations \eqref{eq:F20} and \eqref{eq:F02} as $A = - \bar{\pd} g g^{-1}$, where the group-valued field $g$ becomes the fundamental field of WZW$_{4}$. The remaining ASDYM equation \eqref{eq:F11} becomes $\varpi \wedge \pd ( \bar{\pd} g g^{-1} ) = 0$, which is the equation of motion of WZW$_{4}$, also known as Yang's equation. The more well-known WZW$_{2}$ also arises as a reduction of WZW$_{4}$, and Yang's equation reduces to the familiar holomorphic conservation law characterising this CFT$_{2}$.
Another example is found by solving equations \eqref{eq:F20} and \eqref{eq:F11}, leaving equation \eqref{eq:F02} as the dynamical equation of motion. In this case, the IFT$_{4}$ is known as the LMP model \cite{Leznov:1986mx, Parkes:1992rz}, which gives the pseudo-dual of the principal chiral model after reduction.

Alternatively, motivated by the similarity between Reidemeister moves in knot theory and the Yang-Baxter equation that underpins integrability, Witten suggested \cite{Witten:1989wf} that integrable models might have a description in terms of Chern-Simons theory. The realisation of this idea came some years later, with Costello's understanding \cite{Costello:2013zra, Costello:2013sla} that the gauge theory description should combine the topological nature of Chern-Simons theory with the holomorphic nature of the spectral parameter characterising IFTs. The theory proposed in \cite{Costello:2013zra, Costello:2013sla} was extended and developed in a sequence of papers \cite{Costello:2017dso, Costello:2018gyb, Costello:2019tri} describing a Chern-Simons theory, which we denote by CS$_{4}$, defined over a four-manifold $\Sigma \times C$ with the action
\begin{equation}
S_{\text{CS}_{4}}[A] = \frac{1}{2 \pi \rmi} \int_{ \Sigma\times C } \omega \wedge \Tr \left( A \wedge \dr A + \frac{2}{3} A \wedge A \wedge A \right) \, .
\end{equation}
Here, $\omega$ is a meromorphic differential on the complex curve $C$, which we will take to be $C = \mathbb{CP}^1$.
Specifying boundary conditions at the poles of $\omega$, the dynamics can be `localised' to take place on $\Sigma$, which is identified with the space-time of the IFT$_{2}$, and the curve $C$ is associated to spectral parameter of the Lax connection of the integrable model (see \cite{Lacroix:2021iit} for a pedagogical introduction).

\begin{figure}
\centering
\begin{tikzpicture}
\node at (0,2) {$\mathbf{hCS_6}$};
\node at (-2,0) {$\mathbf{CS_4}$};
\node at (2,0) {$\mathbf{IFT_4}$};
\node at (0,-2) {$\mathbf{IFT_2}$ / $\mathbf{CFT_2}$ };
\draw[->,very thick,decorate, decoration={snake, segment length=12pt, amplitude=2pt}] (-0.4,1.6)--(-1.6,0.4);
\draw[->,very thick] (0.4,1.6)--(1.6,0.4);
\draw[->,very thick,decorate, decoration={snake, segment length=12pt, amplitude=2pt}] (1.6,-0.4)--(0.4,-1.6);
\draw[->,very thick] (-1.6,-0.4)--(-0.4,-1.6);
\end{tikzpicture}
\caption{The diamond correspondence of integrable avatars, in which wavy arrows indicate a descent by reduction and straight arrows involve localisation i.e. integration over $\mathbb{CP}^1$.}
\label{diagram:diamond}
\end{figure}
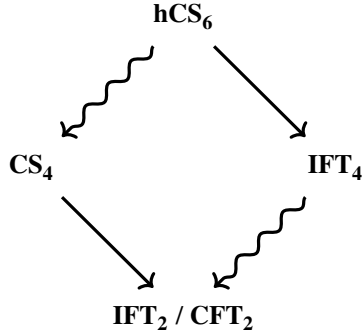

An elegant origin of both the CS$_{4}$ and the ASDYM descriptions was provided in the work of Bittleston and Skinner \cite{Bittleston:2020hfv} in terms of a six-dimensional holomorphic Chern-Simons theory (hCS$_{6}$), first proposed in \cite{Kcostello, Costello:2021bah}. The theory is defined over (the Euclidean slice of) Penrose's twistor space \cite{Penrose:1967wn} with the action functional
\begin{equation}
S_{\text{hCS}_{6}}[\cA] = \frac{1}{2 \pi \rmi} \int_{\mathbb{PT}} \Omega \wedge \Tr \left( \cA \wedge \bar{\partial} \cA + \frac{2}{3} \cA \wedge \cA \wedge \cA \right) \, ,
\end{equation}
in which $\Omega$ is some meromorphic $(3,0)$ form. This action is supplemented by a choice of boundary conditions at the poles of $\Omega$. The various lower-dimensional descriptions follow from exploiting the fibration structure $\mathbb{CP}^1 \hookrightarrow \mathbb{PT} \twoheadrightarrow \mathbb{R}^4$. Reducing along two directions within $\mathbb{R}^4$, hCS$_6$ descends to CS$_{4}$. Alternatively, one may instead first choose to localise over $\mathbb{CP}^1$, and this leads to IFT$_{4}$ of the ASDYM description. Indeed, the integrability properties of ASDYM are fundamentally tied to this twistorial origin and  evidence suggests that at a quantum level this twistor space is the natural arena to consider \cite{Costello:2021bah, Bittleston:2022nfr}.  
 Applying the reduction along $\bR^{4}$ to this IFT$_{4}$ produces an IFT$_{2}$ which may also be recovered by localising the CS$_{4}$ description. In this way, we have a diamond correspondence of theories illustrated in Figure \ref{diagram:diamond}.
Other recent work on hCS$_{6}$ includes \cite{Penna:2020uky, Costello:2021bah, Cole:2023umd}.

Given an IFT$_{2}$ or CFT$_{2}$ it is sometimes possible to obtain another I/CFT$_{2}$ via gauging. Perhaps the most famous example is the GKO $G/H$ coset CFTs \cite{Goddard:1984vk}, which can be given a Lagrangian description by taking a WZW$_{2}$ CFT on $G$ and gauging a (vectorially acting) $H$ subgroup \cite{Gawedzki:1988hq, Karabali:1988au, Bardakci:1990lbc, Witten:1991mm}. This motivates the core question of this work:
\begin{equation*}
\text{\em How can the diamond correspondence be gauged?}
\end{equation*}
Resolving this question dramatically expands the scope of theories that can be given a higher-dimensional avatar. A significant clue is given by the rather remarkable Polyakov-Wiegmann (PW) identity, which shows that the $G/H$ gauged WZW model is actually equivalent to the difference of a $G$ WZW model and an $H$ WZW model. This points towards a general resolution that integrable gauged models might be obtained as differences of ungauged models. This is less obvious than it might first seem; it was noted in \cite{Losev:1995cr} that for a PW identity to apply for WZW$_{4}$ one requires that the gauging is performed by connections with field strength restricted to be type $(1,1)$. The six-dimensional origin of such a constraint is rather intriguing and will be elucidated in this paper. In the context of CS$_{4}$, Stedman recently proposed \cite{Stedman:2021wrw} considering the difference of CS$_{4}$ to give rise to gaugings of IFT$_{2}$. We will recover this construction as a reduction of hCS$_{6}$ theory in the present work, as well as uncovering some additional novelties in the CS$_{4}$ description.

At the top of the diamond, we will consider a theory of two-connections $\mathcal{A} \in \Omega^{1}(\mathbb{PT}) \otimes \fg$ and $\mathcal{B} \in \Omega^{1}(\mathbb{PT}) \otimes \fh$ for a subalgebra $\fh \subset \fg$. The action of this theory is
\begin{equation}
S_{\mathrm{ghCS_6}} = S_{\mathrm{hCS}_{6}}[\mathcal{A}] - S_{\mathrm{hCS}_{6}}[\mathcal{B}] + S_{\mathrm{int}}[\mathcal{A},\mathcal{B} ] \, ,
\end{equation}
in which the term $S_{\mathrm{int}}$ couples the two gauge fields. We will develop this story by means of two explicit examples: choosing $\Omega$ to have two double poles, we will study the diamond relevant to the gauged WZW theory; and with $\Omega$ containing a single fourth-order pole we will study the gauged LMP model. This seemingly simple setup gives rise to a rich story whose results we will now summarise:
\begin{enumerate}
\item Our investigations indicate that general gaugings of the WZW$_{4}$ model break integrability in four dimensions. Integrability is preserved if the gauge field $B$ is constrained to satisfy two of the three anti-self-dual Yang-Mills equations, namely $F^{2,0}[B] = 0$ and $F^{0,2}[B] = 0$.
\item The two gauge fields $\cA$ and $\cB$ of ghCS$_{6}$ source various degrees of freedom in the gauged WZW$_{4}$. In particular, as well as the fundamental field $g$ and the 4d gauge field $B$, auxiliary degrees enter as Lagrange multipliers for $F^{2,0}[B] = 0$ and $F^{0,2}[B] = 0$.
\item Reducing by two dimensions, we recover a variety of IFT$_{2}$ including the special case of gauged WZW$_{2}$. In general, we find a coupled model between a gauged IFT$_{2}$ and a Hitchin system \cite{Hitchin1987} involving the gauge field $B$ and a pair of adjoint scalar fields. These scalars may source a potential for the gauged WZW$_{2}$ in which case we recover the complex sine-Gordon model and more broadly the homogeneous sine-Gordon models \cite{Fernandez-Pousa:1996aoa}. At the special point associated to the 2d PCM, Lagrange multipliers ensure that the gauge field is flat and hence trivial --- this is essential as the gauged PCM is not generically integrable.
\item We also use this formalism to perform an integrable gauging of the LMP model. Just as in the gauging of WZW$_{4}$, the field strength of the gauge field must be constrained to obey two of the anti-self-dual Yang-Mills equations, this time $F^{2,0}[B] = 0$ and $\varpi \wedge F^{1,1}[B] = 0$. It is noteworthy that the two equations which are enforced by Lagrange multipliers agree with the two equations that are identically solved in the ungauged case. This is true for both the WZW$_{4}$ and the LMP model. In addition, we show that the gauged LMP model obeys a PW-like identity such that it may be expressed the difference of an LMP model on $\fg$ and $\fh$.
\end{enumerate}
Let us outline the structure of this paper.
We begin in section \S\ref{sec:ungaugedWZWdiamond} with a review of the diamond correspondence of theories for the ungauged WZW model.
In section \S\ref{sec:gaugedWZWdiamond}, we introduce the gauging of this diamond concentrating in particular on the right hand side. We recover the gauged IFT$_{4}$ and demonstrate that its equations of motion may be rewritten as ASDYM.
The wide array of IFT$_{2}$ are explored in section \S\ref{sec:newIFT} where we also show that they are integrable and provide the associated Lax connection.
Following the gauging of the WZW$_{4}$, section \S\ref{sec:6dCSto4dCS} fleshes out the left hand side of the diamond, connecting to four-dimensional Chern-Simons by first reducing, and then to IFT$_{2}$ by localisation.
Section \S\ref{Gauged LMP action section} describes the diamond in the context of the gauged LMP theory.
We conclude with a brief outlook in section \S \ref{sec:outlook}. Although the subject matter necessarily entails a degree of technical complexity we have endeavoured to keep the main presentation streamlined and complement this with a number of technical appendices.

\section{The ungauged WZW diamond}
\label{sec:ungaugedWZWdiamond}

In this section, we briefly describe the diamond correspondence of theories in which the two-dimensional theory is the WZW$_{2}$ CFT. This is a summary of some analysis first presented in \cite{Bittleston:2020hfv} which will serve to fix conventions and recapitulate key steps relevant to later sections.

\subsection{\texorpdfstring{hCS$_6$}{hCS6} with double poles}

We begin at the top of the diamond with 6d holomorphic Chern-Simons theory (hCS$_6$) whose fundamental field is an algebra-valued connection $\cA \in \Omega^{0,1}(\mathbb{PT}) \otimes \fg$. The six-dimensional action is given by
\begin{equation}\label{eq:hCS6}
S_{\mathrm{hCS}_6}[\cA ] = \frac{1}{2 \pi \rmi} \int_{\bPT} \Omega \wedge \Tr \Big( \cA \wedge \bar{\pd} \cA + \frac{2}{3} \cA \wedge \cA \wedge \cA \Big) \, ,
\end{equation}
in which we have introduced a meromorphic $(3,0)$-form $\Omega$. As a real manifold, there is an isomorphism $\bPT \cong \bR^{4} \times \bCP^{1}$ and we will introduce coordinates $x^{a \dot{a}} \in \bR^{4}$ and $\pi_{a} \in \bCP^{1}$. In these coordinates, the meromorphic $(3,0)$-form (which has two double poles at $\alpha_{a} , \beta_{a} \in \bCP^{1}$) is given by\footnote{Spinor contractions are defined to be $\langle \alpha \beta \rangle = \epsilon^{a b} \alpha_a \beta_b$, see appendix~\ref{sec:spinorconventions} for further details of spinor conventions.}
\begin{equation}\label{eq:Omega}
\Omega = \frac{1}{2} \Phi(\pi) \, \epsilon_{\dot{a}\dot{b}} \, \pi_{a} \dr x^{a \dot{a}} \wedge \pi_b \dr x^{b \dot{b}} \wedge \langle \pi \dr \pi \rangle ~, \qquad
\Phi = \frac{\langle \alpha \beta \rangle^2 }{\langle \pi \alpha \rangle^2 \langle \pi \beta \rangle^2} ~.
\end{equation}
The poles of $\Omega$ in $\bCP^{1}$ play the role of boundaries in hCS$_{6}$ because total derivatives pick up a contribution from $\bar{\pd} \Omega$ which is a distribution with support at these poles. To ensure a well defined variational principal, we impose boundary conditions on the gauge field at these poles given by
\begin{equation} \label{eq:bc1}
\cA \vert_{\pi = \alpha} = 0 \, , \qquad
\cA \vert_{\pi = \beta} = 0 \, .
\end{equation}
Turning to the symmetries of this model, the theory is invariant under gauge transformations acting as
\begin{equation} \label{eq:gt1}
\hat{\gamma}: \quad {\cal A} \mapsto ({\cal A} )^{\hat{\gamma}} = \hat{\gamma }^{-1} \mathcal{A} \hat{\gamma } + \hat{\gamma }^{-1} \bar{\pd} \hat{\gamma } \, ,
\end{equation}
so long as they preserve the boundary conditions. This amounts to restrictions on the allowed transformations at the poles of $\Omega$ which are given by
\begin{equation}
\pi^a \pd_{a \dot{a}} \hat{\gamma} \vert_{\pi = \alpha} = 0 \, , \quad
\pi^a \pd_{a \dot{a}} \hat{\gamma} \vert_{\pi = \beta} = 0 \, .
\end{equation}

\subsection{Localisation of \texorpdfstring{hCS$_6$}{hCS6} with double poles to \texorpdfstring{WZW$_{4}$}{WZW4}}

Surprisingly, all of the physical degrees of freedom in hCS$_{6}$ can be captured by a four-dimensional integrable field theory (IFT$_{4}$). This field theory is derived by localising the hCS$_{6}$ action, integrating out the $\bCP^{1}$ and landing on a theory on $\bR^{4}$. For the choice of meromorphic $(3,0)$-form $\Omega$ and boundary conditions given above, this 4d theory is WZW$_{4}$. This localisation is possible because of the substantial gauge symmetry in Chern-Simons theories. Indeed, the dynamical fields arise precisely where this gauge symmetry is broken, namely at the poles of $\Omega$. Fields capturing these degrees of freedom are known as `edge modes' which enter via the field redefinition
\begin{equation}
\mathcal{A} = ( \mathcal{A}' )^{\hat{g}} = \hat{g}^{-1} \mathcal{A}' \hat{g} + \hat{g}^{-1} \bar{\pd} \hat{g} \, .
\end{equation}
Expressing the action $S_{\mathrm{hCS}_6}[\cA]$ in terms of the fields $\cA'$ and $\hat g$ one obtains
\begin{equation}
\begin{aligned} \label{eq:hcsA'g'}
S_{\mathrm{hCS}_6}[\cA] & = S_{\mathrm{hCS}_6}[\cA^{\prime}]
+ \frac{1}{2 \pi \rmi} \int_{\bPT} \bar{\pd} \Omega \wedge \Tr \big( \cA^\prime \wedge \bar{\pd} \hat{g} \hat{g}^{-1} \big) \\
& \hspace{4em} - \frac{1}{6 \pi \rmi} \int_{\bPT\times[0,1]} \bar \partial\Omega \wedge \Tr \big(\hat{g}^{-1} \dr \hat{g} \wedge \hat{g}^{-1} \dr \hat{g} \wedge \hat{g}^{-1} \dr \hat{g} \big)~,
\end{aligned}
\end{equation}
where, with a slight abuse of notation, we are also denoting by $\hat g$ a smooth homotopy to a constant map in the last term (such abuse will be perpetuated later without further comment).
Notably, the edge mode $\hat{g}$ only appears in this action against the 4-form $\bar{\pd} \Omega$ which is a distribution with support at the poles of $\Omega$. This means that the action only depends on $\hat{g}$ through its value (and $\bCP^{1}$-derivative) at the poles of $\Omega$ which we will denote by
\begin{equation}
\hat{g} \vert_{\pi = \alpha} = g \ , \qquad
\hat{g}^{-1} \pd_{0} \hat{g} \vert_{\pi = \alpha} = u \ , \qquad
\hat{g} \vert_{\pi = \beta} = \tilde{g} \ , \qquad
\hat{g}^{-1} \pd_{0} \hat{g} \vert_{\pi = \beta} = \tilde{u} \ .
\end{equation}
Let us consider the symmetries of the theory in this new parameterisation. The gauge transformation \eqref{eq:gt1} acts trivially on $\cA^{\prime}$ whilst $\hat{g}$ transforms with a right-action as
\begin{equation}\label{eq:gt2}
\hat{\gamma}: \quad \mathcal{A}' \mapsto \mathcal{A}' \, , \qquad
\hat{g} \mapsto \hat{g} \hat{\gamma} \, .
\end{equation}
In addition, the new parameterisation has introduced a redundancy (which we dub an internal gauge symmetry) acting as
\begin{equation} \label{eq:gt3}
\check \gamma: \quad \cA' \mapsto \check \gamma^{-1}\cA' \check \gamma + \check \gamma^{-1}\bar \partial \check \gamma \, , \qquad
\hat g \mapsto \check \gamma^{-1}\hat g \, .
\end{equation}
We can exploit these symmetries to impose gauge fixing conditions on the fields $\cA^{\prime}$ and $\hat{g}$. Let us fix $\cA'$ such that it has no $\mathbb{CP}^1$-leg, and fix the value of $\hat{g}$ at $\pi = \beta$ to the identity.\footnote{At this point, we may further fix the $\mathbb{CP}^1$-derivative of $\hat{g}$ at both $\pi = \alpha$ and $\pi = \beta$ to zero. However, such terms drop out of the action in this ungauged case anyway without specifying this fixing.} The surviving edge mode at the other pole $g = \hat{g} \vert_{\pi = \alpha}$ will become the fundamental field of the WZW$_{4}$.

Returning to the action \eqref{eq:hcsA'g'}, the first term is a genuine six-dimensional bulk term which we eliminate by going on-shell.
The bulk equation of motion imposes holomorphicity of $\cA'$, which may be solved in terms of a series of $\bCP^{1}$-independent components $A'_{a \dot{a}}$ as
\begin{equation}
\cA' = \pi^a A'_{a \dot{a}} \bar{e}^{\dot{a}} \ , \qquad
\bar{e}^{\dot{a}} = \frac{\hat{\pi}_a \dr x^{a \dot{a}}}{\langle \pi \hat{\pi} \rangle } \, .
\end{equation}
In this expression, $\bar{e}^{\dot{a}}$ is a basis $(0,1)$-form on twistor space introduced in appendix~\ref{sec:TwistorSpace}.
This completely specifies the $\bCP^{1}$-dependence of $\cA^{\prime}$, and the boundary conditions eq.~\eqref{eq:bc1} may be solved to determine $A'_{a \dot{a}}$ in terms of $g$,
\begin{equation} \label{eq:A'}
A'_{a \dot{a}} = - \frac{\beta_a \alpha^b }{\langle \alpha \beta \rangle }\pd_{b \dot{a}} g g^{-1} \, .
\end{equation}
From these components, we can construct a 4d connection $A' = A'_{a \dot{a}} \dr x^{a \dot{a}}$, and this parameterisation of $A'$ in terms of $g$ is known in the literature as Yang's parameterisation ($g$ being called Yang's matrix).
This solution for $\cA^{\prime}$ may now be substituted into the action and the integral over $\bCP^{1}$ can be computed explicitly. The second and third term of \eqref{eq:hcsA'g'} localise to a four-dimensional action, and the detailed manipulations are presented in appendix~\ref{appendix on localisation formulae}. We land on the WZW$_{4}$ theory defined by
\begin{equation} \label{eq:WZW4}
\begin{aligned}
S_{\text{WZW}_{4}} & = \frac{1}{2} \int_{\bR^{4}} \, \Tr \big( g^{-1} \dr g \wedge \star g^{-1} \dr g \big)
+ \int_{\bR^{4} \times [0,1]} \omega_{\alpha , \beta} \wedge \mathcal{L}_{\text{WZ}}[g ] \ .
\end{aligned}
\end{equation}
In the second term, we have introduced a 2-form defined by
\begin{equation} \omega_{\alpha,\beta} =\frac{1}{\langle \alpha \beta \rangle } \alpha_a \beta_b \, \epsilon_{\dot a \dot b} \, \dr x^{a \dot a} \wedge \dr x^{b\dot b} \ ,
\end{equation}
and the WZ 3-form
\begin{equation}
\mathcal{L}_{\text{WZ}}[g] = \frac{1}{3} \Tr \big( \tilde{g}^{-1} \dr \tilde{g} \wedge \tilde{g}^{-1} \dr \tilde{g} \wedge \tilde{g}^{-1} \dr \tilde{g} \big) ,
\end{equation}
defined, as is usual, using a suitable extension $\tilde{g}$ of $g$. The equation of motions of this theory are given by
\begin{equation} \label{eq:WZW4EOM}
0 = \dr \left( \star - \omega_{\alpha , \beta} \wedge\right) \dr g g^{-1} \, \quad \Leftrightarrow\, \quad \epsilon^{\dot{a}\dot{b}} \beta^a \pd_{a\dot{a}}\left( \alpha^b \pd_{b \dot{b}} g g^{-1} \right) = 0 \, .
\end{equation}

The six-dimensional gauge transformations (constrained by boundary conditions) descend to semi-local symmetries of this action ($\gamma_L = \hat \gamma \vert_{\beta}$ and $ \gamma_R= \hat \gamma \vert_{\alpha}$) which act as
\begin{equation}
g \rightarrow \gamma_L^{-1} \cdot g\cdot \gamma_R \, , \qquad
\alpha^a \pd_{a \dot{a}} \gamma_R = 0 \, , \qquad
\beta^a \pd_{a \dot{a}} \gamma_L = 0 \ .
\end{equation}
Of particular interest is the case where $\beta = \hat{\alpha}$ (i.e.\ the poles of $\Omega$ are antipodal on $\mathbb{CP}^1$) in which case $\omega_{\alpha,\hat{\alpha}} = \varpi$ is proportional to the K{\"a}hler form on $\bR^4$. Here, we are referring to the K{\"a}hler form with respect to the complex structure $\mathcal{J}_\alpha$ which is defined\footnote{Recall that $\bR^4$ is a hyper-K{\"a}hler manifold which has a $\mathbb{CP}^1$'s worth of complex structures, see appendix~\ref{sec:spinorconventions}, eq.~\eqref{eq:complexStruct} .} by the point $\alpha \in \mathbb{CP}^1$. In this case, the semi-local symmetries can be interpreted as a holomorphic left-action and anti-holomorphic right-action (akin to the two-dimensional WZW current algebra).

\subsection{Interpretation as ADSYM} \label{Yang's Matrix Section}

A 4d Yang-Mills connection $A'$ with curvature $F[A'] = \dr A' + A' \wedge A'$ is said to be anti-self dual if it obeys $F = - {\star} F$. After converting to bi-spinor notation, the anti-self-dual Yang-Mills (ASDYM) equations can be expressed as
\begin{equation}\label{eq:ASDYM}
\pi^a \pi^b F_{a\dot{a} b \dot{b}} = 0 \, , \qquad \forall \, \pi_{a} \in \mathbb{CP}^1 \, .
\end{equation}
This contains three independent equations which can be extracted by introducing some basis spinors $\alpha_{a}$ and $\beta_{a}$ satisfying $\langle \alpha \beta \rangle \neq 0$. The three independent equations are then expressed in terms of contractions with these basis spinors as
\begin{align}
\label{eq:Faa} \alpha^a \alpha^b F_{a\dot{a} b \dot{b}} & = 0 \, ,\\
\label{eq:Fbb} \beta^a \beta^b F_{a\dot{a} b \dot{b}} & = 0 \, , \\
\label{eq:Fab} (\alpha^a \beta^b + \beta^a \alpha^b) F_{a\dot{a} b \dot{b}} & = 0 \, .
\end{align}
The six-dimensional origin of WZW$_4$ (and indeed all such constructed IFT$_{4}$) ensures that the connection $A'$ introduced in the previous section satisfies the ASDYM equation when evaluated on solutions to the WZW$_{4}$ equation of motion. This follows from the six-dimensional equation $\Omega \wedge \mathcal{F}[\mathcal{A}^{\prime}] = 0$ which encodes both the holomorphicity of $\cA^{\prime}$ and eq. \eqref{eq:ASDYM}. To see this explicitly for WZW$_4$ where the connection $A'$ in given by eq.~\eqref{eq:A'}, we note that the $\beta$-contracted eq.~\eqref{eq:Fbb} holds because $\langle \beta \beta \rangle = 0$, and the $\alpha$-contracted eq.~\eqref{eq:Faa} holds due to the Maurer-Cartan identity. The remaining eq.~\eqref{eq:Fab} yields
the equation of motion of WZW$_4$ \eqref{eq:WZW4EOM}.

\subsection{Reduction of \texorpdfstring{WZW$_{4}$}{WZW4} to \texorpdfstring{WZW$_{2}$}{WZW2}}\label{sec:wzw4wzw2}

Next, we will apply a two-dimensional reduction to WZW$_{4}$ specified by two vector fields $V_{i}$ on $\bR^{4}$ with $i = 1, 2$. The idea of reduction is to restrict to field configurations which are invariant under the flow of these vector fields. The two-dimensional dynamics of the reduced theory will be specified by the Lagrangian $\mathcal{L}_{\text{IFT}_2} = (V_1 \wedge V_2) \vee \mathcal{L}_{\text{IFT}_4}$ where $\mathcal{L}_{\text{IFT}_4}$ is the Lagrangian density of the parent theory and we denote the contraction of a vector field $V$ with a differential form $X$ by $V \vee X$.

Let us introduce a pair of unit norm spinors $\gamma_a$ and $\kappa_{\dot a}$ and define the basis of $1$-forms on $\bR^4$
\begin{equation}\label{eq:ComplexCoords}
\dr z = \gamma_a \kappa_{\dot a} \dr x^{a \dot a } ~, \qquad
\dr\bar{z} = \hat{\gamma}_a \hat{\kappa}_{\dot a} \dr x^{a \dot a } ~, \qquad
\dr w= \gamma_a \hat \kappa_{\dot a} \dr x^{a \dot a } ~, \qquad
\dr \bar{w} = - \hat \gamma_a \kappa_{\dot a} \dr x^{a\dot a } ~.
\end{equation}
These are adapted to the complex structure $\mathcal{J}_{\gamma}$ defined by $\gamma_a \in \bCP^{1}$. We choose to reduce along the vector fields dual to $\dr z$ and $\dr \bar z$ by demanding that $\pd_z g = \pd_{\bar{z} }g = 0$.\footnote{In this case for reality we have $\mathcal{L}_{\text{IFT}_2} = \rmi (\partial_z \wedge \partial_{\bar{z}}) \vee \mathcal{L}_{\text{IFT}_4}$. } Then, contracting the WZW$_{4}$ Lagrangian with these vector fields results in the two-dimensional action of a principal chiral model (PCM) plus Wess-Zumino (WZ) term:
\begin{equation} \label{eq:PCM+WZ}
S_{\text{PCM}+ \mathscr k \text{WZ}_2}[g] = \frac{1}{2 } \int_\Sigma \Tr \big( g^{-1}\dr g \wedge \star g^{-1}\dr g\big) ~ + ~ \frac{ \rmi \mathscr k }{3} \int_{\Sigma\times[0,1]} \Tr(\hat g^{-1}\dr \hat g\wedge \hat g^{-1}\dr \hat g\wedge \hat g^{-1}\dr \hat g) \, .
\end{equation}
In this action, the relative coefficient between the WZ-term and the PCM term is given by
\def\ab{\langle \alpha \beta \rangle}
\def\bg{\langle \beta \gamma \rangle}
\def\bgh{\langle \beta \hat \gamma \rangle}
\def\ag{\langle \alpha \gamma \rangle}
\def\ga{\langle \gamma \alpha \rangle}
\def\gb{\langle \gamma \beta\rangle}
\def\agh{\langle \alpha \hat \gamma \rangle}
\def\rma{\mathrm{a}}
\def\rmb{\mathrm{b}}
\def\aa{\mathsf{a}}
\def\bb{\mathsf{b}}
\begin{equation}
\mathscr k = \frac{ \alpha +\beta}{\alpha - \beta} \, , \qquad
\alpha = \frac{\ga}{\agh} \, , \qquad \,
\beta = \frac{\gb}{\bgh} \, .
\end{equation}

Varying the basis spinor $\gamma_{a}$ in these expressions changes the choice of reduction vector fields and interpolates between a family of two-dimensional theories.
The WZW$_{2}$ CFT limit is obtained when $\mathscr k \rightarrow 1$ with $ \alpha \beta $ held fixed. This can be achieved by starting at the K{\"a}hler point in 4d, with $\beta_{a} = \hat{\alpha}_{a}$, and choosing the reduction to be aligned with the complex structure, i.e.\ setting $\gamma_{a} = \alpha_{a}$. An alternative reduction which turns off the WZ term is achieved by setting $\beta = -\alpha $.

For general choices of reduction, the four-dimensional semi-local symmetries descend to a global $G_L\times G_R$ symmetry; this is because, for example, the conditions $\alpha^a \pd_{a \dot{a}} \gamma_R = 0$ and $\pd_z \gamma_R = \pd_{\bar{z}} \gamma_R = 0$ generically contain four independent constraints leaving only constant solutions. However, when the reduction is taken to the CFT point, this system of four constraints is not linearly independent, and chiral symmetries emerge satisfying $\pd_w \gamma_R = 0$ (and vice versa for $\gamma_L$).

\paragraph{Lax connection.}
A virtue of this approach is that a $\mathfrak{g}^\mathbb{C}$-valued Lax connection for the dynamics of the resultant IFT$_{2}$ may be derived from the 4d connection $A'$:
\begin{equation}\label{eq:LaxA}
\begin{aligned}
\mathscr{L}_{\bar{w}} &= \frac{1}{\langle \pi \hat{\gamma } \rangle } \hat{\kappa}^{\dot{a}} \pi^a (\pd_{a \dot{a}} + A'_{a \dot{a}} )= \pd_{\bar{w} } + \frac{\ (\beta-\zeta) }{ ( \alpha-\beta ) }\pd_{\bar{w} } g g^{-1} \, , \\
\mathscr{L}_{w} &= \frac{1}{\langle \pi \gamma \rangle } \kappa^{\dot{a}} \pi^a (\pd_{a \dot{a}} + A'_{a \dot{a}} )= \pd_w + \frac{\alpha (\beta-\zeta) }{ \zeta ( \alpha-\beta ) }\pd_w gg^{-1} \, ,
\end{aligned}
\end{equation}
where the spectral parameter is given as $\zeta = \frac{\langle \gamma\pi \rangle}{\langle \pi \hat{\gamma} \rangle}$. Flatness of this connection for all values of $\zeta$ invokes the field equation of the PCM + WZ theory
\begin{equation}
\alpha \pd_{\bar{w}} ( \pd_w g g^{-1} ) - \beta \pd_w ( \pd_{\bar{w} } g g^{-1} ) = 0 \, \quad \Leftrightarrow \quad \dr (\star- \rmi \mathscr k ) \dr g g^{-1 } = 0 \, .
\end{equation}
Notice that in the CFT limit $\mathscr k \rightarrow 1$ with $\beta \rightarrow \infty$, $\alpha \rightarrow 0$ the Lax connection becomes chiral and spectral parameter independent.

\subsection{Reduction of \texorpdfstring{hCS$_{6}$}{hCS6} to \texorpdfstring{CS$_{4}$}{CS4}}

Instead of first integrating over $\mathbb{CP}^1$ and then reducing to two dimensions, one could instead directly apply the reduction to hCS$_{6}$. This produces CS$_{4}$ with action
\begin{equation}
S_{\text{CS}_{4}}[A] = \frac{1}{2 \pi \rmi} \int_{\Sigma \times \mathbb{CP}^1} \omega \wedge \Tr \left( A \wedge \dr A + \frac{2}{3} A \wedge A \wedge A \right) \, .
\end{equation}
Here $\Sigma$ is the $\bR^2 \subset \bR^4$ with coordinates by $w, \bar{w}$, and the meromorphic 1-form $\omega$ is given by
\begin{equation}
\omega =\rmi (\pd_z \wedge \pd_{\bar{z}}) \vee \Omega \, .
\end{equation}
A crucial feature here is that this contraction introduces zeroes in $\omega$ to complement its poles, as required by the Riemann-Roch theorem. For the case at hand, $\omega$ is given explicitly by
\begin{equation}
\omega =\rmi \frac{\langle \alpha \beta \rangle^2 \langle\pi \gamma \rangle\langle\pi \hat \gamma \rangle
}{
\langle \pi \alpha \rangle^2
\langle \pi \beta \rangle^2} \langle \pi \dr \pi \rangle \, ,
\end{equation}
and the zeros are introduced at the points $\pi_{a} = \gamma_{a}, \hat{\gamma}_{a}$.
The details of the reduction show that, whilst our six-dimensional gauge field was regular, the connection $A$ entering in CS$_{4}$ develops poles at the zeros of $\omega$. In particular, the component $A_w$ will have a simple pole at $\pi_{a} = \gamma_{a}$ and $A_{\bar{w}}$ will have a simple pole at $\pi_{a} = \hat{\gamma}_{a}$. The four-dimensional Chern-Simons connection is subject to the same boundary conditions as its parent, namely it vanishes at the points $\alpha$ and $\beta$ in $\mathbb{CP}^1$. Subsequent localisation of CS$_{4}$ then gives the same PCM+WZ theory derived by reducing WZW$_{4}$.

\section{The gauged WZW diamond} \label{sec:gaugedWZWdiamond}

We now come to the main results of this paper. In this section, we will construct a diamond correspondence of theories which realises the gauged WZW$_{2}$ model, i.e.\ the $G/H$ coset CFT.

\subsection{Gauged WZW Models}
First let us review the gauging of the WZW model and the crucial Polyakov-Wiegmann identity.
Letting $G$ be a Lie group and $g \in C^{\infty}(\Sigma,G)$ a smooth $G$-valued field, the WZW$_2$ action is\footnote{To minimise factors of imaginary units we momentarily adopt Lorentzian signature.
Schematically, we have $S_{\text{Lorentz}} = \rmi S_{\text{Euclid}}|_{\star\to \rmi\star}$.}
\begin{equation}
S_{\text{WZW}_2}[g] = \frac{1}{2} \int_\Sigma \Tr_{\fg} \big( g^{-1}\dr g \wedge \star g^{-1}\dr g\big) ~ + ~ \frac{1}{3} \int_{\Sigma\times[0,1]} \Tr_{\fg} (\hat g^{-1} \dr \hat g \wedge \hat g^{-1} \dr \hat g \wedge \hat g^{-1} \dr \hat g)\, .
\end{equation}
Gauging a vectorial $H$-action of the principal chiral model term is straightforward. We introduce an $\fh$-valued connection $B \in \Omega^{1}(\Sigma) \otimes \fh$ transforming as
\begin{equation}
\ell \in C^{\infty}(\Sigma,H) : \quad B \mapsto \ell^{-1} B \ell + \ell^{-1} \dr \ell \, , \quad g \mapsto \ell^{-1} g \ell \, ,
\end{equation}
with field strength $F[B]= \dr B + B \wedge B$.
The principal chiral term is then gauged by
replacing the exterior derivatives with covariant derivatives $\dr g \to \Dr g = \dr g + [B, g]$.
Less trivially, the gauge completion of the WZ 3-form is \cite{Witten:1991mm, Figueroa-OFarrill:1994uwr,Figueroa-OFarrill:1994vwl,Figueroa-OFarrill:2005vws} 
\begin{equation} \label{eq:gWZWlag}
\mathcal{L}_{\text{gWZ}}[g, B ]= \mathcal{L}_{\text{WZ}}[g ] + \dr \, \Tr_\fg ( g^{-1} \dr g \wedge B + \dr g g^{-1} \wedge B + g^{-1} B g \wedge B ) \, .
\end{equation}
Adding these two pieces together gives the gauged WZW$_2$ action,
\begin{equation} \label{eq:gaugedwzw2}
S_{\text{gWZW}_2}[g,B] = S_{\text{WZW}_2}[g] + \int_\Sigma \Tr_\fg ( g^{-1}\dr g \wedge (1- \star) B + \dr g g^{-1}\wedge (1+\star)B + B \wedge \star B + g^{-1} B g \wedge (1 - \star)B )\, .
\end{equation}
Notice that chiral couplings between currents and gauge fields emerge from combinations of the PCM and WZ contributions.
A remarkable feature,
\begin{equation}
\mathcal{L}_{\text{WZ}}[g_1 g_2] = \mathcal{L}_{\text{WZ}}[g_1 ] + \mathcal{L}_{\text{WZ}}[ g_2] + \dr \, \Tr_\fg \left(\dr g_2 g_2^{-1} \wedge g_1^{-1}\dr g_1 \right) \, ,
\end{equation}
ensures that \eqref{eq:gaugedwzw2} can instead be cast as a difference of two WZW$_2$ models. To see this we choose a parameterisation of the gauge field $B$ in terms of two smooth $H$-valued fields
\begin{equation}
\label{eq:Btoab}
B = \frac{1+\star}{2} a^{-1} \dr a
+ \frac{1-\star}{2}b^{-1}\dr b \, , \quad
a,b\in C^{\infty}(\Sigma,H) \, .
\end{equation}
In two dimensions, this is not a restriction on the field content of the gauge field, but simply a way of parameterising the two independent components of $B$. With such a parameterisation, if we then further define $\tilde{g} = a g b^{-1} \in C^{\infty}(\Sigma,G)$ and $\tilde{h} = a b^{-1}\in C^{\infty}(\Sigma,H)$ the gauged model \eqref{eq:gaugedwzw2} can be written as the difference of two WZW$_2$ models:
\begin{equation}\label{eq:PW2d}
S_{\text{gWZW}_2}[g,B ] = S_{\text{WZW}_2}[\tilde{g}] - S_{\text{WZW}_2}[\tilde{h}] \,.
\end{equation}
This is known as the Polyakov-Wiegmann (PW) identity \cite{Polyakov:1983tt}.

\subsection{Gauging of the \texorpdfstring{WZW$_4$}{WZW4} model} \label{sec:gaugingWZW4}

Let us now consider the four-dimensional WZW$_4$ model, given by eq.~\eqref{eq:WZW4}.
The gauging procedure follows in the exact same manner, producing an analogous gauged WZW$_4$ action,
\begin{equation} \label{eq:gWZW4}
S_{\text{gWZW}_4}^{(\alpha,\beta)}[g,B] = \frac{1}{2} \int_{\mathbb{R}^4} \Tr (g^{-1} \nabla g \wedge \star g^{-1 }\nabla g )
+ \int_{\mathbb{R}^4 \times [0,1] } \omega_{\alpha \beta } \wedge \mathcal{L}_{\text{gWZ}}[g, B] \,.
\end{equation}
Here, we denote the covariant derivative by $\nabla g = \dr g + [B, g]$. A critical difference between two and four dimensions is the applicability of the PW identity, as was pointed out by \cite{Losev:1995cr}. In two dimensions, this mapping relies on the relation \eqref{eq:Btoab}. To extend it to four dimensions, we consider the operator on $1$-forms
\begin{equation}
{\cal J}_{\alpha,\beta} (\sigma ) = - \rmi \star (\omega_{\alpha,\beta } \wedge \sigma) \, .
\end{equation}
One may check that $ {\cal J}_{\alpha,\beta}^2 = -\id$, so that we can introduce useful projectors
\begin{equation}\label{eq:projectors}
P = \frac{1}{2}\left(\id - \rmi{\cal J} \right) \, \qquad \bar{P} = \frac{1}{2}\left(\id + \rmi{\cal J}\right) \ ,
\end{equation}
which furnish a range of identities detailed in appendix~\ref{sec:appendixprojectors}. With these in mind, we can write a four-dimensional analogue to \eqref{eq:Btoab}, 
\begin{equation}\label{eq:B4toab}
B = P \left( a^{-1} \dr a \right) + \bar{P} \left( b^{-1} \dr b \right) \, \quad a, b \in C^{\infty}(\mathbb{R}^4,H)\,.
\end{equation}
With this parameterisation of the gauge field, it is indeed possible to use the composite fields $\tilde g=a g b^{-1}\in C^{\infty}(\mathbb{R}^4,G)$ and $\tilde h=ab^{-1}\in C^{\infty}(\mathbb{R}^4,H)$ to express the gauged WZW$_4$ action in a fashion akin to eq.~\eqref{eq:PW2d} as
\begin{equation}
S_{\text{gWZW}_4}^{(\alpha,\beta)}[g,B] = S_{\text{WZW}_4}^{(\alpha,\beta)}[\tilde{g}] - S_{\text{WZW}_4}^{(\alpha,\beta)}[\tilde{h}] \, .
\end{equation}
However, unlike in two dimensions, this parameterisation of the gauge field eq.~\eqref{eq:B4toab} is not generic. It implies a restriction on the connection, namely that its curvature satisfies
\begin{equation} \label{eq:Bconstraint}
\alpha^a\alpha^b F_{a\dot a b\dot b}[B]=0\,,\quad \beta^{a}\beta^b F_{a \dot{a} b\dot{b}}[B] = 0\, .
\end{equation}
This can be thought of as analogue to imposing that $F$ be strictly a $(1,1)$-form (which indeed this becomes when $\beta =\hat{\alpha}$ and the WZW$_{4}$ is taken at the K{\"a}hler point). It is noteworthy that these constraints on the background gauge field agree with two of the three ASDYM equations; the same two equations that were identically satisfied by the Yang parameterisation of the connection $A'$. In the forthcoming analysis, we will see how this arises from the hCS$_{6}$ construction.

\subsection{A six-dimensional origin} \label{sec:6dorigin}

We now turn to the six-dimensional holomorphic Chern-Simons theory on twistor space that will descend to the above gauged WZW models in two and four dimensions.
Given the factorisation of gWZW$_2$ to the difference of WZW$_2$ models, a natural candidate here is to consider simply the difference of hCS$_6$ theories to generalize the six-dimensional action introduced in \cite{Kcostello, Bittleston:2020hfv, Penna:2020uky, Costello:2021bah}. Indeed, a similar idea was proposed by \cite{Stedman:2021wrw} in the construction of 2d coset models from the difference of CS$_4$ theories. However, how this should work in six dimensions is less clear as the factorisation of gWZW$_4$ requires the curvature of the gauge field to be constrained.

The fundamental fields of our theory are two connections $\cA \in \Omega^{0,1}(\mathbb{PT}) \otimes \fg$ and $\cB \in \Omega^{0,1}(\mathbb{PT}) \otimes \fh$, which appear in the six-dimensional action
\begin{equation} \label{eq:new6daction}
S_{\mathrm{ghCS_6}}[\cA , \cB] = S_{\mathrm{hCS_6}}[\cA] - S_{\mathrm{hCS_6}}[\cB]
- \frac{1}{2\pi \rmi} \int_{\bPT} \bar{\partial} \Omega \wedge \Tr \big( \cA \wedge \cB \big) \, ,
\end{equation}
where the functional $S_{\mathrm{hCS_6}}$ is defined in eq.~\eqref{eq:hCS6}. As well as the bulk hCS$_{6}$ functionals, we have also included a coupling term between the two connections which contributes on the support of $\bar{\pd} \Omega$, i.e.\ at the poles of $\Omega$. We will shortly provide a motivation for this boundary term related to the boundary conditions we will impose on the theory.

This definition is slightly imprecise; strictly speaking, the inner product denoted by `\Tr' should be defined separately for each algebra, i.e.~$\Tr_\fg$ and $\Tr_\fh$. In the coupling term, where $\mathcal{B}$ enters inside $\Tr_\fg$, we should first act on $\cB$ with some Lie algebra homomorphism from $\fh$ to $\fg$, and in principle this homomorphism could be chosen differently at each pole of $\Omega$. We discuss more general gaugings, beyond the vectorial gauging hereby considered, in appendix~\ref{Sec:General gaugings}.

To complete the specification of the theory, we must supply boundary conditions which ensure the vanishing of the boundary term in the variation of \eqref{eq:new6daction},
\begin{equation}
\delta S_{\mathrm{ghCS_6}} \big\vert_{\text{bdry}} = \frac{1}{2 \pi \rmi} \int_{\bPT} \bar{\partial} \Omega \wedge \Tr \big( (\delta \cA + \delta \cB) \wedge (\cA - \cB) \big) \, .
\end{equation}
Since $\bar{\partial} \Omega$ only has support at the poles of $\Omega$, the integral over $\bCP^{1}$ may be computed explicitly in this term. As well as contributions proportional to delta-functions on $\bCP^{1}$, this will also include $\bCP^{1}$-derivatives of delta-functions since the poles in $\Omega$ are second order. Using the localisation formula in the appendix~\ref{appendix on localisation formulae}, we find
\begin{equation}
\begin{aligned}
\delta S_{\mathrm{ghCS_6}} \big\vert_{\text{bdry}} = - \int_{\bR^{4}} \bigg[
& \frac{\alpha_{a} \beta_{b} \Sigma^{ab}}{\langle \alpha \beta \rangle} \wedge \Tr \big( (\delta \cA + \delta \cB) \wedge (\cA - \cB) \big) \\
& \qquad + \frac{1}{2} \alpha_{a} \alpha_{b} \Sigma^{ab} \wedge \pd_{0} \Tr \big( (\delta \cA + \delta \cB) \wedge (\cA - \cB) \big)
\bigg] + \alpha \leftrightarrow \beta \, .
\end{aligned}
\end{equation}
In this expression, we introduce a basis for the self-dual 2-forms defined by $\Sigma^{ab} = \varepsilon_{\dot{a} \dot{b}} \dr x^{a \dot{a}} \wedge \dr x^{b \dot{b}}$.
Let us also introduce an orthogonal decomposition of $\fg$ such that
\begin{equation}
\fg = \fh \oplus \fk \ , \qquad
\Tr \big( X \cdot Y \big) = \Tr \big( X^{\fh} \cdot Y^{\fh} \big) + \Tr \big( X^{\fk} \cdot Y^{\fk} \big) \ .
\end{equation}
To attain the vanishing of the boundary variation, we consider the boundary conditions
\begin{equation} \label{eq:ABboundarycond}
\cA^{\fk} \big\vert_{\alpha , \beta} = 0 \ , \qquad
\cA^{\fh} \big\vert_{\alpha , \beta} = \cB \big\vert_{\alpha , \beta} \ , \qquad
\pd_{0} \cA^{\fh} \big\vert_{\alpha , \beta} = \pd_{0} \cB \big\vert_{\alpha , \beta} \ .
\end{equation}
This completes our definition of the gauged hCS$_{6}$ theory.

One might choose to think of the boundary term in the variation as being a potential for a `symplectic' form\footnote{Precedent in the literature dictates that we denote the symplectic form as $\Omega$; we trust that context serves to disambiguate from the meromorphic differential $\Omega$.} \footnote{This is slightly loose as the 2-form is degenerate, so properly speaking we should restrict to symplectic leaves.}
\begin{equation}
\Theta = \delta S_{\mathrm{ghCS_6}} \big\vert_{\text{bdry}} \ , \qquad
\Omega = \delta \Theta = - \frac{1}{2 \pi \rmi} \int_{\mathbb{PT}} \bar{\partial} \Omega \wedge \Big( \Tr_\fg \big( \delta \cA \wedge \delta \cA \big) - \Tr_\fh \big( \delta \cB \wedge \delta \cB \big) \Big) \ ,
\end{equation}
such that our boundary conditions define a Lagrangian (i.e.\ maximal isotropic) subspace.
We should like to really interpret this as a symplectic form on an appropriate space of fields defined over $\mathbb{R}^{4}$.
Evaluating the integral over $\bCP^{1}$ and writing $\Omega = \Omega_{\cA} - \Omega_{\cB}$, this symplectic form is given by
\begin{equation}
\Omega_{\cA} = \int_{\bR^{4}} \bigg[ \frac{\alpha_{a} \beta_{b} \Sigma^{ab}}{\langle \alpha \beta \rangle} \wedge \Tr_{\fg} \big( \delta \cA \wedge \delta \cA \big) \big\vert_{\alpha} + \frac{1}{2} \alpha_{a} \alpha_{b} \Sigma^{ab} \wedge \pd_{0} \Tr_{\fg} \big( \delta \cA \wedge \delta \cA \big) \big\vert_{\alpha} \bigg] + \alpha \leftrightarrow \beta \ ,
\end{equation}
with an analogous expression for $\Omega_{\cB}$.
Because our boundary conditions are identical at each pole, we concentrate now only on the contributions associated to the pole at $\alpha$. The symplectic form is not sensitive to the entire field configuration $\cA \in \Omega^{1} (\bPT) \otimes \fg$, but rather to the evaluation of $\cA$ at the poles and its first derivatives,
\begin{equation}
\vec{\cA} = \big( \cA \vert_{\alpha} , \pd_{0} \cA \vert_{\alpha} \big) \ .
\end{equation}
This data may be interpreted as defining a 1-form (more precisely a $(0,1)$-form with respect to the complex structure defined by $\alpha$) on $\bR^{4}$ valued in the algebra\footnote{The dimension of $\vec{\fg}$ is $2 \dim(G)$, so it must be isomorphic to $\bR^{\dim(G)} \oplus \bR^{\dim(G)}$ as a vector space. The Lie algebra structure may be derived by considering consecutive infinitesimal gauge transformations. In the CS$_{4}$ literature, these structures have been studied under the name `defect Lie algebra' \cite{Benini:2020skc, Lacroix:2020flf}.} $\vec{\fg} = \fg \ltimes \bR^{\dim(G)}$. With this in mind, it is more accurate to say that the contribution from the pole at $\pi_{a} = \alpha_{a}$ in $\Omega$ is a symplectic form on the space of configurations
\begin{equation}
\big( \vec{\cA} , \vec{\cB} \big) \in \Omega^{0,1} (\bR^{4}) \otimes \big( \vec{\fg} \oplus \vec{\fh} \big) \ .
\end{equation}
This symplectic form may be succinctly written by introducing an inner product on the algebra $\vec{\fg} \oplus \vec{\fh}$, and our boundary conditions describe an isotropic subspace with respect to this inner product.\footnote{ This need not be the case, as our boundary conditions could generically intertwine constraints on the algebra and spacetime components, meaning they could not be captured by a subspace of the algebra alone. They would always, however, define an isotropic subspace of $\Omega^{0,1} (\bR^{4}) \otimes \big( \vec{\fg} \oplus \vec{\fh} \big)$ by definition. Examples of this more general type of boundary condition can be found in \cite{Cole:2023umd}.}

To be explicit we associate $\mathbb{R}^{\dim{G}}$ with the dual $\fg^\ast$ and denote the natural pairing of the algebra and its dual with $\llparenthesis\bullet , \bullet \rrparenthesis$. We let $\vec{X} = (x, \tilde{x}) \in \vec{\fg} $ such that bracket on $\vec{\fg}$ is defined by
\begin{equation}
[\vec{X}, \vec{Y}]_{\vec{\fg}} = ( [x,y] , \ad_x^\ast \tilde{y}-\ad^\ast_y \tilde{x} )\, ,
\end{equation}
where the co-adjoint action is $\llparenthesis x, \ad^\ast_y \tilde{x}\rrparenthesis = \llparenthesis[x,y], \tilde{x}\rrparenthesis $. We equip $\vec{\fg}$ with the inner product
\begin{equation}\label{eq:inner}
\langle \vec{X}, \vec{Y}\rangle_{\vec{\fg}} = \frac{\langle \beta \hat{\alpha} \rangle}{\langle \alpha \beta \rangle \langle \alpha \hat{\alpha} \rangle} \Tr_\fg(x \cdot y)+ \frac{1}{2} \Big( \llparenthesis x, \tilde{y}\rrparenthesis + \llparenthesis y,\tilde{x}\rrparenthesis \Big)
\, ,
\end{equation}
such that the relevant contribution to the symplectic form is given by
\begin{equation}
\Omega_{\cA} = \int_{\bR^{4}} \mu_\alpha \wedge \biprod{\delta \vec{\cA}}{\delta \vec{\cA}}_{\vec{\fg}} \ .
\end{equation}
where $\mu^\alpha= \alpha_{a} \alpha_{b} \Sigma^{ab} $ is the $(2,0)$-form defined by the complex structure associated to $\alpha \in \bCP^{1}$.

In a similar fashion we will let $\vec{U} = (u ,\tilde{u})$ and $\vec{V} = (v, \tilde{v})$ be elements of $\vec{\fh}$ which is equipped with a bracket and pairing via the same recipe. We consider the commuting direct sum $\vec{\fg} \oplus \vec{\fh}$ equipped with pairing and bracket
\begin{equation}
\biprodb{(\vec{X}, \vec{U})}{(\vec{Y}, \vec{V})} = \langle \vec{X}, \vec{Y}\rangle_{\vec{\fg}} - \langle \vec{U}, \vec{V}\rangle_{\vec{\fh}} \, , \qquad \commb{(\vec{X}, \vec{U})}{(\vec{Y}, \vec{V})} = \big( \comm{\vec{X}}{\vec{Y}}_{\vec{\fg}} \, , \comm{\vec{U}}{\vec{V} }_{\vec{\fh}} \big) \,,
\end{equation}
such that the total symplectic form coming from the pole at $\alpha$ is just
\begin{equation}
\Omega = \int_{\bR^{4}} \mu_\alpha \wedge \biprodb{(\delta \vec{\cA} , \delta \vec{\cB})}{(\delta \vec{\cA} , \delta \vec{\cB})} \ .
\end{equation}
Then, our boundary conditions can be expressed as $\big( \vec{\cA} , \vec{\cB} \big) \in \Omega^{0,1} (\bR^{4}) \otimes L$ where we introduce a subspace
\begin{equation}
L = \big\{ (\vec{X}, \vec{U}) \in \vec{\fg} \oplus \vec{\fh} \mid x = u \, , \ P_{\fh}^\ast \tilde{x} = \tilde{u} \big\} \ ,
\end{equation}
in which $P_{\fh}^\ast$ is the dual to the projector $P_\fh$ into the subgroup i.e.\ $\llparenthesis x, P_{\fh}^\ast \tilde{x}\rrparenthesis= \llparenthesis P_{\fh} x , \tilde{x}\rrparenthesis$. As $L$ is defined by $\dim \fg+ \dim \fh$ constraints, it is half-dimensional and it is also isotropic with respect to $\biprodb{\bcdot}{\bcdot}$, hence defining a Lagrangian subspace. Moreover, assuming that $G/H$ is reductive, $L$ is a subalgebra\footnote{If $\fg = \fh + \fk$ is not assumed to be reductive then the stabiliser of $L$ consists of elements of the form 
\begin{equation*}
\textrm{stab}_L = \big\{ (\vec{X}, \vec{U}) \in \vec{\fg} \oplus \vec{\fh} \mid x= u \, , \ P_{\fh}^\ast \tilde{x} = \tilde{u} \, , \ [u, \fk]= 0 \, , \ ([\fh,\fk ] , \tilde{u})=0 \big\} \, .
\end{equation*}
\vspace{-1.75em}
}. Pre-empting the following section, this analysis indicates that there will be a residual $\vec{\fh}$ gauge symmetry associated to the pole at $\alpha$, and similarly at $\beta$.

We can make one further observation\footnote{We thank A. Arvanitakis for this suggestion.} of the role of the boundary contribution from a symplectic perspective that is best illustrated by a finite-dimensional analogy. Recall that the cotangent bundle ${\cal M} = T^*X$ is a symplectic manifold; if we let $\{x^i\}$ be local coordinates on $X$ and $\{ \xi_i\}$ the components of a 1-form $\xi = \xi_i dx^i \in T_x^*X$, then $p=(x^i, \xi_i)$ provide local coordinates for ${\cal M}$ in terms of which the canonical symplectic form is $\Omega= \dr \xi_i \wedge \dr x^i$. The tautological potential (which admits a coordinate free definition in terms of the projection $\pi: T^*X \rightarrow X$) for this is given by $\Theta = \xi_i \dr x^i$.   The zero section, i.e. points $p=(x^i, \xi_i= 0)$ of $T^*X$ is a Lagrangian and notice that $\Theta$ vanishes trivially here.   Now Weinstein's tubular neighbourhood theorem ensures that in the vicinity of a Lagrangian $L$,  any symplectic manifold ${\cal M}$ locally looks like $T^*L$ with $L$ given by the zero section.    In the case at hand, our boundary conditions are of the schematic form $\xi = \cA - \cB  = 0 $, and the effect of including the specific boundary contribution to the Lagrangian ensures that the resultant symplectic potential is the tautological one.

To close this section, let us comment that at the special point for which $\alpha_{a} = \hat{\beta}_{a}$, one of the terms in the inner product eq.~\eqref{eq:inner} vanishes. This allows for a larger class of admissible boundary conditions, even for the ungauged model, including the examples
\begin{equation}\label{natd}
\cA \big\vert_{\hat{\alpha}} = 0 \ , \qquad
\pd_{0} \cA \big\vert_{\alpha} = 0 \ \qquad \textrm{or} \qquad
\pd_{0} \cA \big\vert_{\hat{\alpha}} = 0 \ , \qquad
\pd_{0} \cA \big\vert_{\alpha} = 0 \, .
\end{equation}
We leave these for future development.

\subsection{Localisation to \texorpdfstring{gWZW$_{4}$}{gWZW4}}
\label{sec:localtogwzw4}

The localisation procedure follows in a similar fashion to the ungauged model. However, given that there are now two gauge fields $\cA$ and $\cB$, some care is required to account for degrees of freedom and residual symmetries.

We introduce a new pair of connections $\cA'\in \Omega^{0,1}(\mathbb{PT}) \otimes \fg$ and $\cB'\in \Omega^{0,1}(\mathbb{PT}) \otimes \fh$, along with group valued fields $\hat g \in C^{\infty}(\mathbb{PT}, G)$ and $\hat h \in C^{\infty}(\mathbb{PT}, H)$ related to the original gauge fields by
\begin{equation}
\begin{aligned} \label{eq:redundancyredux}
\cA = \hat{g}^{-1} \cA^\prime \hat{g} + \hat{g}^{-1} \bar{\pd} \hat{g} \equiv \cA^\prime{}^{\hat{g}} ~,\\
\cB = \hat{h}^{-1} \cB^\prime \hat{h} + \hat{h}^{-1} \bar{\pd} \hat{h} \equiv \cB^\prime{}^{\hat{h}} ~.
\end{aligned}
\end{equation}
The redundancy in this parameterisation is given by the action of $\check \gamma \in C^{\infty}(\mathbb{PT},G)$ and $\check \eta \in C^{\infty}(\mathbb{PT},H)$:
\begin{align}
\label{eq:intgamma}
& \cA' \mapsto \check \gamma^{-1} \cA' \check \gamma + \check \gamma^{-1}\bar \partial \check \gamma \,, \qquad
\hat g \mapsto \check \gamma^{-1} \hat g \,,\\
\label{eq:inteta}
& \cB' \mapsto \check \eta^{-1} \cB' \check \eta + \check \eta^{-1}\bar \partial \check \eta \,, \qquad \hspace{0.3em}
\hat h \mapsto \check \eta^{-1} \hat h \,,
\end{align}
which leave $\cA$ and $\cB$ invariant. As before, this is partially used to fix away the $\mathbb{CP}^1$ legs
\begin{equation}
\label{eq:noCP1constraint}
\cA_0^\prime = \cB_0^\prime = 0 \,.
\end{equation}
The localisation procedure will produce a four-dimensional boundary theory with fields given by the evaluations of $\hat g, \hat h$ and their $\mathbb{CP}^1$-derivatives at the poles $\alpha$ and $\beta$ of $\Omega$. Since the $\mathbb{CP}^1$-derivatives will have an important role, we give them names, 
\begin{equation}
\label{ec:uandv}
\hat u = \hat g^{-1} \partial_0 \hat g \,,\quad \hat v = \hat h^{-1} \partial_0 \hat h \,.
\end{equation}
After fixing \eqref{eq:noCP1constraint}, we note that there is still some remaining symmetry given by internal gauge transformations \eqref{eq:intgamma} and \eqref{eq:inteta} which are $\mathbb{CP}^1$-independent. We use this residual symmetry to fix the values
\begin{equation}
\label{eq:ghatinf}
\hat g|_{\beta}=\mathrm{id} \,, \quad \hat h|_{\beta}=\mathrm{id}\,.
\end{equation}
On the other hand, the action \eqref{eq:new6daction} is invariant under gauge transformations acting on $\cA$ and $\cB$ which preserve the boundary conditions \eqref{eq:ABboundarycond}. These are given by smooth maps $\hat \gamma \in C^{\infty}(\mathbb{PT},G)$ and $\hat \eta \in C^{\infty}(\mathbb{PT},H)$ satisfying\footnote{This requires that $G/H$ is reductive meaning $[\fh, \fk] \subset \fk$.}
\begin{equation}
\hat{\gamma }\vert_{\alpha, \beta} = \hat{\eta} \vert_{\alpha, \beta} \,, \quad \partial_0 \hat{\gamma }\vert_{\alpha, \beta} = \partial_0 \hat{\eta} \vert_{\alpha, \beta} \,.
\end{equation}
The induced action of these gauge transformations on the new field content is
\begin{align}
& \quad \cA' \mapsto \cA' \, , \qquad
\hat g \mapsto \hat g \hat \gamma \, , \qquad
\hat u \mapsto \hat \gamma^{-1} \hat u \hat \gamma + \hat \gamma^{-1}\partial_0 \hat \gamma \\
&\quad \cB' \mapsto \cB' \, , \qquad \hspace{0.25em}
\hat h \mapsto \hat h \hat \eta\, , \qquad
\hat v \mapsto \hat \eta^{-1} \hat v\hat \eta + \hat \eta^{-1}\partial_0 \hat \eta\,.
\end{align}
We want to use this symmetry to further fix degrees of freedom. Note that whereas the right action on the fields $\hat g$ and $\hat h$ at $\alpha$ is entirely unconstrained, we would like the action at $\beta$ to preserve the gauge fixing condition \eqref{eq:ghatinf}. This is achieved by performing both an internal and external gauge transformation simultaneously, and requiring $\hat \gamma|_{\beta}=\check \gamma$ and $\hat \eta|_{\beta}=\check \eta$. This results in an induced left action on the fields $\hat g$ and $\hat h$ at $\alpha$. In summary, introducing some notation for simplicity, we have our boundary degrees of freedom
\begin{align}
& \hat g|_{\alpha} \coloneqq g \,, \qquad \hat g|_{\beta} = \mathrm{id}\,,\qquad \hat u|_{\alpha}\coloneqq u \,,\qquad \hat u|_{\beta}\coloneqq \tilde u \\
&
\label{eq:gaugefixh1}\hat h|_{\alpha} \coloneqq h \,, \qquad \hat h|_{\beta} = \mathrm{id}\,,\qquad \hat v|_{\alpha}\coloneqq v \,,\qquad \hat v|_{\beta}\coloneqq \tilde v \,,
\end{align}
and boundary gauge transformations
\begin{align}
& \hat \gamma \vert_{\alpha} = \hat \eta \vert_{\alpha} \coloneqq r \, , \qquad \quad \hspace{0.2em}
\hat{\gamma}^{-1} \partial_0 \hat \gamma \vert_{\alpha} = \hat{\eta}^{-1} \partial_0 \hat \eta \vert_{\alpha} \coloneqq \epsilon \ , \\
& \hat \gamma \vert_{\beta} = \hat \eta \vert_{\beta} \coloneqq \ell^{-1} \,,\qquad
\hat{\gamma}^{-1} \partial_0 \hat \gamma|_{\beta} = \hat{\eta}^{-1} \partial_0 \hat \eta|_{\beta} \coloneqq \tilde \epsilon \ ,
\end{align}
which act on the boundary fields as
\begin{align}
& g \mapsto \ell g r \,,\quad u \mapsto r^{-1}ur+\epsilon \,, \quad \tilde u \mapsto \ell \tilde u \ell^{-1}+\tilde \epsilon \\
&h \mapsto \ell h r \,,\quad v \mapsto r^{-1}vr+\epsilon \,, \quad \tilde v \mapsto \ell \tilde v \ell^{-1}+\tilde \epsilon\,.
\end{align}
with $\ell,r \in C^{\infty}(\mathbb{R}^4,H)$ and $\epsilon, \tilde \epsilon \in C^{\infty}(\mathbb{R}^4, \fh)$. Based on our expectation of a gauge theory containing a $G$-valued field and a vectorial $H$-gauge symmetry, we use the above symmetries to fix
\begin{equation} \label{eq:gaugefixh2}
h = \text{id} \, , \qquad v = \tilde{v} = 0\,.
\end{equation}
We are thus left with a residual symmetry $r = \ell^{-1}$ acting as
\begin{eqnarray}
g \mapsto \ell g \ell^{-1} \, , \quad u \mapsto \ell u \ell^{-1} \, , \quad \tilde{u} \mapsto \ell \tilde{u} \ell^{-1} \, , \quad B \mapsto \ell B \ell^{-1} - \dr \ell \ell^{-1} \, ,
\end{eqnarray}
which will become the gauge symmetry of our 4d theory.

We now proceed with the localisation of the six-dimensional action.
As with the ungauged model, the first step is to write the action in terms of $\cA', \cB'$ and $\hat{g}, \hat{h}$. Given that the localisation formula \eqref{eq:localisationformula} introduces at most one $\partial_0$ derivative, all dependence on $\hat h$ will drop due to our gauge fixing choices \eqref{eq:gaugefixh1} and \eqref{eq:gaugefixh2}. Hence, there will be no contribution from $S_{\mathrm{hCS}_6}[\cB]$ to the four-dimensional action.
As per eq.~\eqref{eq:hcsA'g'} we find that the bulk equations (i.e. contributions to the variation of the action that are not localised to the poles of $\Omega$) enforce $\bar \partial_0 \cA'_{\dot a}=\bar \partial_0 \cB'_{\dot a}= 0$. This implies that the components $\cA'_{\dot a},\cB'_{\dot a}$ are holomorphic, which (combined with the fact that they have homogeneous weight 1) allows us to deduce that
\begin{equation}
\cA'_{\dot a} = \pi^a A'_{a\dot a} \, , \quad \cB'_{\dot a} = \pi^a B'_{a\dot a} \, ,
\end{equation}
in which $A'_{a\dot a},B'_{a\dot a}$ are $\mathbb{CP}^1$-independent. Imposing this bulk equation, and the gauge fixings described above, the remaining contributions in \eqref{eq:new6daction} are given by
\begin{equation} \label{eq:almostlocalised}
\begin{aligned}
& S_{\text{ghCS}_6}[\cA , \cB ] = \frac{1}{2 \pi \rmi} \int_{\bPT} \bar{\pd} \Omega \wedge \Tr \big( \cA^\prime \wedge \bar{\pd} \hat{g} \hat{g}^{-1} - (\hat g^{-1} \cA^{\prime} \hat{g} + \hat{g}^{-1} \bar \partial \hat{g})\wedge \cB^{\prime} \big) \\
& \hspace{9em} - \frac{1}{6 \pi \rmi} \int_{\bPT \times[0,1]} \bar \partial\Omega \wedge \Tr \big(\hat{g}^{-1} \dr \hat{g} \wedge \hat{g}^{-1} \dr \hat{g} \wedge \hat{g}^{-1} \dr \hat{g} \big) \, .
\end{aligned}
\end{equation}

In the ungauged model, the next step was to solve the boundary conditions for $\cA'$ in terms of $\hat{g}$. Here, the boundary conditions do not fully determine $A'_{a\dot a}, B'_{a\dot a}$ and instead relate them as\footnote{The boundary conditions on the $\mathbb{CP}^1$ derivatives of the gauge fields impose
\begin{equation}
\frac{\alpha^{a}}{\langle \alpha \beta \rangle} \big( \nabla_{a \dot{a}} g g^{-1} \big)^{\fh} = - \beta^{a} \nabla_{a \dot{a}} \tilde{u}^{\fh} \ , \qquad
\frac{\beta^{a}}{\langle \alpha \beta \rangle} \big( g^{-1} \nabla_{a \dot{a}} g \big)^{\fh} = - \alpha^{a} \nabla_{a \dot{a}} u^{\fh} \ , 
\end{equation}
however we will not invoke these since they will follow as equations of motion of the 4d theory due to the addition of the boundary term in the gauged hCS$_6$ action~\eqref{eq:new6daction}.
For more details see appendix~\ref{Sec:General gaugings}.}
\begin{equation}
\label{eq:bcsol1}
A'_{a \dot{a}} = B'_{a\dot a} + \Theta_{a\dot a}\coloneqq B'_{a\dot{a}} - \frac{1}{\langle \alpha \beta \rangle} \beta_a \alpha^b \nabla_{b \dot{a}} g g^{-1}\,,
\end{equation}
where the covariant derivative is given by $\nabla_{a\dot a} g g^{-1}=\partial_{a\dot a}g g^{-1} + B'_{a\dot a} - \mathrm{Ad}_g B'_{a\dot a}$.
Equation \eqref{eq:bcsol1} allows us to express \eqref{eq:almostlocalised} entirely in terms of $\cB'$, $\Theta = \pi^a \Theta_{a \dot a} \bar e^{\dot a}$ and $\hat g$. Many of the terms combine to produce a gauged Wess-Zumino Lagrangian contribution (eq.~\eqref{eq:gWZWlag}) with the result
\begin{equation}
S_{\mathrm{ghCS}_6}[\cA , \cB ] = \frac{1}{2 \pi \rmi} \int_{\bPT} \bar{\pd} \Omega \wedge \Tr \big(\Theta \wedge \left( \nabla \hat{g} \hat{g}^{-1} - \cB^\prime \right)\big) - \frac{1}{2 \pi \rmi} \int_{\bPT \times[0,1]} \bar \partial\Omega \wedge \mathcal{L}_{\mathrm{gWZ}}[\hat g,\cB'] \,.
\end{equation}
Given that both $B_{a\dot{a}} $ and $\Theta_{a\dot{a}}$ are $\mathbb{CP}^1$-independent, we have that
\begin{eqnarray}
\int_{\bPT} \bar{\pd} \Omega \wedge \Tr (\Theta \wedge \cB^\prime) = 0 \, ,
\end{eqnarray}
with cancelling contributions from the two end points of the integral. Hence we are left with a manifestly covariant result \begin{equation}
S_{\text{ghCS}_6}[\cA , \cB] = \frac{1}{2 \pi \rmi} \int_{\bPT} \bar{\pd} \Omega \wedge \Tr \big( \Theta \wedge \left( \nabla \hat{g} \hat{g}^{-1} \right)
- {\cal L}_{\mathrm{gWZ}}[\hat{g},\cB^\prime ] \big) \, .
\end{equation}
Application of the localisation formula in the appendix~\eqref{ec:localisationformula} yields the four-dimensional action
\begin{equation}\label{4d action II}
\begin{split}
S_{\text{IFT}_{4}} =& \frac{1}{2} \int_{\bR^{4}} \Tr \big(
\nabla g g^{-1} \wedge {\star} \nabla g g^{-1} \big) + \int_{\bR^{4} \times [0,1]} \omega_{\alpha , \beta} \wedge \cL_{\text{gWZ}} [g, B'] \\ &\qquad - \int_{\bR^{4}}
\mu_\alpha \wedge \Tr ( u \cdot F[B'] ) + \mu_\beta \wedge \Tr ( \tilde{u} \cdot F[B'] )
\ .
\end{split}
\end{equation}
At this point only the $\fh$-components of $u$ and $\tilde u$ contribute to the action, and so henceforth, to ease notation and without loss of generality, we set their $\fk$-components to zero. 

Something rather elegant has occurred; we have found the localisation of the six-dimensional theory returns not only the gauging of the WZW$_{4}$ model, but also residual edge modes serving as Lagrange multipliers that constrain the field strength to obey exactly those conditions of eq.~\eqref{eq:Bconstraint} which ensure the theory can be written as the difference of WZW$_{4}$ models.
The constraints $F^{2,0}=0$ and $F^{0,2}=0$ have also been imposed by Lagrange multipliers in the context of 5d K{\"a}hler Chern-Simons theory \cite{Nair:1990aa,Nair:1991ab}. This theory bears a similar relationship to WZW$_{4}$ as 3d Chern-Simons theory bears to WZW$_{2}$.   This poses a natural question:   what is the direct relationship between this 5d K{\"a}hler Chern-Simons theory  and 6d holomorphic Chern-Simons theory?  We suspect the mechanism here is rather similar to that which relates  CS$_{4}$ and CS$_{3}$  \cite{Yamazaki:2019prm}; we comment further on this in the outlook.

\subsection{Equations of motion and ASDYM} \label{sec:EOMandASDYM}

Making use of the projectors previously introduced in eq.~\eqref{eq:projectors}, the equations of motion then read
\begin{equation}
\label{ec:4deom}
\begin{aligned}
\delta B': &\quad 0 = \bar{P} \nabla g g^{-1}\vert_{\fh} - P g^{-1} \nabla g \vert_{\fh} + \star \left( \mu_\alpha \wedge \nabla u+ \mu_\beta \wedge \nabla \tilde{u} \right) \ , \\
\delta g: &\quad 0 = \nabla \star \nabla g g^{-1} - \omega_{\alpha , \beta}\wedge \nabla (\nabla g g^{-1}) + 2 \omega_{\alpha , \beta}\wedge F[B'] \ , \\
\delta u: &\quad 0 = \mu_\alpha \wedge F[B'] \ , \\
\delta \tilde{u}: &\quad 0 = \mu_\beta \wedge F[B'] \ . \\
\end{aligned}
\end{equation}
We can exploit the projectors to extract from the $B'$ equation of motion the two independent contributions:
\begin{equation}
\delta B': \quad \begin{aligned}
0 = \bar{P} \left( \nabla g g^{-1} \vert_{\fh} + \star( \mu_\beta \wedge \nabla \tilde{u} ) \right) \\
0 = P \left( g^{-1}\nabla g \vert_{\fh} - \star( \mu_\alpha\wedge \nabla u ) \right) \label{eq:Beqm}
\end{aligned} \ . 
\end{equation}
In fact, these are exactly the conditions that arise from the $\mathbb{CP}^1$ derivative components of the boundary condition $\pd_0 \cA^\fh \vert_{\alpha, \beta} = \pd_0 \cB \vert_{\alpha, \beta}$.

Making use of the identity
\begin{equation}
\nabla( \omega_{\alpha \beta }\wedge \star( \mu_\beta \wedge \nabla \tilde{u} ) ) = \nabla( \mu_\beta \wedge \nabla \tilde{u} ) = \mu_\beta \wedge F[B'] \cdot \tilde{u} \ , 
\end{equation}
we obtain an on-shell integrability condition for the first of eq.~\eqref{eq:Beqm}, namely that
\begin{equation}
\nabla ( \omega_{\alpha \beta }\wedge \bar{P} (\nabla g g^{-1} \vert_{\fh} ) ) = 0 \, .
\end{equation}
Hence, using the projection of the $\delta g$ equation of motion into $\fh$, we have that $\omega_{\alpha,\beta} \wedge F[B']= 0$ follows on-shell.

Let us turn back to the ASDYM equations which we can recast as
\begin{equation}
\mu_\alpha \wedge F = 0 \, , \quad \mu_\beta \wedge F = 0\, , \quad \omega_{\alpha, \beta} \wedge F = 0 \, .
\end{equation}
In differential form notation, the solution of the boundary condition eq.~\eqref{eq:bcsol1} can be written as
\begin{equation}
A' = B' - \bar{P}( \nabla g g^{-1}) \, .
\end{equation}
By virtue of the identities obeyed by the projectors, eq. \eqref{eq:projectorid}, and the covariant Maurer-Cartan identity obeyed by $R^\nabla = \nabla g g^{-1} $,
\begin{equation}
\nabla R^\nabla - R^\nabla \wedge R^\nabla = (1-\Ad_g) F[B'] \, ,
\end{equation}
one can readily establish
\begin{eqnarray}
\mu_\beta \wedge F[A'] &=& \mu_\beta \wedge F[B'] \, , \\
\mu_\alpha \wedge F[A'] &=&\mu_\alpha \wedge \Ad_g F[B'] \, , \\
2\omega_{\alpha \beta } \wedge F[A'] &=& 2 \omega_{\alpha \beta } \wedge F[B'] + 2 \omega_{\alpha \beta } \wedge \nabla \bar{P} (R^\nabla ) \\
&=& 2 \omega_{\alpha \beta } \wedge F[B'] - \nabla (\star \nabla g g^{-1}) + \omega_{\alpha \beta } \wedge \nabla (\nabla g g^{-1}) \, .
\end{eqnarray}
Hence we conclude that the $\delta g, \delta u, \delta \tilde{u}$ equations of motion are equivalent to the ASDYM equations for the connection $A'$. Demanding that the $B'$ connection is also ASD requires in addition that $\omega_{\alpha,\beta}\wedge F[B']= 0$, and as shown above this is indeed a consequence of the $B'$ equations of motion.

\subsection{Constraining then reducing}

We now proceed to the bottom of the diamond by reduction of the IFT$_4$. In this section, we shall first implement the constraints imposed by the Lagrange multipliers $u,\tilde{u}$ in the 4d theory and then reduce.
While not the most general reduction, this will allow us to directly recover the gauged WZW coset CFT.
In section \ref{sec:newIFT}, we will investigate more general reductions, in particular what happens if we reduce without first imposing constraints.

Imposing the reduction ansatz that $\pd_z = \pd_{\bar{z}} =0$ in the complex coordinates of eq.~\eqref{eq:ComplexCoords}, we have that the solution to the constraints $B= P(a^{-1} \dr a) + \bar{P}(b^{-1} \dr b) $ becomes
\begin{equation}\label{eq:Bincomplex}
\begin{aligned}
B' = B'_{a \dot{a}} \dr x^{a \dot{a}}&=\frac{1}{\alpha - \beta } \left(\alpha
b^{-1}\partial_{w} b - \beta a^{-1} \partial_{w} a \right) \, \dr w - \frac{1}{\alpha - \beta } \left(\beta b^{-1} \partial_{\bar{w}} b - \alpha a^{-1} \partial_{\bar{w}} a \right) \, \dr \bar{w} \\
& \qquad + \frac{1}{\alpha - \beta } \left( b^{-1}\pd_{\bar{w}} b - a^{-1}\pd_{\bar{w}}a \right) \dr z + \frac{\alpha \beta }{\alpha -\beta } \left( b^{-1}\pd_{w} b - a^{-1}\pd_{w} a \right) \dr \bar{z} \ .
\end{aligned}
\end{equation}
For simplicity, let us first consider the K\"ahler point and align the reduction to the complex structure (implemented simply by taking $\alpha \rightarrow 0$ and $\beta \rightarrow \infty$). In this scenario, the reduction ansatz enforces that $B'_z = 0$ and $B'_{\bar{z}} = 0$ with the remaining components of $B'$ parameterising a generic two-dimensional gauge field. Effectively, we can simply ignore the constraints altogether but impose $B'_z =0 $ and $B'_{\bar{z}} =0 $ as part of the specification of a reduction ansatz. This could be interpreted as demanding $D_z = D_{\bar{z}} = 0 $ acting on fields. In this case it is immediate that the 4d gauged WZW reduces to a 2d gauged WZW.

Away from the K\"ahler point and aligned reduction, i.e. not fixing $\alpha$ and $\beta$, one must keep account of contributions coming from $B'_z$ and $B'_{\bar{z}}$. We can still view $B'_w $ and $B'_{\bar{w}}$ components of eq.~\eqref{eq:Bincomplex} as a parametrisation of a generic 2d gauge field, but there is no way in which we can view the $B'_z$ and $B'_{\bar{z}}$ as a local combination of the $B'_w$ and $B'_{\bar{w}}$. We forced to work with the variables $a$ and $b$ rather than a 2d gauge field. Fortunately, however, the reduction can still be performed immediately if we use the composite fields $\tilde g=a g b^{-1}$ and $\tilde h=ab^{-1}$.
These composite variables are invariant under the $\fh$-gauge symmetry, but a new would-be-affine symmetry emerges under $a \rightarrow \ell a$, $b \rightarrow b r^{-1}$ with $\alpha^b \partial_{b \dot{b}} r = \beta^{b}\partial_{b \dot{b}} \ell = 0$. These leave $B', g, h$ invariant but act as $\tilde{g} \rightarrow \ell\tilde{g} r $ and $\tilde{h} \rightarrow \ell \tilde{h} r $. At the K\"ahler point and aligned reduction, these symmetries descend to affine symmetries, but in general descend only to global transformations. Recall that the 4d gWZW becomes
\begin{equation}
S_{\text{gWZW}_4}^{(\alpha,\beta)}[g,B'] = S_{\text{WZW}_4}^{(\alpha,\beta)}[\tilde{g}] - S_{\text{WZW}_4}^{(\alpha,\beta)}[\tilde{h}] \, .
\end{equation}
It is then immediate that this reduces to the difference of PCM+WZ theories of eq.~\eqref{eq:PCM+WZ} with WZ coefficient $\mathscr{k}$:
\begin{equation}
S_{ \textrm{IFT}_{2}}[ \tilde{g} ,\tilde{h}] = S_{\text{PCM}+ \mathscr k \text{WZ}_2}[\tilde{g}] - S_{\text{PCM}+ \mathscr k \text{WZ}_2}[\tilde{h} ]\,.
\end{equation}
Away from the CFT point, $\mathscr{k} = 1$, this cannot be recast in terms of a deformation of the gauged WZW expressed as a local function of $B',g$.

\paragraph{Lax formulation.}
To obtain the Lax of the resultant IFT$_{2}$ we first note that the four-dimensional gauge fields, upon solving the constraints on $B'$, are gauge equivalent to
$$
A'_{a\dot{a}} = -\frac{1}{\ab} \beta_a \alpha^b \pd_{b\dot{a}} \tilde{g} \tilde{g} ^{-1} \, , \quad B'_{a\dot{a}} = -\frac{1}{\ab} \beta_a \alpha^b \pd_{b\dot{a}} \tilde{h} \tilde{h} ^{-1}\,.
$$
Thus, we may simply follow the construction of the Lax from the ungauged model of eq.~\eqref{eq:LaxA}, with the connection $A'$ producing a Lax for the $S_{\text{PCM}+ \mathscr k \text{WZ}_2}[\tilde{g}]$ and the $B'$ producing one for $S_{\text{PCM}+ \mathscr k \text{WZ}_2}[\tilde{h}]$.

\section{More General \texorpdfstring{IFT$_{2}$}{IFT2} from \texorpdfstring{IFT$_{4}$}{IFT4}: Reducing then Constraining}
\label{sec:newIFT}

In the previous section, we reduced from the gauged WZW$_{4}$ model to an IFT$_2$, but prior to reduction we enforced the constraints imposed by the Lagrange multiplier fields.
These constraints determine implicit relations between the components of the gauge field as per eq.~\eqref{eq:Bincomplex}.
In the simplest case, where we work at the K\"ahler point and align the reduction directions with the complex structure, the constraints enforce $B'_z = B'_{\bar{z}}= 0$.
However, if we do not impose the constraints in 4d, the standard reduction ansatz would only require that $B'_z$ and $B'_{\bar{z}}$ are functionally independent of $z$ and $\bar{z}$, a weaker condition. 

\def\PPu{\Phi}
\def\PPb{\bar\Phi}
In this section, we shall explore the consequences of reducing without first constraining.
Denoting the reduction with $\rightsquigarrow$ we anticipate that the lower-dimensional description will include additional fields as\footnote{Note, we are dropping the prime on the 2d gauge field $B$.}
\begin{equation}
\label{ec:2dfieldvar}
\begin{aligned}
&B'_w(w,\bar w,z,\bar z) \rightsquigarrow B_w(w,\bar w)\,,\quad B'_{\bar w}(w,\bar w,z,\bar z) \rightsquigarrow B_{\bar w}(w,\bar w)\,,\\
&B'_z(w,\bar w,z,\bar z) \rightsquigarrow \PPb(w,\bar w) \,,\quad B'_{\bar z}(w,\bar w,z,\bar z) \rightsquigarrow \PPu (w,\bar w)\,,
\end{aligned}\end{equation}
where $\PPu$ and $\PPb$ will be adjoint scalars in the lower-dimensional theory (sometimes called Higgs fields in the literature).
These will enter explicitly in the lower-dimensional theory through the reduction of covariant derivatives
\begin{equation}
\nabla_{z}g {g}^{-1} \rightsquigarrow \PPb - g \PPb {g}^{-1} \,, \qquad \nabla_{\b{z}}g {g}^{-1} \rightsquigarrow \PPu - g \PPu {g}^{-1}\,.
\end{equation}
On-shell the 4d gauge field $B'$ is ASD and couples to matter in the gWZW$_{4}$ model.
It is well-known that the reduction of an ASDYM connection leads to the Hitchin system, and we shall see this feature in the lower-dimensional dynamics below.

The two-dimensional Lagrangian that arises from reducing eq.~\eqref{4d action II} without first constraining is
\unskip\footnote{2d Lagrangians are defined as $S_{\text{IFT}_2} = 2 \rmi \int_{\mathbb{R}^2} \dr w \wedge \dr \bar{w} \, L_{\text{IFT}_2}$.
We denote
\begin{equation*}L_{\text{gWZ}} = L_{\text{WZ}}(g) + \Tr\big(( g^{-1}\partial_w g + \partial_w g g^{-1})B_{\bar w} - (g^{-1}\partial_{\bar w} g + \partial_{\bar w} g g^{-1})B_{w} + B_w\Ad_g B_{\bar w} - B_w \Ad_g^{-1} B_{\bar w}\big),
\end{equation*}
where
$
\int \dr w \wedge \dr \bar{w} \, L_{\text{WZ}}(g) = \int_{\mathbb{R}^2\times[0,1]} \mathcal{L}_{\text{WZ}}(\hat g) = \frac{1}{3}\int_{\mathbb{R}^2\times[0,1]}\Tr\big(\hat g^{1}\dr \hat g \wedge \hat g^{1}\dr \hat g \wedge \hat g^{1}\dr \hat g\big)
$.}
\def\aa{\mathsf{a}}
\def\bb{\mathsf{b}}
\begin{equation}
\begin{split}\label{eq:2daction}
L_{\text{IFT}_2}
&= \frac{1}{2}\Tr\big( g^{-1}D_w g g^{-1}D_{\bar w}g\big)+\frac{1}{2}\frac{\alpha + \beta }{\alpha - \beta } L_{\text{gWZ}}
+\Tr\big( \PPu\PPb + \frac{\alpha }{\alpha - \beta } \PPu\Ad_g\PPb -\frac{\beta }{\alpha - \beta } \PPu\Ad_g^{-1}\PPb\big) \\
& \qquad + \frac{1}{\alpha - \beta }\Tr\big(\PPu\,(g^{-1}D_{\bar w}g + D_{\bar w}gg^{-1}) + \alpha \beta \, \PPb\,(g^{-1}D_{w}g +D_{w}gg^{-1})\big) \\
& \qquad + \Tr\big(\tilde u ( F_{\bar w w} - \beta ^{-1} D_{\bar w}\PPu - \beta D_{w}\PPb -[\PPb,\PPu])\big) + \Tr\big( u ( F_{\bar w w} -\alpha ^{-1}  D_{\bar w}\PPu -\alpha  D_{w}\PPb  -[\PPb,\PPu])\big) \,,
\end{split}
\end{equation}
where we denote the 2d covariant derivative as $ D = \dr + \operatorname{ad}_B$ and note that we have rescaled $\tilde u \to \frac{\tilde u}{\langle \beta\gamma\rangle\langle\beta\hat\gamma\rangle}$ and $u \to \frac{u}{\langle \alpha\gamma\rangle\langle\alpha\hat\gamma\rangle}$.
The fields of the IFT$_2$ are $g\in G$ and $B_{w,\bar{w}},\PPu,\PPb,u,\tilde{u} \in \mathfrak{h}$.
\def\kk{\mathscr k}
In addition to the overall coupling, the IFT$_2$ eq.~\eqref{eq:2daction} only depends on a single parameter.
This can be seen by introducing
\unskip\footnote{Here, we have implicitly assumed that $\alpha\beta \geq 0$, which implies that $|\kk| \geq 1$.
The other regime of interest, $\alpha \beta \leq 0$ and $|\kk|\leq1$ is related by an analytic continuation $\kk' \to - \rmi \kk'$.}
\begin{equation}
\kk = \frac{\alpha +\beta }{\alpha - \beta }\,, \qquad
\kk' = - \frac{2\sqrt{\alpha \beta}}{\alpha - \beta}\,, \qquad
\kk^2 - \kk'{}^2 = 1\,,
\end{equation}
rescaling $\PPu \to \sqrt{\alpha \beta }\PPu$ and $\PPb \to \frac{1}{\sqrt{\alpha \beta }} \PPb$, and defining
$X^- = \kk'{}^{-1}(u + \tilde u)$ and $\tilde X^+ = \kk'{}^{-1}(u - \tilde u)$.
The Lagrangian eq.~\eqref{eq:2daction} can be rewritten as
\def\Op{\mathcal{O}}
\begin{equation}\label{eq:actk}
\begin{split}
L_{\text{IFT}_2}
&= \frac{1}{2}\Tr\big( g^{-1}D_w gg^{-1}D_{\bar w}g\big)+\frac{\kk}{2} L_{\text{gWZ}} + \Tr\big(\PPu \Op \PPb + \PPu \,V_{\bar w} + \PPb \, V_{w}\big)
\\ & \qquad
+ \Tr\big( X^- ( \kk' ( F_{\bar w w} - [\PPb,\PPu] ) + \kk (D_w\PPb + D_{\bar w} \PPu))\big)
+ \Tr\big( \tilde X^+ (D_{w}\PPb - D_{\bar w}\PPu)\big)\,,
\end{split}
\end{equation}
where
\begin{equation}\begin{gathered}
\Op = 1 - \frac{\kk+1}{2} \Ad_g + \frac{\kk-1}{2} \Ad_g^{-1} \,,
\qquad
V_{w,\bar{w}} = -\frac{\kk'}{2} (g^{-1}D_{w,\bar{w}} g + D_{w,\bar{w}} gg^{-1}) \,.
\end{gathered}\end{equation}
Note that the CFT points $\kk = 1$ or $\kk = - 1$ correspond to taking $\gamma_{a} \to \hat{\alpha}_{a}$ or $\gamma_{a} \to \alpha_{a}$, i.e.\ when the zeroes of the twist function coincide with the poles.

By construction, as the reduction of gWZW$_{4}$, the equations of motion of this theory are equivalent to the zero curvature of Lax connections, whose components are given by the $\dr w$ and $\dr\bar{w}$ legs of the 4d gauge fields.
Explicitly, these Lax connections are given by
\begin{align}
&\begin{aligned}\label{eq:laxka}
\mathcal{L}^{(A)}_w & = \partial_w + B_w -\frac{\kk+1}{2} K_w - \frac{1}{\zeta} \big(\PPu + \frac{\kk'}{2}K_w\big) \,,
\\
\mathcal{L}^{(A)}_{\bar w} & = \partial_{\bar w} + B_{\bar w} + \frac{\kk-1}{2} K_{\bar w} + \zeta \big(\PPb + \frac{\kk'}{2}K_{\bar w}\big) \,,
\end{aligned}
\\
&\begin{aligned}\label{eq:laxkb}
\mathcal{L}^{(B)}_w & = \partial_w + B_w - \frac{1}{\zeta} \PPu \,,
\qquad \qquad
\mathcal{L}^{(B)}_{\bar w} = \partial_{\bar w} + B_{\bar w} + \zeta \PPb \,,
\end{aligned}\end{align}
where we have also redefined the spectral parameter $\zeta \to \sqrt{\alpha\beta}\zeta$ compared to section \ref{sec:wzw4wzw2}
and we have introduced the currents
\begin{equation}
K_w = D_w g g^{-1} + \frac{\kk-1}{\kk'}(1 - \Ad_g )\PPu \,,
\qquad
K_{\bar w} = D_{\bar w} g g^{-1} - \frac{\kk+1}{\kk'}(1 - \Ad_g) \PPb \,.
\end{equation}

\subsection{Lax formulation}
Before analysing the Lagrangian eq.~\eqref{eq:actk} in more detail, let us show explicitly that its equations of motion are indeed equivalent to the zero-curvature condition for the Lax connections eq.~\eqref{eq:laxka} and eq.~\eqref{eq:laxkb}.
The equations of motion that follow from the Lagrangian eq.~\eqref{eq:actk} varying $\tilde{X}^+$, $X^-$ and $g$ are
\begin{equation}\begin{aligned}\label{eq:xpxmg}
& \delta \tilde X^+: & & \mathcal{E}_+ \equiv D_{w}\PPb - D_{\bar w}\PPu = 0 \,,
\\
& \delta X^-: & & \mathcal{E}_- \equiv\kk' \big(F_{\bar w w} - [\PPb,\PPu]\big) + \kk \big(D_w\PPb + D_{\bar w} \PPu) = 0\,,
\\
&\delta gg^{-1}: & & \mathcal{E}_g \equiv \frac{\kk-1}{2}\Big(D_w K_{\bar w} + \frac{\kk+1}{\kk'}[\PPb,K_w]\Big)
- \frac{\kk+1}{2}\Big(D_{\bar w} K_{w} - \frac{\kk-1}{\kk'}[\PPu,K_{\bar w}]\Big)
\\ &&& \hspace{200pt} + \frac{\kk }{\kk'}\mathcal{E}_- - \frac{1}{\kk'} \big(D_w\PPb + D_{\bar w} \PPu\big) = 0 \,.
\end{aligned}
\end{equation}
We also have the Bianchi identity following from the zero-curvature of the Maurer-Cartan form $\dr g g^{-1}$
\begin{equation}\begin{split}\label{eq:bianchi}
&\mathcal{Z} \equiv D_w K_{\bar w} + \frac{\kk+1}{\kk'} [\PPb,K_w] - D_{\bar w} K_w + \frac{\kk-1}{\kk'} [\PPu,K_{\bar w}] + [K_{\bar w},K_w] +
\frac{1}{\kk'}(1-\Ad_g)(\mathcal{E}_- + \mathcal{E}_+) = 0\,.
\end{split}\end{equation}

The zero curvature of the A-Lax eq.~\eqref{eq:laxka} gives rise to three equations that are linear combinations of the equations of motion eq.~\eqref{eq:xpxmg} and the Bianchi identity eq.~\eqref{eq:bianchi}:
\begin{equation}\begin{aligned}
0 & = \frac{\kk-1}{2} \mathcal{Z}' - \mathcal{E}_g + \frac{\kk}{\kk'}\mathcal{E}_- - \frac{1}{\kk'}\mathcal{E}_+ \,,\\
0 & = \kk'^2 \mathcal{Z}' - 2 \kk \mathcal{E}_g + 2\kk'\mathcal{E}_- \,,
\\
0 & = \frac{\kk+1}{2} \mathcal{Z}' - \mathcal{E}_g + \frac{\kk}{\kk'}\mathcal{E}_- + \frac{1}{\kk'}\mathcal{E}_+\,,
\end{aligned}\end{equation}
where we have defined $\mathcal{Z}' \equiv \mathcal{Z} - \frac{1}{\kk'}(1-\Ad_g) (\mathcal{E}_- + \mathcal{E}_+ )$.
On the other hand, the zero curvature of the B-Lax~\eqref{eq:laxkb} defines the Hitchin system:
\begin{equation}\begin{aligned}
\label{eq:HitchinSystem}
0 &= D_{\bar{w }} \PPu \,, \qquad
0 = F_{\bar{w} w} - [\PPb, \PPu]\,, \qquad
0 = D_w \PPb \, ,
\end{aligned}\end{equation}
which can be rewritten as the three equations $\mathcal{E}_\pm = 0$ and $\mathcal{E}_0 \equiv D_{w}\PPb + D_{\bar w}\PPu = 0$.
Therefore, the two Lax connections give rise to five independent equations, which are linear combinations of the equations of motion~\eqref{eq:xpxmg}, the Bianchi identity~\eqref{eq:bianchi}, and the additional equation $\mathcal{E}_0 = 0 $.

To recover this final equation from the equations of motion, let us consider the variational equations for $B_w$, $B_{\bar w}$, $\PPb$ and $\PPu$
\begin{equation}\begin{aligned}
& \delta B_w: & & \mathcal{E}_B \equiv \kk' D_{\bar w} X^- - [\PPb,\tilde X^+ + \kk X^-] + \frac{\kk-1}{2}P_{\mathfrak{h}}K_{\bar w} + \frac{\kk+1}{2}P_{\mathfrak{h}}\Ad_g^{-1} K_{\bar w} - \frac{\kk+1}{\kk'}P_{\mathfrak{h}}(1-\Ad_g^{-1})\PPb = 0 \,,
\\
& \delta B_{\bar w}: & & \mathcal{E}_{\bar B} \equiv \kk' D_{ w} X^- - [\PPu,\tilde X^+ - \kk X^-] + \frac{\kk+1}{2}P_{\mathfrak{h}}K_{w} + \frac{\kk-1}{2}P_{\mathfrak{h}}\Ad_g^{-1} K_{w} - \frac{\kk-1}{\kk'}P_{\mathfrak{h}}(1-\Ad_g^{-1})\PPu = 0 \,,
\\
&\delta \PPu: & & \mathcal{E}_{\PPu} \equiv D_{\bar w} (\tilde X^+ - \kk X^-) + \kk' [\PPb,X^-] - \frac{\kk'}{2} P_{\mathfrak{h}}(1+\Ad_g^{-1})K_{\bar w} + P_{\mathfrak{h}}(1-\Ad_g^{-1})\PPb =0 \,,
\\
&\delta \PPb: & & \mathcal{E}_{\PPb} \equiv D_w (\tilde X^+ + \kk X^-) + \kk' [\PPu,X^-] + \frac{\kk'}{2}P_{\mathfrak{h}}(1+\Ad_g^{-1})K_w - P_{\mathfrak{h}}(1-\Ad_g^{-1})\PPu =0 \,.
\end{aligned}\end{equation}
These can be understood as a first-order system of equations for $\tilde{X}^+$ and $X^-$.
Consistency of the system implies that they should satisfy the integrability conditions $[D_{\bar w},D_w] \tilde{X}^+ = [F_{\bar w w},\tilde{X}^+]$ and $[D_{\bar w},D_w] X^- = [F_{\bar w w},X^-]$.
We find that
\begin{equation}\begin{split}
&\kk' [D_{\bar w},D_w] X^- - \kk' [F_{\bar w w},X^-] = [X^+ ,\mathcal{E}_+]
+[X^-,\mathcal{E}_-] + P_{\mathfrak{h}} (1-\Ad_g^{-1}) \mathcal{E}_g + \kk
P_{\mathfrak{h}} \Ad_g^{-1} \mathcal{Z} \,,
\end{split}\end{equation}
hence, using the Bianchi identity~\eqref{eq:bianchi}, this vanishes on the equations of motion for $\tilde X^+$, $X^-$ and $g$~\eqref{eq:xpxmg}.
On the other hand, we have
\begin{equation}
\kk' [D_{\bar w},D_w] \tilde X^+ - \kk' [F_{\bar w w},\tilde X^+] =
[X^+,\mathcal{E}_-] + [X^-,\mathcal{E}_+]
+ \frac{2\kk}{\kk'}\mathcal{E}_-
- \frac{2}{\kk'}\mathcal{E}_0
-P_{\mathfrak{h}} (1+\Ad_g^{-1})\mathcal{E}_g + \kk P_{\mathfrak{h}}\Ad_g^{-1}\mathcal{Z} \, .
\end{equation}
Here we see that in addition to the Bianchi identity~\eqref{eq:bianchi} and equations of motion~\eqref{eq:xpxmg}, we also require $\mathcal{E}_0 = 0$, recovering the final equation from the Lax system.

\subsection{Relation to known models}

As we will shortly see, if we take $H$ to be abelian, the Lagrangian~\eqref{eq:actk} can be related to known models, including the homogeneous sine-Gordon models and the PCM plus WZ term.
However, for non-abelian $H$~\eqref{eq:actk} has not been considered before, and defines a new integrable field theory in two dimensions.
Moreover, by integrating out $\PPu$, $\PPb$ and the gauge field $B_{w,\bar{w}}$, it leads to an integrable sigma model for the fields $g$, $\tilde X^+$ and $X^-$.
We leave the study of these models for future work.

To recover a sigma model from the Lagrangian \eqref{eq:actk} for abelian $H$, in addition to integrating out $B_w$ and $B_{\bar w}$, we have two options.
The first is to integrate out $\PPu$ and $\PPb$.
The second is to solve the constraint imposed by the Lagrange multiplier $\tilde X^+$.
For abelian $H$ the Lagrangian~\eqref{eq:actk} simplifies to
\begin{equation}\label{eq:actkab}
\begin{split}
L_{\text{IFT}_2}^{\text{ab}}
&= \frac{1}{2}\Tr\big( g^{-1}D_w g g^{-1}D_{\bar w}g\big)+\frac{\kk}{2}L_{\text{gWZ}} + \Tr\big(\PPu\Op \PPb + \PPu\,V_{\bar w} + \PPb \, V_{w}\big)
\\ & \qquad + \Tr\big(( X^- ( \kk' F_{\bar w w} + \kk (\pd_w\PPb + \pd_{\bar w} \PPu ))\big)
+ \Tr\big( \tilde X^+ ( \partial_{w}\PPb - \partial_{\bar w}\PPu)\big) \,.
\end{split}
\end{equation}
This takes the form of the first-order action in the Buscher procedure, and it follows that the two sigma models will be T-dual to each other with dual fields $X^+$ and $\tilde X^+$.
Explicitly the Lagrangians, before integrating out $B_w$ and $B_{\bar w}$, are
\begin{equation}\label{eq:actkintphi}
\begin{split}
L_{\text{IFT}_2}^{\tilde{X}}
&= \frac{1}{2}\Tr\big( g^{-1}D_w g g^{-1}D_{\bar w}g\big)+\frac{\kk}{2} L_{\text{gWZ}}
+ \kk' \Tr\big( X^- F_{\bar w w} \big)
\\ & \qquad + \Tr\big( (\pd_w \tilde X^+ - V_w + \kk \pd_w X^- ) \Op^{-1} (\pd_{\bar w} \tilde X^+ + V_{\bar w} - \kk \pd_{\bar w} X^-) \big) \,,
\end{split}
\end{equation}
and
\begin{equation}\label{eq:actk2}
\begin{split}
L_{\text{IFT}_2}^{X}
&= \frac{1}{2}\Tr\big( g^{-1}D_w gg^{-1}D_{\bar w}g\big) +\frac{\kk}{2} L_{\text{gWZ}} + \kk' \Tr\big( X^- F_{\bar w w} \big)
\\ & \qquad + \frac14\Tr\big(\partial_w X^+\Op \partial_{\bar w} X^+ + 2\partial_w X^+ ( V_{\bar w} - \kk \partial_{\bar w} X^-) + 2\partial_{\bar w} X^+ ( V_{w} - \kk \partial_w X^-)\big)\,,
\end{split}
\end{equation}
where in the second we have locally solved the constraint imposed by the Lagrange multiplier $\tilde{X}^+$ by setting
\begin{equation}\label{eq:locsol}
\PPu = \frac12 \partial_w X^+ \,, \qquad
\PPb = \frac12\partial_{\bar w} X^+ \,,\qquad X^+ \in \mathfrak{h} \,.
\end{equation}

As mentioned above, the first approach can also be straightforwardly applied for non-abelian $H$.
Generalising the second approach is more subtle.
The constraint imposed by the Lagrange multiplier $\tilde{X}^+$ in the Lagrangian~\eqref{eq:actk} implies that
\begin{equation}\label{eq:covflat}
D_{w}\PPb - D_{\bar w}\PPu = 0.
\end{equation}
Typically the full solution to this equation would be expressed in terms of path-ordered exponentials of $B_w$ and $B_{\bar w}$.
To avoid non-local expressions, we can restrict $\PPu$ and $\PPb$ to be valued in the centre of $\mathfrak{h}$, denoted $\mathcal{Z}(\mathfrak{h})$.
Note that this is not a restriction if $H$ is abelian.
With this restriction, the Lagrangian~\eqref{eq:actk} again simplifies to~\eqref{eq:actkab}, and the constraint~\eqref{eq:covflat} then becomes $\partial_{ w}\PPb - \partial_{\bar w}\PPu = 0$, which we can again locally solve by~\eqref{eq:locsol} now with $X^+ \in \mathcal{Z}(\mathfrak{h})$, similarly leading to the Lagrangian~\eqref{eq:actk2}.

\medskip

\paragraph{Relation to PCM plus WZ term.}

Taking $H$ to be abelian, we can relate the Lagrangian~\eqref{eq:actkab} to that of the PCM + WZ term for $G \times H$ through a combination of T-dualities and field redefinitions.
We start by parametrising
\begin{equation}
g = e^{\frac12\tau} g e^{\frac12\tau}\,, \qquad \tau \in \mathfrak{h}\,,
\end{equation}and setting $\partial_{\bar w,w}\tau \to 2C_{w,\bar{w}}$.
We also integrate by parts and set $\partial_{w}X^- \to 2\Psi$ and $\partial_{\bar w}X^- \to 2\bar \Psi$.
To maintain equivalence with the Lagrangian that we started with, we add $\Tr\big(\tilde{\tau} (\partial_w C_{\bar w} - \partial_{\bar w} C_w)\big) + \Tr\big(\tilde{X}^- (\partial_w \bar\Psi - \partial_{\bar w} \Psi)\big)$, i.e., the Lagrange multipliers $\tilde \tau$ and $\tilde X^-$ locally impose $C_{w,\bar w} = \frac12\partial_{\bar w,w}\tau$, $\Psi = \frac12\partial_w X^-$ and $\bar\Psi = \frac12\partial_{\bar w}X^-$.
We can then redefine the fields as
\unskip\footnote{To arrive at this field redefinition, we first look for the shifts of $B_{w,\bar w}$, $C_{w,\bar w}$, $\Psi$ and $\bar \Psi$ that decouple $\Phi$ and $\bar \Phi$ from all other fields apart from $\tilde{X}^+$.
Since both $C_{w}$ and $C_{\bar w}$ transform in the same way, as do $\Psi$ and $\bar \Psi$, we can then easily compute the transformation of $\tilde\tau$, $\tilde X^-$ and $\tilde X^+$ by demanding that the triplet of terms $\Tr\big( \tilde{\tau}F_{w\bar{w}}(C) + \tilde{X}^-F_{w\bar{w}}(\Psi) + \tilde{X}^+F_{w\bar{w}}(\Phi)\big)$ is invariant up to a simple rescaling, i.e., it becomes $\Tr\big( \tilde{\tau}F_{w\bar{w}}(C) + \kk' \tilde{X}^-F_{w\bar{w}}(\Psi) + \frac{1}{\kk'}\tilde{X}^+F_{w\bar{w}}(\Phi)\big)$.}
\begin{equation}\begin{aligned}\label{tildetrans}
B_w & \to B_w - \frac{\kk}{\kk'}\PPu \,, \qquad &
C_w & \to C_w - \frac{1}{\kk'}\PPu \,, \qquad &
\Psi & \to \Psi + \frac{\kk}{\kk'^2} \PPu \,,
\\ B_{\bar w} & \to B_{\bar w} + \frac{\kk}{\kk'}\PPb \,, \qquad &
C_{\bar w} & \to C_{\bar w} - \frac{1}{\kk'}\PPb\,, \qquad &
\bar\Psi & \to \bar\Psi + \frac{\kk}{\kk'^2} \PPb \,,
\\
\tilde{X}^+ & \to \frac{1}{\kk'}\tilde{X}^+ - \frac{\kk}{\kk'} \tilde{X}^- + \frac{1}{\kk'}\tilde{\tau}\,, \qquad &
\tilde{X}^- & \to \kk' \tilde{X}^-\,, \qquad &
\tilde{\tau} & \to \tilde{\tau}\,.
\end{aligned}\end{equation}
Doing so, we arrive at the following Lagrangian
\begin{equation}\begin{split}
L_{\text{IFT}_2}^{\text{ab}} & = \frac{1}{2}\Tr\big( g^{-1}\pd_w g g^{-1}\pd_{\bar w} g \big)
+ \frac{\kk}{2} L_{\text{WZ}}(g)
\\ & \qquad + \frac{1-\kk}{2}\Tr\big(g^{-1}\pd_w g (C_{\bar w} - B_{\bar w})
+ \pd_{\bar w} g g^{-1} ( C_{w} + B_{w})
+ (C_{w} + B_{w}) \Ad_g(C_{\bar w} - B_{\bar w}) \big)
\\ & \qquad + \frac{1+\kk}{2}\Tr\big(g^{-1}\pd_{\bar w} g( C_{w} - B_{w})
+ \pd_{w} g g^{-1}(C_{\bar w} + B_{\bar w})
+ (C_{w} - B_{w}) \Ad_g^{-1} (C_{\bar w} + B_{\bar w}) \big)
\\
& \qquad + \Tr\big( B_w B_{\bar w}
+ C_w C_{\bar w}
+ \kk C_w B_{\bar w}
- \kk B_w C_{\bar w} \big)
\\ & \qquad + \Tr\big( \tilde{\tau} (\partial_w C_{\bar w} - \partial_{\bar w} C_w)\big)
+ \kk' \Tr\big( \tilde{X}^- (\partial_w \bar \Psi - \partial_{\bar w} \Psi)\big)
+ 2\kk' \Tr\big( \Psi B_{\bar w} - B_w,\bar\Psi\big)
\\ & \qquad + \frac{1}{\kk'} \Tr\big( \tilde{X}^+ (\partial_w \bar \Phi - \partial_{\bar w} \Phi)\big)
- \frac{2}{\kk'^2}\Tr\big(\Phi\bar\Phi\big) \,.
\end{split}\end{equation}
The final steps are to integrate out $\tilde \tau$, $\Psi$ and $\bar \Psi$, and $\Phi$ and $\bar \Phi$, leading us to set
\begin{equation}
C_{w,\bar w} = \frac12\partial_{w,\bar{w}}\tau, \qquad B_{w, \bar w} = - \frac12\partial_{w,\bar{w}}\tilde X^-, \qquad \Phi = - \frac{\kk'}{2}\pd_w\tilde{X}^+, \quad \bar\Phi = \frac{\kk'}{2}\pd_{\bar w}\tilde{X}^+.
\end{equation}
Redefining $g \to e^{-\frac12(\tau + \tilde{X}^-)} g e^{-\frac12(\tau - \tilde{X}^-)}$, we find the difference of the PCM plus WZ term Lagrangians for $G$ and $H$
\begin{equation}\begin{split}\label{eq:pcmpluswz}
L_{\text{PCM} + \kk \text{WZ}_2} & = \frac{1}{2}\Tr\big(g^{-1}\pd_w g g^{-1}\pd_{\bar w} g \big)
+ \frac{\kk}{2} L_{\text{WZ}}(g)
- \frac12 \Tr\big(\partial_w \tilde{X}^+ \partial_{\bar w} \tilde{X}^+ \big),
\end{split}\end{equation}
where we recall that for abelian $H$ the WZ term vanishes.

To summarise, starting from the sigma model~\eqref{eq:actk2} we T-dualise in $\tau$, $X^+$ and $X^-$, we then perform a GL$(3)$ transformation on the dual coordinates, and finally T-dualise back in $\tau$ to recover~\eqref{eq:pcmpluswz}, the difference of the PCM plus WZ term Lagrangians for $G$ and $H$.
This relation may have been anticipated since this is the model we would expect to find starting from the ghCS$_6$ action~\eqref{eq:new6daction} and instead imposing the boundary conditions $\mathcal{A}|_{\alpha,\beta} = \mathcal{B}|_{\alpha,\beta} = 0$.

\medskip

\paragraph{\texorpdfstring{$\kk \to 1$}{k → 1} limit.}
As we have seen, the $\kk \to 1$ limit is special since if we first constrain and then reduce we recover the gauged WZW coset CFT.
By first reducing and then constraining, we can recover massive integrable perturbations of these theories.
We consider the setup where $\PPu$ and $\PPb$ are restricted to lie in $\mathcal{Z}(\mathfrak{h})$ and solve the constraint imposed by the Lagrange multiplier $\tilde{X}^+$ by~\eqref{eq:locsol}.
Taking $\kk\to 1$ the Lagrangian~\eqref{eq:actk2} simplifies further to
\begin{equation}\label{eq:ppwave}
\begin{split}
L_{\text{IFT}_2}
&= \frac{1}{2}\Tr\big( g^{-1}D_w gg^{-1}D_{\bar w}g\big)+\frac{1}{2} L_{\text{gWZ}}
\\ & \qquad + \frac14\Tr\big(\partial_w X^+(1-\Ad_g) \partial_{\bar w} X^+ - 2\partial_w X^+ \partial_{\bar w} X^- - 2\partial_{\bar w} X^+ \partial_w X^-\big)\,,
\end{split}
\end{equation}
This is reminiscent of a sigma model for a pp-wave background, with the kinetic terms for the transverse fields described by the gauged WZW model for the coset $G/H$, except that the would-be light-cone coordinates $X^+$ and $X^-$ have $\dim \mathcal{Z}(\mathfrak{h})$ components.
Nevertheless, we still have the key property that the equation of motion for $X^-$ is $\partial_w\partial_{\bar w}X^+ = 0$, whose general solution is $X^+ = Y(w) + \bar Y(\bar w)$. Substituting into the Lagrangian~\eqref{eq:ppwave} we find
\begin{equation}\label{eq:gsg}
\begin{split}
L_{\text{IFT}_2}
&= \frac{1}{2}\Tr\big( g^{-1}D_w g g^{-1}D_{\bar w}g\big)+\frac{1}{2} L_{\text{gWZ}} + \frac14 \Tr\big( Y'\bar Y' - Y'\Ad_g\bar Y'\big) \,.
\end{split}
\end{equation}
In the special case that $Y = w \Lambda$ and $\bar Y = \bar{w}\bar\Lambda$ this is the gauged WZW model for the coset $G/H$ deformed by a massive potential $V = \Tr(\Lambda \Ad_g \bar \Lambda) - \Tr(\Lambda\bar\Lambda)$ as studied in \cite{Park:1994bx}.
Taking the limit $\mathscr{k}\rightarrow1$ directly at the level of the Lax connection given by eq.~\eqref{eq:laxka}, keeping track of the definitions of the currents $K_w,K_{\bar{w}}$ which depend on $\kk$, we find
\begin{equation}
\mathcal{L}_w \rightarrow \partial_w + B_w - D_w g g^{-1} + \frac{1}{2\zeta} \Lambda \, ,\qquad
\mathcal{L}_{\bar{w}} = \partial_{\bar{w}} + B_{\bar w} - \frac{\zeta}{2} \Ad_g \bar\Lambda \,,
\end{equation}
recovering the Lax given in \cite{Park:1994bx,Fernandez-Pousa:1996aoa}.

When $G$ is compact and $H = U(1)^{\mathrm{rk}_G}$, $\Lambda$ and $\bar\Lambda$ can be chosen such that these models have a positive-definite kinetic term and a mass gap.
These are known as the homogeneous sine-Gordon models \cite{Fernandez-Pousa:1996aoa}.
For $G=SU(2)$ and $H=U(1)$ the homogeneous sine-Gordon model becomes the complex sine-Gordon model after integrating out the gauge fields $B_w$ and $B_{\bar w}$. Note that if $\mathcal{Z}(\mathfrak{h})$ is one-dimensional and $Y(w)$ and $\bar Y(\bar w)$ are both non-constant then we can always use the classical conformal symmetry to reach $Y = w \Lambda$ and $Y' = \bar{w}\Lambda$, hence recovering a constant potential.
For higher-dimensional $\mathcal{Z}(\mathfrak{h})$, this is not the case.

\subsection{Example: \texorpdfstring{$SL(2)/U(1)_V$}{SL(2)/U(1)V}}

To illustrate the features of this construction, let us consider the example of $SL(2)/U(1)_V$ for which the 2d gauged WZW describes the trumpet CFT. To be explicit we use $\mathfrak{sl}(2)$ generators
\begin{equation}
T_1= \left(
\begin{array}{cc}
1 & 0 \\
0 & -1 \\
\end{array}
\right) \, , \quad T_2 = \left(
\begin{array}{cc}
0 & 1 \\
1 & 0 \\
\end{array}
\right) \, , \quad T_3 = \left(
\begin{array}{cc}
0 & 1 \\
-1 & 0 \\
\end{array}
\right)\, ,
\end{equation}
and parametrise the group element as
\begin{equation}
g= \left(
\begin{array}{cc}
\cos (\theta ) \sinh (\rho )+\cosh (\rho ) \cos (\tau ) &
\sin (\theta ) \sinh (\rho )+\cosh (\rho ) \sin (\tau )
\\
\sin (\theta ) \sinh (\rho )-\cosh (\rho ) \sin (\tau ) &
\cosh (\rho ) \cos (\tau )-\cos (\theta ) \sinh (\rho )
\\
\end{array}
\right) \, .
\end{equation}
We choose the $U(1)$ vector action generated by $T_3$ such that
\begin{equation}
\delta g = \epsilon [g, T_3] \quad \Rightarrow \quad \delta \rho=\delta \tau = 0 \, , \quad \delta \theta = \epsilon,
\end{equation}
hence we gauge fix by setting $\theta = 0 $. The analysis here is simplified by the observation that there is no WZ term since there are no 3-forms on the two-dimensional target space.

\paragraph{The CFT point.}
For orientation, we first work at the CFT point corresponding to $\kk=1$. Recall from the discussion in \S\ref{sec:gaugedWZWdiamond}, that first constraining in 4d and then reducing, enforces $\PPb = \PPu = 0 $ and the Lagrange multiplier sector vanishes.
This gives the conventional gauged WZW model described by a target space geometry
\begin{equation}
\dr s^2 = \dr \rho^2 + \coth^2\rho \, \dr \tau^2 \, .
\end{equation}
Let us now consider the IFT$_{2}$ that results from taking the same reduction that would lead to the CFT, but now in our reduction ansatz set $\PPu = \frac{m}{2} T_3 $ and $\PPb = -\frac{m}{2} T_3 $.
The Lagrangian that follows is
\begin{equation}\label{eq:csg}
L_{\text{CsG}} = \pd_w \rho \pd_{\bar{w}} \rho + \coth^2 \rho \, \pd_w \tau \pd_{\bar{w} } \tau - m^2 \sinh^2 \rho \, .
\end{equation}
This theory is well known as the complex sinh-Gordon model, a special case of the integrable massive deformations of $G/H$ gauged WZW models known as the homogeneous sine-Gordon models \cite{Park:1994bx,Fernandez-Pousa:1996aoa}.

\paragraph{Unconstrained reduction: integrating out \texorpdfstring{$\PPu$, $\PPb$ and $B_{w,\bar{w}}$}{Phi, Phi-bar, and B}.}
We now turn to the more general story, away from the CFT point, by considering the reduction without first imposing constraints. Taking the IFT$_{2}$~\eqref{eq:actkab} and integrating out $\PPu$, $\PPb$ and the gauge field $B_{w,\bar{w}}$ while retaining $X^-$ and $\tilde{X}^+$, results in the sigma model with target space metric and B-field
\begin{equation}\begin{aligned}\label{eq:sl2backorig}
\dr s^2 &= \dr \rho^2 + \coth^2\rho \, \dr \tau^2 + \textrm{csch}^2\rho \left( \dr \tilde{X}^+{}^2 - \dr X^-{}^2 \right) \, \\
B_2 &= \mathcal{V} \wedge \dr \tilde{X}^+ \, ,\qquad \mathcal{V} = \kk \textrm{csch}^2\rho \, \dr X^- + \kk' \coth^2\rho \, \dr \tau \,.
\end{aligned}\end{equation}

\paragraph{Unconstrained reduction: the dual.}
On the other hand, if we solve the constraint imposed by the Lagrange multiplier $\tilde{X}^+$ setting $\PPu = \frac12\pd_{w} X^+$ and $\PPb = \frac12\pd_{\bar{w}} X^+$, we find the sigma model with target space geometry
\begin{equation}\begin{aligned}\label{sl2back}
\dr s^2 &= \dr \rho^2 + \coth^2\rho \, \dr \tau^2 - \textrm{csch}^2\rho \, \dr X^-{}^2 + \sinh^2\rho \, (\dr X^++\mathcal{V})^2  \, ,  \\
B_2 &= 0\, .
\end{aligned}
\end{equation}
This can of course be recognised as the T-dual of~\eqref{eq:sl2backorig} along $\tilde{X}^+$.
In the limit $\kk \to 1$~\eqref{sl2back} becomes the pp-wave background
\begin{equation}\begin{aligned}
\dr s^2 &= \dr \rho^2 + \coth^2\rho \, \dr \tau^2 + \sinh^2\rho \, \dr X^+{}^2 +2\dr X^+ \dr X^-  \, ,
\\
B_2 &= 0\, ,
\end{aligned}\end{equation}
and if we light-cone gauge fix, $X^+ = m (w-\bar w)$, in the associated sigma model we recover the complex sinh-Gordon Lagrangian~\eqref{eq:csg} as expected.

\paragraph{Relation to PCM plus WZ term.}

Finally we demonstrate a relation between the models above and the PCM plus WZ term.
Let us start with the metric and B-field for the PCM plus WZ term for $G = \text{GL}(2)$
\begin{equation}\begin{aligned}
\dr s^2 & = \dr \tilde{X}^+{}^2 + \dr \rho^2 - \cosh^2\rho \dr \tau^2 + \sinh^2 \rho \dr \tilde X^-{}^2 \, ,
\\
B & = \kk \cosh^2\rho \, \dr \tau \wedge \dr \tilde X^- \, .
\end{aligned}\end{equation}
Note that $\dr B = \kk \sinh 2\rho \, \dr\rho\wedge\dr \tau \wedge \dr \tilde X^-$, which is proportional to the volume for $SL(2)$.
We first T-dualise $\tau \to \tilde \tau$, and then perform the following field redefinition
\begin{equation}
\tilde X^+ \to \kk' \tilde X^+ + \frac{\kk}{\kk'}\tilde X^- - \tilde \tau \ , \qquad
\tilde X^- \to \frac{1}{\kk'} \tilde X^- \ .
\end{equation}
It is straightforward to check that this is the inverse transformation to~\eqref{tildetrans}.
Finally, T-dualising back, $\tilde X^+ \to X^+$, $\tilde X^- \to X^-$ and $\tilde \tau \to \tau$, we precisely recover the background~\eqref{sl2back}, demonstrating that it can be understood as a generalised TsT transformation of the PCM plus WZ term.

\subsection{The LMP limit}
The PCM plus WZ term admits a limit in which it becomes the 2d analogue of the LMP model, otherwise known as the pseudodual of the PCM \cite{Zakharov:1973pp}, see, e.g.~\cite{Hoare:2018jim}.
It is possible to generalise this limit to the gauged model~\eqref{eq:actk} by setting
$g = \exp(\varepsilon U)$, $\kk = \varepsilon^{-1} \ell$, $\tilde X^+ \to \varepsilon^2 \tilde X^+$, $X^- \to \varepsilon^3 X^- - \varepsilon P_{\mathfrak{h}} U$ rescaling the Lagrangian by $\varepsilon^{-2}$, and taking $\varepsilon \to 0$.
Implementing this limit in~\eqref{eq:actk} we find
\begin{equation}\label{eq:actklmp}
\begin{split}
L_{\text{IFT}_2}^{\text{LMP}}
&= \frac{1}{2}\Tr\big( D_w U D_{\bar w} U + [\PPu,U][\PPb,U]\big) - \frac{\ell}{6} \Tr\big((D_w U + [\PPu,U][U,(D_{\bar w}U - [\PPb,U]]\big)
\\ & \qquad
+ \ell\Tr\big(X^- ( F_{\bar w w} - [\PPb,\PPu] + D_w\PPb + D_{\bar w} \PPu)\big)
+ \Tr\big((\tilde X^+ ( D_{w}\PPb - D_{\bar w}\PPu)\big)
\\ & \qquad + \frac{1}{2\ell} \Tr\big(U ( F_{\bar w w} - [\PPb,\PPu] -D_w\PPb - D_{\bar w} \PPu)\big) \,.
\end{split}
\end{equation}
Similarly we can take the limit in the Lax connections~\eqref{eq:laxka} and~\eqref{eq:laxkb}.
The B-Lax~\eqref{eq:laxkb} is unchanged, while the A-Lax~\eqref{eq:laxka} becomes
\begin{equation}
\begin{aligned}\label{eq:laxkalmp}
\mathcal{L}^{(A)}_w & = \partial_w + B_w -\frac{\ell}{2} K^{\text{LMP}}_w- \frac{1}{\zeta} \big(\PPu + \frac{\ell}{2}K^{\text{LMP}}_w\big) \,,
\\
\mathcal{L}^{(A)}_{\bar w} & = \partial_{\bar w} + B_{\bar w} + \frac{\ell}{2} K^{\text{LMP}}_{\bar w} + \zeta \big(\PPb + \frac{\ell}{2}K^{\text{LMP}}_{\bar w}\big) \,,
\end{aligned}
\end{equation}
where
\begin{equation}
K^{\text{LMP}}_w = D_w U + [\PPu,U] \,,
\qquad
K^{\text{LMP}}_{\bar w} = D_{\bar w} U - [\PPb,U] \,.
\end{equation}
As we will see in \S\ref{Gauged LMP action section} this model can also be found directly from 6d hCS and 4d CS by considering a twist function with a single fourth-order pole.

As in the gauged WZW case, we can again find an integrable sigma model from~\eqref{eq:actklmp} by integrating out $\PPu$, $\PPb$ and the gauge field $B_{w,\bar w}$.
For abelian $H$ we can also construct the dual model by solving the constraint imposed by the Lagrange multiplier $\tilde{X}^+$ and integrating out $B_w$ and $B_{\bar w}$.
For $SL(2)/U(1)_V$ the resulting backgrounds can be found by taking the LMP limit
\begin{equation}\begin{aligned}
& \rho \to \varepsilon \rho - \frac16 \varepsilon^3 \rho\tau^2 \,, \qquad
\tau \to \varepsilon \tau -\frac13 \varepsilon^3 \rho^2 \tau \,, \qquad
(\dr s^2, B_2) \to \varepsilon^{-2} (\dr s^2, B_2) \,, \qquad \kk \to \varepsilon^{-1}\ell \,,
\\ & X^- \to \varepsilon^3 X^- - \varepsilon \tau \,, \qquad
\tilde{X}^+ \to \varepsilon^2 \tilde{X}^+ \,, \qquad
X^+ \to X^+ \,, \qquad \varepsilon \to 0 \,,
\end{aligned}\end{equation}
in eqs.~\eqref{eq:sl2backorig} and~\eqref{sl2back}.
This limit breaks the manifest global symmetry given by shifts of the coordinate $\tau$.
This is in agreement with the fact that the Lagrangian~\eqref{eq:actklmp} is not invariant under $U \to U + H_0$ ($H_0 \in \mathfrak{h}$), while its gauged WZW counterpart~\eqref{eq:actk} is invariant under $g \to h_0 g h_0$ ($h_0 \in H$) for abelian $H$.

Curiously, we can actually take a simplified LMP limit
\begin{equation}\begin{aligned}
& \rho \to \varepsilon \rho \,, \qquad
\tau \to \varepsilon \tau \,, \qquad
(\dr s^2, B_2) \to \varepsilon^{-2} (\dr s^2, B_2) \,, \qquad \kk \to \varepsilon^{-1}\ell \,,
\\ & X^- \to \varepsilon^3 X^- - \varepsilon \tau \,, \qquad
\tilde{X}^+ \to \varepsilon^2 \tilde{X}^+ \,, \qquad
X^+ \to X^+ \,, \qquad \varepsilon \to 0 \,,
\end{aligned}\end{equation}
in the backgrounds~\eqref{eq:sl2backorig} and~\eqref{sl2back} that preserves this global symmetry. Taking this limit in eq.~\eqref{eq:sl2backorig} we find
\begin{equation}\begin{aligned}\label{eq:lmporig}
\dr s^2 & = \dr \rho^2 + \dr \tau^2 + \frac{1}{\rho^{2}}\dr \tilde X^+{}^2 + \frac{2}{\rho^{2}} \dr X^- \dr \tau \,,
\\ B_2 &= \mathcal{V} \wedge \dr\tilde{X}^+ \,, \qquad
\mathcal{V} = \frac{\ell}{\rho^2}\dr X^- + \left(\ell - \frac{1}{2\ell\rho^2} \right)\dr \tau \,,
\end{aligned}\end{equation}
while the limit of eq.~\eqref{sl2back} is
\begin{equation}\begin{aligned}
\dr s^2 & = \dr \rho^2 + \dr \tau^2 + \rho^{2}(\dr \tilde X^+ + \mathcal{V} )^2 + \frac{2}{\rho^{2}} \dr X^- \dr \tau \,,
\\
B_2 &= 0\, .
\end{aligned}\end{equation}
As for the gauged WZW case these two backgrounds above can also be constructed as a generalised TsT transformation of the background for the LMP model on GL$(2)$
\begin{equation}\begin{aligned}
\dr s^2 & = \dr \tilde X^+{}^2 + \dr \rho^2 - \dr \tau^2 +\rho^{2}\dr \tilde X^-{}^2\,,
\\
B_2 &= \ell \rho^2 d\tau \wedge \dr \tilde X^-\, .
\end{aligned}\end{equation}
Explicitly, if we first T-dualise $\tau\to\tilde\tau$, then perform the following field redefinition
\begin{equation}
\tilde{X}^+ \to \ell \tilde{X}^+ + \frac{1}{2\ell^2}\tilde{X}^- - \tilde{\tau} \,, \qquad \tilde{X}^- \to \frac{1}{\ell} \tilde{X}^- \,,\qquad\tilde\tau \to \tilde \tau - \frac{1}{2\ell^2}\tilde X^- \,,
\end{equation}
and finally T-dualise back,
\unskip\footnote{Note that here the order of T-dualities matters. In particular, we cannot first T-dualise $\tilde\tau$ after the coordinate redefinition since it turns out to be a null coordinate.}
$\tilde X^+ \to X^+$, $\tilde X^- \to X^-$ and $\tilde \tau \to \tau$, we recover the background~\eqref{eq:lmporig}.

\section{Reduction to \texorpdfstring{gCS$_4$}{gCS4} and localisation}
\label{sec:6dCSto4dCS}

Having discussed the right hand side of the diamond, we briefly describe the left hand side that follows from first reducing to obtain a gauged 4d Chern-Simons theory on $\mathbb{R}^2\times \mathbb{CP}^1$ and then integrating over $\mathbb{CP}^1$ to localise to a two-dimensional field theory on $\mathbb{R}^2$. We show that the resulting IFT$_2$ matches \eqref{eq:2daction}.

We recall the six-dimensional coupled action
\begin{equation}
\label{ec:recall6daction}
S_{\text{ghCS}_6}[\cA , \cB] = S_{\text{hCS}_6}[\cA] - S_{\text{hCS}_6}[\cB] - \frac{1}{2\pi \rmi }\int_{\mathbb{PT} } \bar{\partial} \Omega \wedge \Tr\left( \cA \wedge \cB \right) \, .
\end{equation}
We note that the three terms in the action are invariant under the transformations $\cA\mapsto \hat \cA = \cA + \rho^{\cA}_{\dot a} e^{\dot a}+\rho^{\cA}_{0}e^{0}$ and $\cB\mapsto \hat \cB=\cB + \rho^{\cB}_{\dot a} e^{\dot a}+\rho^{\cB}_{0}e^{0}$, given that both $\Omega$ and $\bar\partial \Omega$ are top forms in the holomorphic directions. By choosing $\rho^{\cA}$ and $\rho^{\cB}$ appropriately, we can ensure that neither $\hat \cA$ nor $\hat \cB$ have $\dr z$ or $\dr \bar z$ legs, so
\begin{align}
\hat \cA &= \hat \cA_{w} \dr w + \hat \cA_{\bar w} \dr \bar w+\cA_0 \bar e^{0} \quad \text{with}\quad \hat \cA_{w}=-\frac{[\cA \kappa]}{\langle \pi \gamma\rangle} \,, \,\hat \cA_{\bar w}=-\frac{[\cA \hat\kappa]}{\langle \pi \hat\gamma\rangle} \ , \\
\hat \cB &= \hat \cB_{w} \dr w + \hat \cB_{\bar w} \dr \bar w+\cB_0 \bar e^{0} \quad \text{with}\quad \hat \cB_{w}=-\frac{[\cB \kappa]}{\langle \pi \gamma\rangle} \,, \,\hat \cB_{\bar w}=-\frac{[\cB \hat\kappa]}{\langle \pi \hat\gamma\rangle} \,.
\end{align}
To perform the reduction we follow the procedure outlined in \S \ref{sec:wzw4wzw2}. Namely, we contract the six-dimensional Lagrangian of \eqref{ec:recall6daction} with the vector fields $\partial_z$ and $\partial_{\bar z}$, and restrict to gauge connections which are invariant under the flow of these vector fields. Thus, since the shifted gauge fields $\hat \cA$ and $\hat \cB$ manifestly have no $\dr z$ or $\dr \bar z$ legs, and we are restricting to field configurations satisfying $L_{\partial_z}\hat \cA=L_{\partial_z}\hat \cB=L_{\partial_{\bar z}}\hat \cA=L_{\partial_{\bar z}}\hat \cB=0$, the contraction by $\partial_z$ and $\partial_{\bar z}$ only hits $\Omega$ in the first two terms, and $\bar \partial \Omega$ in the third. In particular, we find
\begin{equation}
(\partial_z \wedge \partial_{\bar z})\vee \Omega = \frac{\langle \alpha \beta\rangle^2}{2}\frac{\langle \pi \gamma\rangle \langle \pi \hat \gamma\rangle}{\langle \pi \alpha\rangle^2\langle \pi \beta\rangle^2}e^0\,, \qquad (\partial_z \wedge \partial_{\bar z})\vee \bar \partial \Omega=- \frac{\langle \alpha \beta\rangle^2}{2}\bar\partial_0\bigg(\frac{\langle \pi \gamma\rangle \langle \pi \hat \gamma\rangle}{\langle \pi \alpha\rangle^2\langle \pi \beta\rangle^2}\bigg)e^0 \wedge \bar e^{0}\,.
\end{equation}
Hence, the six-dimensional action reduces to a four-dimensional coupled Chern-Simons action
\begin{equation}
\label{ec:4dgCS}
S_{\text{gCS}_4}[\hat A,\hat B]=\int_{X}\omega \wedge \mathrm{CS}[\hat A]-\int_{X}\omega \wedge \mathrm{CS}[\hat B]-\frac{1}{2\pi i}\int_{X}\bar \partial \omega \wedge \langle \hat A, \hat B\rangle\,,
\end{equation}
where $X = \mathbb{CP}^1 \times \bR^{2}$,
\begin{equation}
\omega=\frac{\langle \alpha \beta\rangle^2}{2}\frac{\langle \pi \gamma\rangle \langle \pi \hat \gamma\rangle}{\langle \pi \alpha\rangle^2\langle \pi \beta\rangle^2}e^0 \ ,
\end{equation}
and $\hat A$ and $\hat B$ are the restrictions of $\hat \cA$ and $\hat \cB$ to $X$. Similarly, the boundary conditions \eqref{eq:ABboundarycond} descend to analogous boundary conditions on $\hat A$ and $\hat B$. The action \eqref{ec:4dgCS} has been considered before in \cite{Stedman:2021wrw}, albeit not with the choice of $\omega$ hereby discussed.

With the gauged 4d Chern-Simons action at hand, we may now localize. The procedure is entirely analogous to the one depicted in \S \ref{sec:localtogwzw4} so we shall omit some of the details. We begin by reparametrising our four-dimensional gauge fields $\hat A$ and $\hat B$ in terms of a new pair of connections $\hat A', \hat B'$ and smooth functions $\hat g \in C^{\infty}(X,G)$ and $\hat h \in C^{\infty}(X,H)$. We use the redundancy in the reparametrisation to fix $\hat A'_0 = \hat B'_0=0$. The boundary degrees of freedom of the resulting IFT$_2$ will be \emph{a priori} be given by the evaluation of $\hat g$, $\hat h$, $\hat u $ and $\hat v$ at $\alpha$ and $\beta$. However, as in the 6d setting, we have some residual symmetry we can use to fix $\hat g|_{\beta}=\mathrm{id}$, $\hat h|_{\alpha,\beta}=\mathrm{id}$, and similarly, $\hat v|_{\alpha,\beta}=0$. We are thus left with
\begin{equation}
\hat g|_{\alpha}\coloneqq g \,,\quad \hat u|_{\alpha}\coloneqq u\,, \quad \hat u|_{\beta}=\tilde u\,.
\end{equation}
In terms of these variables, the bulk equations of motion of gCS$_4$ theory imply
\begin{equation}
\label{ec:4dCSbulkeom}
\bar{\pd}_{0} \hat A^{\prime}_{i} = 0 \ , \qquad
\bar{\pd}_{0} \hat B^{\prime}_{i} = 0 \ ,
\end{equation}
away from the zeroes of $\omega$, namely $\gamma$ and $\hat \gamma$.
The on-shell gCS$_4$ action can be thus written as
\begin{equation} \label{eq:almostlocalised4d}
\begin{aligned}
& S_{\text{gCS}_4}[\hat A' , \hat B' ] = \frac{1}{2 \pi \rmi} \int_{X} \bar{\pd} \omega \wedge \Tr \big( \hat A^\prime \wedge \bar{\pd} \hat{g} \hat{g}^{-1} -(\hat g^{-1} \hat A^{\prime} \hat{g} + \hat{g}^{-1} \bar \partial \hat{g})\wedge \hat B^{\prime} \big) \\
& \hspace{9em} - \frac{1}{6 \pi \rmi} \int_{X\times[0,1]} \bar \partial\omega \wedge \Tr \big(\hat{g}^{-1} \dr \hat{g} \wedge \hat{g}^{-1} \dr \hat{g} \wedge \hat{g}^{-1} \dr \hat{g} \big) \, .
\end{aligned}
\end{equation}
To obtain the IFT$_2$ we begin by looking at the bulk equations of motion \eqref{ec:4dCSbulkeom}. Liouville's theorem shows that the only bounded, holomorphic functions on $\bCP^{1}$ are constant functions. We are after something a little more general than this, however, as we do not require the components of our gauge field to be bounded at the zeroes of $\omega$.
Indeed, we allow the $w$-component to have a pole at $\pi \sim \gamma$ and the $\bar{w}$-component to have a pole at $\pi \sim \hat{\gamma}$.
With this analytic structure in mind, we can parameterise the solution of the bulk equation for $\hat B^{\prime}$ by
\begin{equation}
\label{ec:expforB}
\hat B^{\prime}_{w} = B_{w} + \frac{\langle \pi \hat{\gamma} \rangle}{\langle \pi \gamma \rangle} \, \Phi \ , \qquad
\hat B^{\prime}_{\bar{w}} = B_{\bar{w}} - \frac{\langle \pi \gamma \rangle}{\langle \pi \hat{\gamma} \rangle} \, \bar \Phi \,,
\end{equation}
where we have conveniently used the field variables introduced in \eqref{ec:2dfieldvar} to ease comparison with \eqref{eq:2daction} after localisation to the IFT$_2$. In particular, under $\pi$-independent gauge transformations $B_{w}, B_{\bar w}$ have the transformation of 2d gauge fields, whilst $\Phi$ and $\bar \Phi$ transform as adjoint scalars.

Note that in the singular piece of these solutions, we have chosen to align the zero of each with the pole of the other. Notice that this choice is also completely general, since moving the zeros in the singular pieces amounts to field redefinitions between $B_w$ and $\Phi$, respectively, $B_{\bar w}$ and $\bar \Phi$.
This is convenient since the flatness condition on $\hat B^{\prime}$ reproduces Hitchin's equations,
\begin{equation}
F_{w \bar{w}}[\hat B^{\prime}] = F_{w \bar{w}}[B] - [ \Phi , \bar \Phi] - \frac{\langle \pi \gamma \rangle}{\langle \pi \hat{\gamma} \rangle} D_{w} \bar \Phi - \frac{\langle \pi \hat{\gamma} \rangle}{\langle \pi \gamma \rangle} D_{w} \Phi \ .
\end{equation}
On the other hand, for the $\hat A^{\prime}$ gauge field a convenient choice of parameterisation when solving the bulk equation of motion \eqref{ec:4dCSbulkeom} is
\begin{equation}
\label{ec:expforA}
\hat A^{\prime}_{i} = \frac{\langle \pi \alpha \rangle}{\langle \pi \gamma \rangle} \frac{\langle \beta \gamma \rangle}{\langle \beta \alpha \rangle} \, U_{i} + \frac{\langle \pi \beta \rangle}{\langle \pi \gamma \rangle} \frac{\langle \alpha \gamma \rangle}{\langle \alpha \beta \rangle} \, V_{i} \,,\quad i=w,\bar w\,.
\end{equation}
This parametrisation, in which we have chosen the coefficients such that one term vanishes at $\pi \sim \alpha$ while the other vanishes at $\pi \sim \beta$, is adapted to the boundary conditions which can be solved for $U_{i}$ and $V_{i}$ to yield
\begin{equation}
\label{ec:Aprimefinal}
\begin{aligned}
\hat A^{\prime}_{w} & = \hat B^{\prime}_{w}- \frac{\langle \pi \beta \rangle}{\langle \pi \gamma \rangle} \frac{\langle \alpha \gamma \rangle}{\langle \alpha \beta \rangle} \bigg( D_{w} g g^{-1} + \frac{\langle \alpha \hat{\gamma} \rangle}{\langle \alpha \gamma \rangle}(1- \Ad_{g} )\Phi \bigg)   \, \\
\hat A^{\prime}_{\bar{w}} & = \hat B^{\prime}_{\bar{w}} - \frac{\langle \pi \beta \rangle}{\langle \pi \hat{\gamma} \rangle} \frac{\langle \alpha \hat{\gamma} \rangle}{\langle \alpha \beta \rangle} \bigg(  D_{\bar{w}} g g^{-1} - \frac{\langle \alpha \gamma \rangle}{\langle \alpha \hat{\gamma} \rangle}(1- \Ad_{g} ) \bar\Phi \bigg)  \ .
\end{aligned}
\end{equation}
Replacing \eqref{ec:expforB} and \eqref{ec:Aprimefinal} in \eqref{eq:almostlocalised4d} and integrating\footnote{To do so, we use the localisation formula in homogeneous coordinates
\begin{equation}
\label{ec:4d2dlocalformula}
\frac{1}{2 \pi \rmi} \int_{X} \bar{\pd} \omega \wedge Q = - \frac{1}{2}\int_{\mathbb{R}^2}
\bigg[
\frac{
\langle \alpha \gamma\rangle\langle \beta \hat \gamma \rangle + \langle \alpha \hat \gamma\rangle \langle \beta \gamma\rangle
}{
\langle \alpha \beta \rangle
} Q \vert_{\alpha}
+ \langle \alpha \gamma\rangle\langle \alpha \hat \gamma\rangle (\pd_{0} Q) \vert_{\alpha} \bigg]
\quad + \quad \alpha \leftrightarrow \beta \ .
\end{equation}
for any $Q \in \Omega^2(X)$.} along $\mathbb{CP}^1$
we recover the IFT$_2$ given in \eqref{eq:2daction}.

\section{Gauged LMP action} \label{Gauged LMP action section}

In the previous sections, we analysed a ghCS$_6$ setup where the meromorphic $(3,0)$-form $\Omega$ had two double poles, showing that such a theory leads to a gauged WZW$_4$ upon localisation to $\mathbb{R}^{4}$. To highlight some of the universal features of this procedure, we will now focus on another example in which the meromorphic $(3,0)$-form will have a single fourth order pole. Such a configuration of the ungauged hCS$_{6}$ was shown in \cite{Bittleston:2022cmz} to lead to the LMP action for ASDYM \cite{Leznov:1986mx}, \cite{Parkes:1992rz}.

\subsection{LMP action from \texorpdfstring{hCS$_6$}{hCS6}}
Let us start by reviewing the ungauged localisation of hCS$_{6}$ with a fourth order pole. We start with the action and $(3,0)$-form defined by
\begin{equation}\label{hCS Action}
S_{\mathrm{hCS_6}}[\cA] = \frac{1}{2 \pi \rmi} \int_{\mathbb{PT}} \Omega \wedge \text{CS}(\cA) \; , \qquad
\Omega = k \, \frac{{e}^{0} \wedge {e}^{\d{a}} \wedge {e}_{\d{a}}}{{\la \pi \alpha \ra }^{4}} \; .
\end{equation}
As is usual in hCS$_{6}$, we must impose boundary conditions on the gauge field $\cA$ to ensure the vanishing of the boundary variation
\begin{equation}
\delta S_{\mathrm{hCS_6}} \big\vert_{\text{bdry}} = \frac{1}{2 \pi \rmi} \int_{\mathbb{PT}} \b{\partial} \Omega \wedge \text{tr}(\delta \mc{A} \wedge \mc{A} ) \; .
\end{equation}
Evaluating the above integral is achieved by making use of the localisation formula (see appendix~\ref{appendix on localisation formulae})
\begin{equation} \label{4th order pole locln}
\frac{1}{2 \pi \rmi} \int_{\bPT} \bar{\pd} \Omega \wedge Q = \frac{k}{6}
\int_{\bR^{4}} \alpha_{a} \alpha_{b} \Sigma^{ab}\wedge \pd_{0}^{3} Q \big\vert_{\alpha} \ .
\end{equation}
Then, one finds that the boundary variation vanishes if we impose the boundary conditions
\begin{equation}
{\mc{A}}{|}_{\pi = \alpha} = 0 \; \; \; \; \text{and} \; \; \; \; {\partial}_{0} {\mc{A}} {|}_{\pi = \alpha} = 0 \; .
\end{equation}

\paragraph{Admissible gauge transformations.}
We now check which residual gauge symmetries survive with the preceding choice of boundary conditions.
We proceed in a familiar fashion, introducing a new parameterisation of our gauge field $\mc{A}$ as
\begin{equation}\label{parameterisation of cA}
\mc{A} = \hat{g}^{-1} \mc{A}' \hat{g} + \hat{g}^{-1} \bar{\partial} \hat{g} \; , \qquad
{\mc{A}}'_{0}=0 \ .
\end{equation}
This parameterisation has both external and internal gauge symmetries which act as
\begin{equation}
\begin{aligned}
\text{External} \quad \hat{\gamma} :& \quad \mc{A} \mapsto \mc{A}^{\hat{\gamma}}  \, , \quad \mc{A}^\prime \mapsto \mc{A}^\prime \, , \quad    \hat{g} \mapsto \hat{g} \hat{\gamma} \ , \\
\text{Internal} \quad \check{\gamma} :& \quad \mc{A} \mapsto \mc{A}\, , \quad \mc{A}^\prime  \mapsto \mc{A}^\prime{}^{\check{\gamma}} \, , \quad    \hat{g} \mapsto \check{\gamma}^{-1} \hat{g} \ .
\end{aligned}
\end{equation}
The internal gauge transformations must satisfy $\b{\partial}_{0}\c{\gamma} = 0$ to preserve the condition ${\mc{A}}'_{0}=0$. These transformations leave the value of $\mc{A}$ invariant and as such they are fully compatible with the boundary conditions. We will use the internal gauge symmetry to fix $\hat{g}{|}_{\pi = \alpha} = \id$. The story for the external gauge symmetries is slightly different, under external gauge transformations $\mc{A} \mapsto \mc{A}^{\hat{\gamma}}$ and so the value of $\mc{A}$ at the poles is not generically invariant. As such we must proceed with caution: we require our boundary conditions to be invariant under external gauge transformations, imposing constraints on the admissible symmetries at $\pi = \alpha$. This limits the amount of symmetry available for gauge fixing. The gauge transformation of the first boundary condition reads
\begin{equation}
{\mc{A}}^{\hat{\gamma}}{|}_{\pi = \alpha } = \l( \hat{\gamma}^{-1} \mc{A} \hat{\gamma} + \hat{\gamma}^{-1} \b{\partial} \hat{\gamma} \r){|}_{\pi = \alpha } = 0 \quad \Longrightarrow \quad
{\gamma}^{-1} \alpha^{a} \partial_{a \d{a}} \gamma = 0 \; ,
\end{equation}
where we have defined
$$\hat{\gamma}{|}_{\pi = \alpha} = \gamma \; .$$
Here, we have derived the fact that at $\pi = \alpha$ we restrict our gauge transformations such that they are holomorphic on $\mathbb{R}^{4}$ with respect to the complex structure given by the point $\pi = \alpha$. Another way of stating this is that our admissible external gauge symmetries on $\mathbb{PT}$ localise to semi-local symmetries in the effective theory on $\mathbb{R}^{4}$. However this restriction is derived from only one half of the boundary conditions. Introducing the notation
$$\hat{\mathbf{\Gamma}}:=\hat{\gamma}^{-1} \partial_{0} \hat{\gamma} \; ,$$
the gauge transformation of the second boundary condition reads
\begin{equation}
\begin{aligned}
\partial_{0} \mc{A}^{\hat{\gamma}} {|}_{\pi = \alpha} =& \; \partial_{0}\l( \hat{\gamma}^{-1} \mc{A} \hat{\gamma} + \hat{\gamma}^{-1} \b{\partial} \hat{\gamma} \r){|}_{\pi = \alpha }= 0 \; , \\
=& \; \left( \l[ \hat{\gamma}^{-1} \mc{A} \, \hat{\gamma}, \, \hat{\mathbf{\Gamma}} \r] + \hat{\gamma}^{-1} \partial_{0} {\mc{A}} \hat{\gamma} + \b{\partial} \hat{\mathbf{\Gamma}} + \l[ \hat{\gamma}^{-1} \b{\partial} \hat{\gamma} , \hat{\mathbf{\Gamma}} \r] +\hat{\gamma}^{-1} \partial_{\dot{a}} \hat{\gamma} \; \b{e}^{\dot{a}} \right){|}_{\pi = \alpha} \ . \\
\end{aligned}
\end{equation}
Imposing the original boundary conditions we arrive at the constraint equation
\begin{equation}
{\alpha}^{a} {\partial}_{a \dot{a}} {\mathbf{\Gamma}} + {\gamma}^{-1}\hat{\alpha}^{a} \partial_{a \dot{a}} \gamma = 0 \; ,
\end{equation}
where we have used $\langle \alpha \hat{\alpha} \rangle = 1$ and defined
$$\hat{\mathbf{\Gamma}}{|}_{\pi= \alpha} = \mathbf{\Gamma} \; .$$
One solution is that the external gauge transformations are global symmetries of the localised effective theory $\dr_{\mathbb{R}^{4}} \gamma = 0$, and $\mathbf{\Gamma}$ is holomorphic on $\mathbb{R}^{4}$ with respect to choice of complex structure given by the point $\alpha \in \mathbb{CP}^{1}$.

Tentatively, our localised theory should have 4 degrees of freedom, known as ‘edge modes',
\begin{equation}
\underline{\mathbf{u}}:= ( g , \mathbf{u}^{1} , \mathbf{u}^{2}, \mathbf{u}^{3}) \; .
\end{equation}
where
\begin{equation}
g = \hat{g} {|}_{\pi = \alpha} \; , \; \; \; \mathbf{u}^{1} : = {\hat{g}}^{-1} \pd_{0}\hat{g} {|}_{\pi = \alpha} \; , \; \; \; \mathbf{u}^{2} : = \hat{g}^{-1} \pd^{2}_{0} \hat{g} {|}_{\pi = \alpha}\; , \; \; \; \mathbf{u}^{3} : = \hat{g}^{-1} \pd^{3}_{0} \hat{g} {|}_{\pi = \alpha}\; .
\end{equation}
However, some of these fields are spurious and can be gauged fixed away using the admissible gauge symmetries. We have already used the internal gauge symmetry to fix $g= \mathrm{id}.$. Furthermore, the second and third $\pd_{0}$-derivatives of the external gauge transformations are unconstrained by the boundary conditions, so they can be used to gauge fix $\mathbf{u}^{2} = \mathbf{u}^{3} = 0$. This leaves us with one dynamical degree of freedom in the localised theory on $\mathbb{R}^{4}$, namely $\mathbf{u}^{1}: \mathbb{R}^{4} \rightarrow \tf{g}$ which we will now denote by $\mathbf{u}$ for brevity. In conclusion, after gauge fixing we have
\begin{equation}\label{Gauge fixing u}
\underline{\mathbf{u}}= ( \id , \mathbf{u} , 0,0 ) \; .
\end{equation}

\paragraph{Solving the boundary conditions.}
Using the boundary conditions, we will solve for $\mc{A} '$ in the parametrisation \eqref{parameterisation of cA} in terms of the edge modes.
The first boundary condition tells us
\begin{equation}
\mc{A}'{|}_{\pi=\alpha} = 0 \quad \Rightarrow \quad
\alpha^{a}{A}_{a \d{a}} = 0 \quad \Rightarrow \quad
A_{a \d{a}} = \alpha_{a} C_{\dot{a}} \ .
\end{equation}
Then, the second boundary condition equation is written as
\begin{equation}
\partial_{0} \mc{A}' {|}_{\pi=\alpha} + \b{\partial} \mathbf{u} = 0 \; ,
\end{equation}
which allows us to conclude that
\begin{equation}
{C}_{\d{\alpha}} = \alpha^{a} \partial_{\alpha \d{\alpha}} \mathbf{u} \; .
\end{equation}
We now have all the ingredients to localise the hCS$_{6}$ action to $\mathbb{R}^{4}$.

\paragraph{Localisation to \texorpdfstring{$\mathbb{R}^{4}$}{R4}.}
We can write the action \eqref{hCS Action} in the new variables as
\begin{equation}
S = \frac{1}{2 \pi \rmi} \int_{\mathbb{PT}} \b{\partial} \Omega \wedge \Tr ( \mc{A}' \wedge \b{\partial} \hat{g} \hat{g}^{-1} )
- \frac{1}{6 \pi \rmi} \int_{\mathbb{PT} \times [0,1]} \bar{\pd} \Omega \wedge \Tr \l( (\hat{g}^{-1} \dr \hat{g})^{3} \r) \; ,
\end{equation}
where in the second term we have extended $\mathbb{PT}$ to the 7-manifold $\mathbb{PT} \times [0,1]$, whose boundary is a disjoint union of two copies of $\mathbb{PT}$. We have also extended our fields via a smooth homotopy $\hat{g} \rightarrow \hat{g}(t)$ so that $\hat{g}(0) = \id$ and $\hat{g}(1) = \hat{g}$. Applying the localisation formula \eqref{4th order pole locln} and the choice of gauge fixing \eqref{Gauge fixing u}, we arrive at the spacetime action
\begin{equation}\label{LMP Action}
S_{\text{LMP}}[\mathbf{u}] = \frac{k}{3} \int_{\mathbb{R}^{4}} \frac{1}{2} \Tr ( \dr \mathbf{u} \; \wedge \star \dr \mathbf{u} ) + \frac{1}{3} \alpha_{a} \alpha_{b} \Sigma^{ab} \wedge \Tr( \, \mathbf{u} \l[ \dr \mathbf{u} , \dr \mathbf{u} \r] \, ) \; .
\end{equation}
We identify the action \eqref{LMP Action} as the LMP model for ASDYM. Upon performing reduction to $\mathbb{R}^{2}$ the above action becomes the pseudo-dual of the PCM \cite{Liniado:2023uoo}.

\subsection{Gauged LMP action from \texorpdfstring{ghCS$_{6}$}{ghCS6}}
In the previous subsection, we derived the LMP action from hCS$_6$. Next, we shall consider the same fourth order pole structure for gauged hCS$_{6}$. The starting point is to calculate the boundary variation and make a choice of isotropic subspace such that it vanishes.

\paragraph{Boundary conditions.}
Starting from the action
\begin{equation} \label{4th pole ghCS6}
S_{\mathrm{ghCS_6}}[\cA , \cB] = S_{\mathrm{hCS_6}}[\cA] - S_{\mathrm{hCS_6}}[\cB]
- \frac{1}{2\pi \rmi} \int_{\bPT} \bar{\partial} \Omega \wedge \Tr \big( \cA \wedge \cB \big) \, ,
\end{equation}
the boundary variation is given by
\begin{equation}
\delta S_{\mathrm{ghCS_6}} \big\vert_{\text{bdry}} = \frac{1}{2 \pi \rmi} \int_{\mathbb{PT} } \bar{\partial} \Omega \wedge \Tr \big( \delta \cA \wedge (\cA - \cB) - \delta \cB \wedge (\cB - \cA) \big) \; .
\end{equation}
Following in a parallel fashion to the hCS$_{6}$ case, one finds a suitable choice of boundary conditions is given by
\begin{equation}
\cA {|}_{\pi = \alpha} = \cB{|}_{\pi = \alpha} \; , \quad \partial_{0}\cA {|}_{\pi = \alpha} = \partial_{0}\cB{|}_{\pi = \alpha} \; , \quad \partial^{2}_{0}\cA^{\tf{h}} {|}_{\pi = \alpha} = \partial^{2}_{0}\cB{|}_{\pi = \alpha}
\; , \quad \partial^{3}_{0}\cA^{\tf{h}} {|}_{\pi = \alpha} = \partial^{3}_{0}\cB{|}_{\pi = \alpha} \;.
\end{equation}

\paragraph{Gauge fixing.}
Gauge fixing will once again prove dividends in completing the localisation calculation, as such, we will consider the set of admissible gauge transformations respecting our boundary conditions. Performing a gauge transformation on the first boundary condition, one arrives at
\begin{equation}
\left( \hat{\gamma}^{-1}\cA \hat{\gamma} + \hat{\gamma}^{-1} \b{\pd} \hat{\gamma} \right){|}_{\pi = \alpha} = \left( \hat{\eta}^{-1}\cB \hat{\eta} + \hat{\eta}^{-1} \b{\pd} \hat{\eta} \right){|}_{\pi = \alpha} \ ,
\end{equation}
from which one concludes that the admissible gauge transformations must obey
$\hat{\gamma}{|}_{\alpha} = \hat{\eta}{|}_{\alpha} $. Running through systematically, the second boundary condition requires
\begin{equation}
\begin{aligned}
\left( \l[ \hat{\gamma}^{-1} \cA \, \hat{\gamma} + \hat{\gamma}^{-1} \b{\pd} \hat{\gamma}, \, \hat{\mathbf{\Gamma}} \r] + \right. & \left. \hat{\gamma}^{-1} \pd_{0} {\mc{A}} \hat{\gamma} + \b{\partial} \hat{\mathbf{\Gamma}} +\hat{\gamma}^{-1} \partial_{\dot{a}} \hat{\gamma} \; \b{e}^{\dot{a}} \right){|}_{\pi = \alpha} \\
&= \left( \l[ \hat{\eta}^{-1} \cB \, \hat{\eta} + \hat{\eta}^{-1} \b{\pd} \hat{\eta}, \, \hat{\mathbf{N}} \r] + \hat{\eta}^{-1} \pd_{0} {\mc{B}} \hat{\eta} + \b{\partial} \hat{\mathbf{N}} +\hat{\eta}^{-1} \partial_{\dot{a}} \hat{\eta} \; \b{e}^{\dot{a}} \right){|}_{\pi = \alpha} \; ,
\end{aligned}
\end{equation}
where we have denoted $\hat{\mathbf{\Gamma}} = \hat{\gamma}^{-1} \pd_{0} \hat{\gamma}$ and $\hat{\mathbf{N}} = \hat{\eta}^{-1} \pd_{0} \hat{\eta}$. Making use of the original boundary condition and the constraint $\hat{\gamma}{|}_{\alpha} = \hat{\eta}{|}_{\alpha} $, we conclude that admissible gauge transformations must also obey $\hat{\mathbf{\Gamma}}{|}_{\pi = \alpha} = \hat{\mathbf{N}}{|}_{\pi = \alpha}$.
In a similar fashion, from the third boundary condition
we conclude that $\hat{\mathbf{\Gamma}}^{(2)}_{\tf{h}}{|}_{\alpha} = \hat{\mathbf{N}}^{(2)}{|}_{\alpha}$ where $\hat{\mathbf{\Gamma}}^{(2)} := \hat{\gamma}^{-1} \pd_{0}^{2} \hat{\gamma}$ and $\hat{\mathbf{N}}^{(2)} := \hat{\eta}^{-1} \pd_{0}^{2} \hat{\eta}$.
Finally, from the fourth boundary condition
we find $\hat{\mathbf{\Gamma}}^{(3)}_{\tf{h}}{|}_{\alpha} = \hat{\mathbf{N}}^{(3)}{|}_{\alpha}$ where $\hat{\mathbf{\Gamma}}^{(3)}:= \hat{\gamma}^{-1} \pd_{0}^{3} \hat{\gamma}$ and $\hat{\mathbf{N}}^{(3)} := \hat{\eta}^{-1} \pd_{0}^{3} \hat{\eta}$.

Armed with the admissible gauge symmetries of our theory, we set ourselves the task of gauge fixing our degrees of freedom. Naively, we would think there are 8 degrees of freedom in our theory,
\begin{equation}
\begin{aligned}
\underline{\mathbf{u}} &:= \l( g , \mathbf{u} , \mathbf{u}^{2}, \mathbf{u}^{3} \r) \; , \\
\underline{\mathbf{v}} &:= \l( h , \mathbf{v}^{1} , \mathbf{v}^{2}, \mathbf{v}^{3} \r) \ . \\
\end{aligned}
\end{equation}
First, one considers the internal gauge symmetries of $\cA$ and $\cB$ which one can use to set both $g$ and $h$ to the identity. Next, one should note that the $H$-valued external gauge transformations of $\cB$ parameterised by $\hat{\eta}$ are unconstrained at the point $\pi = \alpha$. As such, one can gauge fix $\mathbf{v}^{i} = 0$ for $i = 1,2,3$. Now, as our external gauge transformations of $\cA$ parameterised by $\hat{\gamma}$ are constrained to coincide with $\hat{\eta}$ at $\pi = \alpha$, and we have used these symmetries in our choice of gauge fixing, we find that we are unable to gauge fix $\mathbf{u}^{i}$. As such, each of these degrees of freedoms will appear as fields in our effective theory on $\mathbb{R}^{4}$. In summary, after gauge fixing one has,
\begin{equation}\label{Gauge fixed u and v}
\begin{aligned}
\underline{\mathbf{u}} &= \l( \id , \mathbf{u} , \mathbf{u}^{2}, \mathbf{u}^{3} \r) \; , \\
\underline{\mathbf{v}} &= \l( \id , 0 , 0, 0 \r) \ . \\
\end{aligned}
\end{equation}

\paragraph{Solving the boundary conditions.}
The first boundary condition reads
\begin{equation}
\l( \hat{g}^{-1} \cA ' \hat{g} + \hat{g}^{-1} \b{\pd} \hat{g} \r){|}_{\pi = \alpha} = ( \hat{h}^{-1} \cB ' \hat{h} + \hat{h}^{-1} \b{\pd} \hat{h} ){|}_{\pi = \alpha} \ .
\end{equation}
Given our choice of gauge fixing \eqref{Gauge fixed u and v} and the explicit solutions e.g.\ $\cA^{\prime}_{\dot{a}} = \pi^{a} A_{a \dot{a}}$, this implies
\begin{equation}\label{1st BC}
\cA ' {|}_{\pi = \alpha}= \cB'{|}_{\pi = \alpha} \quad \Rightarrow \quad
\alpha^{a}{A}_{a \d{a}} = \alpha^{a} {B}_{a \d{a}} \quad \Rightarrow \quad
{A}_{a \d{a}} = {B}_{a \d{a}} -{\alpha}_{a}{Q}_{\d{a}} \;.
\end{equation}
We can then use the second boundary condition to solve for ${Q}_{\d{a}}$,
\begin{equation}
\partial_{0} \cA {|}_{\pi = \alpha} = \partial_{0} \cB {|}_{\pi = \alpha} \; , \qquad \implies \qquad
{Q}_{\d{a}} = -{\alpha}^{a} \l( \l[ {B}_{a \d{a}} , \mathbf{u} \r] + \pd_{a \dot{a}} \mathbf{u} \r) = - \alpha^{a} \nabla_{a \d{a}} \mathbf{u} \; .
\end{equation}
These two boundary conditions are sufficient to solve for $A_{a \dot{a}}$ in terms of the other degrees of freedom,
\begin{equation}
{A}_{a \d{a}} = {B}_{a \d{a}} + {\alpha}_{a} \alpha^{b} \nabla_{b \d{a}} \mathbf{u} \; .
\end{equation}

\paragraph{Localisation to \texorpdfstring{$\mathbb{R}^{4}$}{R4}.}
Writing the action \eqref{4th pole ghCS6} in terms of the new field variables,
one can see the only terms that will contribute to the effective action given our choice of gauge \eqref{Gauge fixed u and v} will be
\begin{equation}
\begin{aligned}
S_{\mathrm{ghCS_6}} = \frac{1}{2 \pi i} \int_{\mathbb{PT}} \b{\pd} \Omega \wedge \Tr( \cA' \wedge \b{\pd} \hat{g} \hat{g}^{-1} - \hat{g}^{-1} \cA' \hat{g} \wedge \cB' - {\hat{g}^{-1} \b{\pd} \hat{g} \wedge \cB'} ) - \frac{1}{6\pi i}\int_{\mathbb{PT} \times [0,1]} \b{\pd}\Omega \wedge \Tr\l((\hat{g}^{-1} \dr \hat{g} )^{3} \r) \; .
\end{aligned}
\end{equation}
The localisation calculation of the gauged model is slightly more involved than the ungauged case due to the additional degrees of freedom appearing. However, as seen in the calculations in previous sections we expect $\mathbf{u}^{2}$ and $\mathbf{u}^{3}$ to appear only as Lagrange multipliers, in particular imposing self-duality type constraints for our gauge field $B$. With this tenet in mind, one can show that the 4d theory is given by
\begin{equation}\label{Gauged LMP action}
\begin{aligned}
S_{\text{gLMP}}[\mathbf{u},B] = k \int_{\mathbb{R}^{4}} \vol_{4} \frac{1}{2} \Tr \, (\nabla^{a \d{a}} \mathbf{u} \nabla_{a \d{a}} \mathbf{u}&) + \frac{1}{3} \epsilon^{\d{a} \d{b}} \, \Tr \, ( \mathbf{u}  \l[ \alpha^{a} \nabla_{a \d{a}} \mathbf{u} , \alpha^{b} \nabla_{b \d{b}} \mathbf{u}  \r] ) +  \mathbf{u} \; \epsilon^{\d{a} \d{b}} \hat{\alpha}^{a} \hat{\alpha}^{b} {F}_{a \d{a} b \d{b}} (B) \\
& + \frac{1}{2}\mathbf{u}^{2} \epsilon^{\d{a} \d{b}} \l( {\alpha}^{a} \hat{\alpha}^{b} + \hat{\alpha}^{a} \alpha^{b} \r) {F}_{a \d{a} b \d{b}}(B) + \tilde{\mathbf{u}}^{3} \epsilon^{\d{a} \d{b}} \alpha^{a} \alpha^{b} {F}_{a \d{a} b \d{b}}(B) \; ,
\end{aligned}
\end{equation}
where we have performed a field redefinition $\mathbf{u}^{3} \rightarrow \tilde{\mathbf{u}}^{3}:= \frac{1}{6}( \mathbf{u}^{3} + 2\l[ \mathbf{u} , \mathbf{u}^{2} \r])$. Upon reducing along a particular $\mathbb{R}^{2}$ subgroup, and appropriately performing redefinitions of our fields and parameters, one finds that the gauged LMP action matches the two-dimensional action eq.~\eqref{eq:actklmp}.

\paragraph{Implementing the Lagrange multipliers.}
In section \ref{Yang's Matrix Section} we reviewed how solutions to the ASDYM can be formulated in terms of a Yang's matrix after a partial gauge fixing of the ASD connection. In this section we will look to integrate out our Lagrange multiplier fields present in the action eq.~\eqref{Gauged LMP action} by solving the self duality constraints they impose in a similar vein.
Indeed, one may understand the LMP equations of motion as the remaining ASDYM equation after these two constraints have been solved. This is analogous to the statement that the WZW$_{4}$ equation of motion is the remaining ASDYM equation for Yang's matrix.

The equation of motion found by varying $\tilde{\mathbf{u}}^{3}$ is an integrability condition along the 2-plane defined by $\alpha^{a}$, and it maybe be solved by
\begin{equation}
\epsilon^{\d{a} \d{b}} \alpha^{a} \alpha^{b} {F}_{a \d{a} b \d{b}}(B) = 0 \qquad \implies \qquad
\alpha^{a}{B}_{a \d{a}} = {h}^{-1} \alpha^{a} \partial_{a \d{a}}{h} \; ,
\end{equation}
where $h \in C^{\infty}(\mathbb{R}^{4}) \otimes H$. It is helpful to parameterise the remaining degrees of freedom in $B_{a \dot{a}}$ in terms of a new field $C_{\dot{a}}$, defined by the relation
\begin{equation}
B_{a \dot{a}} = h^{-1} \pd_{a \dot{a}} h - \alpha_{a} \, h^{-1} C_{\dot{a}} h \ .
\end{equation}
Then, the $\mathbf{u}^{2}$ equation of motion becomes
\begin{equation}
\epsilon^{\d{a} \d{b}} \l( {\alpha}^{a} \hat{\alpha}^{b} + \hat{\alpha}^{a} \alpha^{b} \r) {F}_{a \d{a} b \d{b}}(B) = 0 \qquad \Longleftrightarrow \qquad
\epsilon^{\d{a} \d{b}} \alpha^{a} \pd_{a \dot{a}} C_{\dot{b}} = 0 \ .
\end{equation}
This may be solved explicitly by ${C}_{\d{a}} = \alpha^{a} \partial_{a \d{a}} f$ for $f \in C^{\infty}(\mathbb{R}^{4}) \otimes \tf{h}$, such that the gauge field $B$ is given by
\begin{equation}\label{on-shell gauge field B}
{B}_{a\d{a}} = {h}^{-1} \partial_{a \d{a}} h + {h}^{-1} X_{a \d{a}} h \; , \; \text{where} \; \; {X}_{a \d{a}} = - {\alpha}_{a} {\alpha}^{b} \partial_{{b} \d{a}} f \; .
\end{equation}
Reinserting this expression into the action eq.~\eqref{Gauged LMP action}, the resulting theory may be written as a difference of two LMP actions. This can be done by performing a field redefinition $h \mathbf{u} {h}^{-1} = v - f$, for $v \in C^{\infty}(\mathbb{R}^{4}) \otimes \tf{g}$, such that one arrives at the action
\begin{equation}
\begin{aligned}
S_{\text{gLMP}}[\mathbf{u},B] = \mathscr{k} \int_{\mathbb{R}^{4}} \frac{1}{2} \Tr ( \dr v \wedge \star \dr v ) + & \frac{1}{3} \alpha_{a} \alpha_{b} \Sigma^{ab} \wedge \Tr(v \l[ \dr v , \dr v \r] ) \\
& - \mathscr{k} \int_{\mathbb{R}^{4}} \frac{1}{2} \Tr ( \dr f \wedge \star \dr f ) + \frac{1}{3} \alpha_{a} \alpha_{b} \Sigma^{ab} \wedge \Tr(f \l[ \dr f , \dr f \r] ) \; .
\end{aligned}
\end{equation}
This demonstrates the conclusion
\begin{equation}
S_{\text{gLMP}}[\mathbf{u},B] = S_{\text{LMP}}[v] - S_{\text{LMP}}[f] \; .
\end{equation}

\section{Outlook}\label{sec:outlook}

The construction presented in this work has led us to new integrable field theories in both four and two dimensions.
We conclude by highlighting a number of interesting future directions prompted by these results.

Motivated by the observation that the gauged WZW model on the coset $G/H$ in two dimensions can be written as the difference of WZW models for the groups $G$ and $H$, we took the difference of two hCS$_6$ theories as our starting point.
The boundary conditions~\eqref{eq:ABboundarycond} led us to add a boundary term resulting in the action~\eqref{eq:new6daction}.
It is worth highlighting that the boundary variation vanishes on the boundary conditions~\eqref{eq:ABboundarycond} whether or not the boundary term is included, and the contribution of the boundary term to the IFT$_4$ vanishes if we invoke all the boundary conditions.
However, while the algebraic boundary conditions, $\mathcal{A}^{\mathfrak{k}}|_{\alpha,\beta}=0$ and $\mathcal{A}^{\mathfrak{h}}|_{\alpha,\beta}=\mathcal{B}_{\alpha,\beta}$ can be straightforwardly solved, this is not the case for the differential one $\partial_0\mathcal{A}^{\mathfrak{h}}|_{\alpha,\beta}=\partial_0\mathcal{B}_{\alpha,\beta}$.
Therefore, we relaxed this condition meaning that the contribution of the boundary term no longer vanishes.
Remarkably, for the specific boundary term added in~\eqref{eq:new6daction}, the constraints implied by the differential boundary condition now follow as on-shell equations of motion, leading to fully consistent IFT$_4$ and IFT$_2$.

There are compelling reasons to follow this strategy, including that the symplectic potential becomes tautological upon including the boundary term.
However, it remains to understand why the differential boundary condition can be consistently dropped for this particular choice of boundary term, and a systematic interpretation of this is an open question.
To address this, it would be appropriate to pursue a more formal study, complementing a homotopic analysis (along the lines done for CS$_4$ in~\cite{Benini:2020skc}) with a symplectic/Hamiltonian study of the 6d holomorphic Chern-Simons theory (similar to \cite{Vicedo:2019dej}  in the context of CS$_4$).  

A second arena for formal development is the connection between 6d holomorphic Chern-Simons and five-dimensional K\"ahler Chern-Simons (KCS$_{5}$) theory \cite{Nair:1990aa,Nair:1991ab}.  This should mirror the relationship between CS$_{4}$ and CS$_{3}$ theories described by Yamazaki \cite{Yamazaki:2019prm}.  To make this suggestion precise in the present context one may consider a Kaluza-Klein expansion around the $U(1)$ rotation in the $\mathbb{CP}^1$ that leaves fixed the location of the double poles, retaining the transverse coordinate as part of the bulk five-manifold of KCS$_{5}$. The details of this are left for future study.

It would also be interesting to explore the new integrable IFT$_4$ and IFT$_2$ that we have constructed.
$G/H$ coset CFTs in two dimensions have a rich spectrum of paraferminonic operators \cite{Bardakci:1990lbc,BARDAKCI1991439}. It would be very interesting to establish the lift or analogue of these objects in the context of the IFT$_4$.
The natural framework for this is likely to involve the study of co-dimension one defects and associated higher-form symmetries.

For abelian $H$ we find IFT$_2$ that, in the $\kk \to 1$ limit, are related to massive integrable perturbations of the $G/H$ gauged WZW models known as homogeneous sine-Gordon models~\cite{Park:1994bx,Fernandez-Pousa:1996aoa}.
These include the sine-Gordon and complex sine-Gordon models as special cases, two of the most well-understood IFT$_2$.
There is nothing in our construction that prohibits non-abelian $H$ and it would be instructive to study the resulting models in more detail.
 
An important class of IFT$_2$ are the symmetric space sigma models.  These can be constructed either by restricting fields to parameterise $G/H$ directly or by gauging a left action of H on the PCM.
These theories have been realised in CS$_{4}$ through branch cut defects \cite{Costello:2019tri} and recently in hCS$_{6}$ \cite{LewisPeter}.
One might explore the realisation the gauging construction of such models within the current framework, and generalise to $\mathbb{Z}_4$ graded semi-symmetric spaces (relevant for applications of CS$_4$ to string worldsheet theories \cite{Costello:2020lpi,PhysRevD.109.106015}).      

When $G/H$ is a symmetric space, an alternative class of massive integrable perturbations of the $G/H$ gauged WZW model are known as the symmetric space sine-Gordon models \cite{Bakas:1995bm,Fernandez-Pousa:1996aoa}.
In the landscape of IFT$_2$ these are related to the $\lambda\to0$ limit~\cite{Hollowood:2014rla,Hoare:2015gda} of the $\lambda$-deformation of the symmetric space sigma model~\cite{Sfetsos:2013wia}.
Note that $\kk \to 1$ and $\lambda \to 0$ both correspond to conformal limits and it would be instructive to explore the relation between the two constructions.
More generally, it would be interesting to generalise the construction in this work to deformed models, in particular splitting one or both double poles in the meromorphic (3,0)-form $\Omega$ into simple poles, or dual models, for example considering the alternative boundary conditions~\eqref{natd}.

Finally, recently novel approaches to constructing IFT$_3$ using higher Chern-Simons theory in 5d has been explored in \cite{Schenkel:2024dcd,Chen:2024axr}. 
Given that there is an overlap between the models that can be obtained from these constructions and from hCS$_6$, or more precisely its reduction to five dimensions, CS$_5$ on the mini-twistor correspondence space $\mathbb{PN}$ ~\cite{Bittleston:2020hfv}, it would be exciting to understand the link between the two, and investigate the existence of categorical generalisations of hCS$_6$.

\section*{Acknowledgements}
The authors would like to thank Alex Arvanitakis, Roland Bittleston, Falk Hassler, Tim Hollowood, Sylvain Lacroix, Marc Magro and Benoit Vicedo for interesting discussions.
This work was first presented by D.C.T. at the Workshop on Integrable Sigma-Models held in ETH Zurich in May 2024.
B.H and D.C.T. would also like to thank the organisers and participants for numerous stimulating and lively discussions.
B.H. is supported by a UKRI Future Leaders Fellowship (grant number MR/T018909/1). The work of J.L. is supported by CONICET.
D.C.T. is supported by the STFC grant ST/X000648/1, and by the Royal Society through a University Research Fellowship `Generalised Dualities in String Theory and Holography' URF 150185. For the purpose of open access, the authors have applied a Creative Commons Attribution (CC BY) licence to any Author Accepted Manuscript version arising. No new data were generated for this manuscript.

\appendix
\section{Spinor and Differential Form Conventions }
\label{sec:spinorconventions}

We work on $\mathbb{R}^4$ and define coordinates in bispinor notation by
\begin{equation}
x^{a \dot{a}} = \frac{1}{\sqrt{2}} \begin{pmatrix}
x_0 +\rmi x_1 & x_2 +\rmi x_3 \\ -x_2 + \rmi x_3 & x_0 - \rmi x_1
\end{pmatrix} \, .
\end{equation}
We fix orientation such that $\star 1 = \vol_4= \dr x_0 \wedge \dr x_1 \wedge \dr x_2 \wedge\dr x_3$. For 1-forms $\sigma = \sigma_{a \dot{a}} \dr x^{a \dot{a}} $ and $\tau = \tau_{a \dot{a}} \dr x^{a \dot{a}} $ we have
\begin{equation}
\star^2 \sigma = - \sigma \, , \quad \sigma \wedge \star \tau = - \star \sigma \wedge \tau = \vol_4 \epsilon^{ab} \epsilon^{\dot{a} \dot{b}} \sigma_{a \dot{a}}\tau_{b \dot{b}} \, , \quad \dr \star \sigma = \vol_4 \epsilon^{ab} \epsilon^{\dot{a} \dot{b}} \pd_{a \dot{a}}\sigma_{b \dot{b}} \, .
\end{equation}
Contraction of spinors is given by
\begin{equation}
\langle \alpha \beta \rangle = \alpha_1 \beta_2 - \alpha_2 \beta_1 = \alpha^a \beta_a
\end{equation}
and raising is achieved by
\begin{equation}
\alpha^a = \epsilon^{ab }\alpha_b \, \quad \epsilon^{12} = - \epsilon^{21} =-1 \, .
\end{equation}
We define $\epsilon_{12} = + 1$ such that $\epsilon^{a b} \epsilon_{b c} = \delta^a_{c}$. The (quaternionic) conjugation of a spinor $\alpha_a = (\alpha_1, \alpha_2)$ is defined to be $\hat{\alpha}_a = (- \bar{\alpha}_2 , \bar{\alpha}_1) $. Identical definitions hold for the anti-chiral spinors with dotted indices and contraction denoted with square brackets though these don't enter in this work.

A basis for self dual two forms is given by
\begin{equation}
\Sigma^{ab} = \epsilon_{\dot{a} \dot{b}} \dr x^{a \dot{a}} \wedge \dr x^{b\dot{b}} \, ,
\end{equation}
from which given any two spinors we can define self dual forms
\begin{equation}
\Sigma_{\alpha,\beta} = \alpha_a \beta_b \Sigma^{ab} = \alpha_a \beta_b \epsilon_{\dot{a} \dot{b}} \dr x^{a \dot{a}} \wedge \dr x^{b\dot{b}} \, , \qquad \star \Sigma_{\alpha,\beta} = \Sigma_{\alpha,\beta} \, .
\end{equation}
As they will play key roles let we denote
\begin{equation}
\omega_{\alpha, \beta} = \frac{1}{\langle \alpha \beta \rangle } \Sigma_{\alpha,\beta} \, , \quad \mu_\alpha = \Sigma_{\alpha,\alpha }\, , \quad \mu_\beta = \Sigma_{\beta,\beta} \, .
\end{equation}

$\mathbb{R}^4$ is equipped with a Hyper K\"ahler structure and has a $\mathbb{CP}^1$'s worth of complex structures. We can compactly express the corresponding complex structure to a spinor $\gamma_a$ as
\begin{equation}\label{eq:complexStruct}
{\mathcal J}_\gamma = - i (\gamma^a \pd_{a\dot{a}} ) \otimes (\hat{\gamma}_b \dr x^{b \dot{a}} )- i (\hat{\gamma}^a \pd_{a\dot{a}} ) \otimes (\gamma_b \dr x^{b \dot{a}} ) \, ,
\end{equation}
to which adapted complex coordinates are given by
\begin{equation}
\begin{aligned}
\dr z = \gamma_a \kappa_{\dot a} \dr x^{a \dot a } ~, \qquad &
\dr\bar{z} = \hat{\gamma}_a \hat{\kappa}_{\dot a} \dr x^{a \dot a } ~, \qquad
\dr w = \gamma_a \hat \kappa_{\dot a} \dr x^{a \dot a } ~, \qquad
\dr \bar{w} = - \hat \gamma_a \kappa_{\dot a} \dr x^{a\dot a } ~.
\end{aligned}
\end{equation}
With these coordinates we have that
\begin{align}
\mu_\alpha = -2 \langle \alpha \gamma\rangle^2 \dr \bar{w} \wedge \dr \bar{z}- 2 \langle \alpha \gamma\rangle \langle \alpha \hat{\gamma}\rangle ( \dr z \wedge \dr \bar{z} +\dr w \wedge \dr \bar{w} )-2 \langle \alpha \hat{\gamma}\rangle^2 \dr w \wedge \dr z \\
\omega_{\alpha, \beta} = -2 \frac{\langle\alpha \hat\gamma \rangle \langle\beta \hat{ \gamma} \rangle }{\langle\alpha \beta \rangle } \dr w \wedge \dr z -2 \frac{\langle\alpha \gamma \rangle \langle\beta \gamma \rangle }{\langle\alpha \beta \rangle } \dr \bar{w} \wedge \dr \bar{z} - \frac{ \langle\alpha \gamma \rangle \langle\beta \hat{ \gamma} \rangle +\langle\alpha \hat\gamma \rangle \langle\beta \gamma \rangle }{\langle\alpha \beta \rangle} ( \dr z \wedge \dr \bar{z} +\dr w \wedge \dr \bar{w} )
\end{align}
Notice that if we align the spinor $\alpha$ to $\gamma$ and $\beta$ to $\hat{\gamma}$ then $\omega_{\gamma, \hat{\gamma}}$ is (proportional to) the corresponding K\"ahler form $\varpi$ of type $(1,1)$ and $\mu_\gamma$ is a holomorphic $(2,0)$-form and $\mu_{\hat{\gamma}}$ is $(0,2)$-form.

\section{Twistor Space} \label{sec:TwistorSpace}
In this work we will be working on the Euclidean slice of Penrose's twistor space, $\mathbb{PT}_{\mathbb{E}}$. Starting from the twistor space of complexified Minkowski space,
\begin{equation}
\mathbb{PT} = \mathbb{CP}_3 / \mathbb{CP}_1 = \{ Z^\alpha = (\omega^{\dot{a}} , \pi_a ) \} \vert \pi_a \neq 0 \, , Z^\alpha \sim r Z^\alpha ~~ r\in \mathbb{C}^\times \} \, ,
\end{equation}
we obtain $\mathbb{PT}_{\mathbb{E}}$ by making a choice of reality conditions, in particular by selecting the slice of $\mathbb{PT}$ invariant under the anti-holomorphic (quartic)-involution acting on the holomorphic coordinates as ${Z}^{\alpha} \mapsto \hat{Z}^{\alpha} = (\hat{\omega}^{\d{a}} , \hat{\pi}_{a})$. This choice of reality conditions induces a double fibration and we find that in particular Euclidean twistor space can be viewed as the holomorphic vector bundle $\mathbb{PT}_{\mathbb{E}} \cong \mc{O}(1) \oplus \mc{O}(1) \rightarrow \mathbb{CP}^{1}$, where the holomorphic coordinates along the fibre direction are given by the incidence relations $\omega^{\d{a}} = {x}^{a \d{a}} {\pi}_{a}$. With this we choose a basis of $(1,0)$-forms and $(0,1)$-forms
\begin{equation}
\begin{aligned}
& e^0 = \langle \pi \dr \pi \rangle ~, &&
& e^{\dot{a}} = \pi_{a} \dr x^{a \dot{a} } ~, \\
& \bar{e}^0 = \frac{\langle \hat{\pi} \dr \hat{\pi} \rangle}{\langle \pi \hat{\pi} \rangle^2} ~, &&
& \bar{e}^{\dot{a}} = \frac{\hat{\pi}_{a} \dr x^{\dot{a} a }}{\langle \pi \hat{\pi} \rangle} ~.
\end{aligned}
\end{equation}
and their dual vector fields
\begin{equation}
\label{eq:twistorderivatives}
\begin{aligned}
& \pd_0 = \frac{\hat{\pi}_{a}}{\langle \pi \hat{\pi} \rangle} \frac{\pd}{\pd \pi_{a}} ~, &&
& \pd_{\dot{a}} = - \frac{\hat{\pi}^{a} \pd_{a \dot{a} }}{\langle \pi \hat{\pi} \rangle} ~, \\
& \bar{\pd}_0 = - \langle \pi \hat{\pi} \rangle \, \pi_{a} \frac{\pd}{\pd \hat{\pi}_{a}} ~, &&
& \bar{\pd}_{\dot{a}} = \pi^{a} \pd_{a \dot{a} } ~.
\end{aligned}
\end{equation}
It is important to note that this basis of $1$-forms, and their duals, enjoy the structure equations,
\begin{equation}
\begin{aligned}
\bar{\pd} e^{\d{a}} = e^0 \wedge \bar{e}^{\d{a}} ~, \quad
\pd \bar{e}^{\d{a}} = e^{\d{a}} \wedge \bar{e}^0 ~,\\
\comm{\bar{\pd}_0}{\pd_{\d{a}}} = \bar{\pd}_{\d{a}} ~, \quad
\comm{\bar{\pd}_{\d{a}}}{\pd_0} = \pd_{\d{a}} ~.
\end{aligned}
\end{equation}

\subsection{Homogeneous and Inhomogeneous Coordinates}

Homogeneous coordinates on $\bCP^{1}$ will be denoted by $\pi_{a} = (\pi_{1} , \pi_{2})$ which are defined up to the equivalence relation $\pi_{a} \sim s \, \pi_{a}$ for any non-zero $s \in \bC^{\ast}$.
These have the advantage of being globally defined on $\bCP^{1}$ but can lead to technical challenges in certain calculations.
It can also be useful to work with inhomogeneous coordinates on two patches covering $\bCP^{1} \cong S^{2}$.
Introducing an arbitrary spinor $\gamma_{a}$ which satisfies $\langle \gamma \hat{\gamma} \rangle = 1$, the two patches covering $\bCP^{1}$ will be defined by
\begin{equation}
U_{1} = \{ \pi_{a} \mid \langle \pi \hat{\gamma} \rangle \neq 0 \} \ , \qquad
U_{2} = \{ \pi_{a} \mid \langle \pi \gamma \rangle \neq 0 \} \ .
\end{equation}
Inhomogeneous coordinates may be defined on each patch by
\begin{equation}
\zeta = \frac{
\langle \gamma \pi \rangle
}{
\langle \pi \hat{\gamma} \rangle
} \ , \qquad
\xi = \frac{
\langle \pi \hat{\gamma} \rangle
}{
\langle \gamma \pi \rangle
} \ , \qquad
\xi = \zeta^{-1} \ .
\end{equation}
In this section, we will restrict our attention to $U_{1}$ and the inhomogeneous coordinate $\zeta$, knowing that an analogous discussion may be had on the other patch.
The complex conjugate of the inhomogeneous coordinate $\zeta$ is
\begin{equation}
\bar{\zeta} = - \frac{
\langle \hat{\pi} \hat{\gamma} \rangle
}{
\langle \gamma \hat{\pi} \rangle
} \ .
\end{equation}

Forms and vector fields on $\bCP^{1}$ written in these coordinates are related to one another by
\begin{equation}
\begin{aligned}
\dr \zeta & = \frac{
e^{0}
}{
\langle \pi \hat{\gamma} \rangle^{2}
} \ , \quad & \quad
\dr \bar{\zeta} & = \frac{
\langle \pi \hat{\pi} \rangle^{2}
}{
\langle \gamma \hat{\pi} \rangle^{2}
} \bar{e}^{0} \ , \\
\pd_{\zeta} & = \langle \pi \hat{\gamma} \rangle^{2} \pd_{0} \ , \quad & \quad
\pd_{\bar{\zeta}} & = \frac{
\langle \gamma \hat{\pi} \rangle^{2}
}{
\langle \pi \hat{\pi} \rangle^{2}
} \bar{\pd}_{0} \ .
\end{aligned}
\end{equation}
It is also helpful to defined a weight zero basis of $(1,0)$-forms on $\bR^{4} \subset \bPT$ by
\begin{equation}
\theta^{\dot{a}} = \frac{
e^{\dot{a}}
}{
\langle \pi \hat{\gamma} \rangle
}
= \dr x^{a \dot{a}} \, \gamma_{a} + \zeta \, \dr x^{a \dot{a}} \, \hat{\gamma}_{a} \ .
\end{equation}
Likewise the weight zero basis of $(0,1)$-forms on $\mathbb{R}^{4} \subset \mathbb{PT}$ are defined by
\begin{equation}
\b{\theta}^{\d{a}} = \la \pi \hat{\gamma} \ra \, \b{e}^{\d{a}} = \frac{1}{1+\zeta \b{\zeta}}\l( dx^{a \d{a}} \, \hat{\gamma}_{a} - \b{\zeta} \, d x^{a \d{a}} \, \gamma_{a} \r) \; .
\end{equation}
Given a point on $\bCP^{1}$ defined by $\alpha_{a}$ in homogeneous coordinates, we denote the corresponding point in the inhomogeneous coordinate $\zeta$ by
\begin{equation}
\alpha = \frac{
\langle \gamma \alpha \rangle
}{
\langle \alpha \hat{\gamma} \rangle
}
= \zeta \vert_{\pi_{a} \sim \alpha_{a}} \ .
\end{equation}
We also have the relationship
\begin{equation} \label{eq:hom_inhom_identity}
\frac{
\langle \pi \alpha \rangle
}{
\langle \pi \hat{\gamma} \rangle
\langle \hat{\gamma} \alpha \rangle
}
= (\zeta - \alpha) \ .
\end{equation}

\section{Projector Technology}
\label{sec:appendixprojectors}
We consider the operator on $1$-forms on $\mathbb{R}^4$ given by
\begin{equation}
{\cal J}_{\alpha,\beta} (\sigma ) = - \rmi \star (\omega_{\alpha,\beta } \wedge \sigma) \, , \quad {\cal J}_{\alpha,\beta}^2 = -\id \,,
\end{equation}
which allows us to define projectors
\begin{equation}
P = \frac{1}{2}\left(\id - \rmi{\cal J} \right) \, \qquad \bar{P} = \frac{1}{2}\left(\id + \rmi{\cal J}\right) \ .
\end{equation}
This is suggestive of a complex structure and indeed if we take $\alpha = \gamma$ and $\beta =\hat{\gamma}$ then ${\cal J}_{\gamma,\hat{\gamma}}$ is the complex structure ${\cal J}_\gamma$ (e.q. \eqref{eq:complexStruct}. The projectors $P$ and $\bar{P}$ project onto $(1,0)$ and $(0,1)$ components thus realising the Dolbeault complex.

These projectors enjoy a range of identities that we deploy in calculation:
\begin{equation} \label{eq:projectorid}
\bar{P} ( \star (\mu_\alpha\wedge \sigma) ) = 0 \, , \quad P( \star (\mu_\beta \wedge \sigma) ) = 0 \,,\quad \mu_\beta \wedge \bar{P}(\sigma) = 0 \, , \quad \mu_\alpha \wedge P(\sigma) = 0 \, ,
\end{equation}
\begin{equation} \label{eq:projectoridII}
\omega_{\alpha, \beta }\wedge \bar{P}(\sigma) = -\star \bar{P}(\sigma) \, , \quad \omega_{\alpha, \beta }P(\sigma) = \star P(\sigma) \,
\end{equation}
\begin{equation} \label{eq:projectoridIII}
\omega_{\alpha, \beta }\wedge \bar{P}(\sigma) \wedge \tau = \omega \wedge \sigma \wedge P(\tau) \, , \quad \omega_{\alpha, \beta }\wedge \bar{P}(\sigma) \wedge \bar{P}(\tau) = 0 \, .
\end{equation}
To move between form and component notation is useful to observe that
\begin{equation}\label{eq:projecttocomp}
P(\sigma)_{a \dot{a}} = - \frac{1}{\ab} \alpha_a \beta^b \sigma_{b \dot{a}} \, , \quad \bar{P} (\sigma)_{a \dot{a}} = \frac{1}{\ab} \beta_a \alpha^b \sigma_{b \dot{a}} \, .
\end{equation}
Further relations, used for processing the $\mathbb{CP}_1$ derivative boundary conditions, are
\begin{align}\label{eq:CP1compid}
\alpha^{a}\sigma_{a\dot{a}} \bar{e}^{\dot{a}}\vert_{\alpha} = 2 \star ( \mu_\alpha \wedge a) \, , \quad \beta^{a}\tau_{a\dot{a}} \bar{e}^{\dot{a}}\vert_{\alpha} = -\langle \alpha \beta \rangle P(\tau) \, , \\
\beta^{a}\sigma_{a\dot{a}} \bar{e}^{\dot{a}}\vert_{\beta} = 2 \star ( \mu_\beta\wedge a) \, , \quad \alpha^{a}\tau_{a\dot{a}} \bar{e}^{\dot{a}}\vert_{\beta} = \langle \alpha \beta \rangle \bar{P}(\tau) \, .
\end{align}

As an application of this projector technology let us consider the (ungauged) WZW$_4$ model for which the equation of motion can be cast in terms of the right-invariant Maurer Cartan form $R= \dr g g^{-1}$ that obeys $\dr R = R\wedge R $ as
\begin{equation}
\dr \star \bar{P}(R) = \frac{1}{2} \dr\; (\star - \omega_{\alpha,\beta} \wedge)\; \dr g g^{-1} = 0 \, .
\end{equation}
We consider now a Yang-Mills connection $A = - \bar{P}(X) $. The equations for this to be anti-self dual are
\begin{equation}
\mu_\beta F[A] = 0 \, , \quad \mu_\alpha F[A]= 0 \, , \quad \omega_{\alpha,\beta} \wedge F[A] = 0\, .
\end{equation}
The first of these vanishes identically by virtue of the fact that $\mu_\beta \wedge A = 0 $. Since $ \mu_\alpha \wedge A = - \mu_\alpha \wedge X$, the second provides a Bianchi identity
\begin{equation}
\mu_\alpha F[A]= - \mu_\alpha \wedge \left( \dr X - X \wedge X\right)
\end{equation}
and hence solved with $X=R$. The final equation returns the desired equation of motion as
\begin{equation}
\omega_{\alpha,\beta} \wedge F[A] = -\dr( \omega_{\alpha,\beta}\wedge \bar{P}(R) ) +\omega_{\alpha,\beta}\wedge \bar{P}(R) \wedge \bar{P}(R) = \dr \star \bar{P}(R) \, .
\end{equation}

At the K{\"a}hler point $\beta = \hat{\alpha} = \hat{\gamma}$, we can simply cast the ASDYM equations as
\begin{equation}
F^{2,0} = 0 \ , \qquad
F^{0,2} = 0 \ , \qquad
\varpi \wedge F^{1,1} = 0 \ .
\end{equation}
In this case, the connection is given by $A = - \bar{\pd} g g^{-1}$ is of type $(0,1)$, hence $F^{2,0}=0$ automatically, $F^{0,2}=0$ is zero by Bianchi identity and the equation of motion of WZW$_{4}$ is
\begin{equation}
\varpi \wedge \pd ( \bar{\pd} g g^{-1} ) = 0 \, .
\end{equation}

\section{Derivation of Localisation Formulae}
\label{appendix on localisation formulae}

In this work we are required to evaluate integrals of the form
\begin{equation}\label{General form of integral}
I = \frac{1}{2 \pi \rmi} \int_{\bPT} \bar{\pd} \Omega \wedge Q \ , \qquad
Q \in \Omega^{0,2} (\bPT) \ .
\end{equation}
In this appendix, we will derive general formulae for these integrals for the cases in which $\Omega$ has either two double poles or a single fourth order pole as used in the paper. To compute these integrals efficiently we will move to inhomogeneous coordinates and make use of the identity
\begin{equation} \label{eq:inhom_pole_der}
\pd_{\bar{\zeta}} \bigg( \frac{1}{\zeta - \alpha} \bigg) = - 2 \pi \rmi \, \delta^{2}(\zeta - \alpha) \ , \qquad
\int_{\bCP^{1}} \dr \zeta \wedge \dr \bar{\zeta} \, \delta^{2}(\zeta - \alpha) \, f(\zeta) = f(\alpha) \ .
\end{equation}

\subsection{Two double poles}
We consider the $(3,0)$-form given by
\begin{equation}
\Omega = \frac{1}{2}
\frac{
\langle \alpha \beta \rangle^{2}
}{
\langle \pi \alpha \rangle^{2}
\langle \pi \beta \rangle^{2}
}
e^0 \wedge e^{\dot{a}} \wedge e_{\dot{a}} = \frac{1}{2}
\frac{
(\alpha - \beta)^{2}
}{
(\zeta - \alpha)^{2}
(\zeta - \beta)^{2}
}
\dr \zeta \wedge \theta^{\dot{a}} \wedge \theta_{\dot{a}} \ .
\end{equation}
Substituting this into our integral gives
\begin{equation}
I = - \frac{1}{2} \frac{1}{2 \pi \rmi}
\int_{\bPT} \dr \zeta \wedge \dr \bar{\zeta} \, \pd_{\bar{\zeta}}
\bigg( \frac{
(\alpha - \beta)^{2}
}{
(\zeta - \alpha)^{2}
(\zeta - \beta)^{2}
} \bigg)
\wedge \theta^{\dot{a}} \wedge \theta_{\dot{a}} \wedge Q \ .
\end{equation}
Then, using the identity \eqref{eq:inhom_pole_der} gives
\begin{equation}
I = - \frac{(\alpha - \beta)^{2}}{2}
\int_{\bPT} \dr \zeta \wedge \dr \bar{\zeta} \,
\bigg[ \frac{
\pd_{\zeta} \delta(\zeta - \alpha)
}{
(\zeta - \beta)^{2}
}
+ \frac{
\pd_{\zeta} \delta(\zeta - \beta)
}{
(\zeta - \alpha)^{2}
} \bigg]
\wedge \theta^{\dot{a}} \wedge \theta_{\dot{a}} \wedge Q \ .
\end{equation}
Since the integral is symmetric under $\alpha \leftrightarrow \beta$, we will only compute the first term explicitly.
Integration by parts and evaluating the integral over $\bCP^{1}$ gives
\begin{equation} \label{eq:checkpoint}
I = \frac{(\alpha - \beta)^{2}}{2}
\int_{\bR^{4}} \pd_{\zeta}
\bigg( \frac{
\theta^{\dot{a}} \wedge \theta_{\dot{a}} \wedge Q
}{
(\zeta - \beta)^{2}
} \bigg) \bigg\vert_{\alpha}
\quad + \quad \alpha \leftrightarrow \beta \ .
\end{equation}
We will first distribute the $\pd_{\zeta}$ derivative, leaving the 2-form $Q$ completely general, resulting in
\begin{equation}
\begin{aligned}
I = \frac{(\alpha - \beta)^{2}}{2}
\int_{\bR^{4}}
\bigg[
& \frac{-2}{(\zeta - \beta)^{3}} \theta^{\dot{a}} \wedge \theta_{\dot{a}} \wedge Q
+ \frac{2}{(\zeta - \beta)^{2}} \hat{\gamma}_{a} \dr x^{a \dot{a}} \wedge \theta_{\dot{a}} \wedge Q \\
& \qquad + \frac{
\theta^{\dot{a}} \wedge \theta_{\dot{a}}
}{
(\zeta - \beta)^{2}
} \wedge \pd_{\zeta} Q
\bigg] \bigg\vert_{\alpha}
\quad + \quad \alpha \leftrightarrow \beta \ .
\end{aligned}
\end{equation}
The overall factor of $(\alpha - \beta)^{2}$ outside the integral cancels with the denominators in the integrand.
We will also make use of \eqref{eq:hom_inhom_identity} to return to spinor notation, and introduce self-dual 2-forms defined by $\Sigma^{ab} = \varepsilon_{\dot{a} \dot{b}} \dr x^{a \dot{a}} \wedge \dr x^{b \dot{b}}$.
\begin{equation}
\begin{aligned}
I = \frac{1}{2}
\int_{\bR^{4}}
\bigg[
& \frac{
-2 \langle \hat{\gamma} \beta \rangle
}{
\langle \alpha \beta \rangle
\langle \alpha \hat{\gamma} \rangle
} \alpha_{a} \alpha_{b} \Sigma^{ab} \wedge Q \vert_{\alpha}
+ \frac{2}{
\langle \alpha \hat{\gamma} \rangle
}
\hat{\gamma}_{a} \alpha_{b} \Sigma^{ab} \wedge Q \vert_{\alpha} \\
& \qquad + \alpha_{a} \alpha_{b} \Sigma^{ab} \wedge \frac{
\pd_{\zeta} Q
}{
\langle \pi \hat{\gamma} \rangle^{2}
} \bigg\vert_{\alpha} \bigg]
\quad + \quad \alpha \leftrightarrow \beta \ .
\end{aligned}
\end{equation}
Expanding $\alpha_{a}$ in the basis formed by $\hat{\gamma}_{a}$ and $\beta_{a}$, we see that one component of the first term cancels the entire second term, and only a term proportional to $\alpha_{a} \beta_{b} \Sigma^{ab}$ survives.
In the third term of the action, we recognise the combination $\pd_{0}$ acting on $Q$ and make this replacement.
In conclusion, we have the general formula
\begin{equation} \label{eq:localisationformula}
\frac{1}{2 \pi \rmi} \int_{\bPT} \bar{\pd} \Omega \wedge Q = \int_{\bR^{4}}
\bigg[
\frac{
\alpha_{a} \beta_{b} \Sigma^{ab}
}{
\langle \alpha \beta \rangle
} \wedge Q \vert_{\alpha}
+ \frac{1}{2} \alpha_{a} \alpha_{b} \Sigma^{ab} \wedge (\pd_{0} Q) \vert_{\alpha} \bigg]
\quad + \quad \alpha \leftrightarrow \beta \ ,
\end{equation}
or in differential form notation
\begin{equation}
\frac{1}{2 \pi \rmi} \int_{\bPT} \bar{\pd} \Omega \wedge Q = \int_{\bR^{4}}
\bigg[
\omega_{\alpha,\beta} \wedge Q \vert_{\alpha}
+ \frac{1 }{2} \mu_\alpha \wedge (\pd_{0} Q) \vert_{\alpha} \bigg]
\quad + \quad \alpha \leftrightarrow \beta \ .
\end{equation}
It is also helpful to specialise to 2-forms of the form $Q = \pi^{a} \pi^{b} Q_{a \dot{a} b \dot{b}} \bar{e}^{\dot{a}} \wedge \bar{e}^{\dot{b}}$ which we will often encounter in practice.
In this case, we may make use of the identity
\begin{equation} \label{eq:es_to_vol4}
e^{\dot{c}} \wedge e_{\dot{c}} \wedge \bar{e}^{\dot{a}} \wedge \bar{e}^{\dot{b}} = - 2 \, \vol_{4} \, \varepsilon^{\dot{a} \dot{b}} \ .
\end{equation}
and its generalisation valid for any spinors $\alpha_{a}$ and $\beta_{a}$
\begin{equation}
\alpha_{a} \beta_{b} \Sigma^{ab} \wedge \bar{e}^{\dot{a}} \wedge \bar{e}^{\dot{b}} = - 2 \, \vol_{4} \, \frac{\langle \alpha \hat{\pi} \rangle \langle \beta \hat{\pi} \rangle}{\langle \pi \hat{\pi} \rangle^{2}} \, \varepsilon^{\dot{a} \dot{b}} \ .
\end{equation}
Using these identities on the above formula in the case $Q = \pi^{a} \pi^{b} Q_{a \dot{a} b \dot{b}} \bar{e}^{\dot{a}} \wedge \bar{e}^{\dot{b}}$ gives
\begin{equation}
\label{ec:localisationformula}
\frac{1}{2 \pi \rmi} \int_{\bPT} \bar{\pd} \Omega \wedge Q = - \int_{\bR^{4}} \vol_{4} \,
\bigg[
\frac{
\varepsilon^{\dot{a} \dot{b}} (\alpha^{a} \beta^{b} + \beta^{a} \alpha^{b})
}{
\langle \alpha \beta \rangle
} Q_{a \dot{a} b \dot{b}} \vert_{\alpha}
+ \varepsilon^{\dot{a} \dot{b}} \alpha^{a} \alpha^{b}
(\pd_{0} Q_{a \dot{a} b \dot{b}}) \vert_{\alpha} \bigg]
\quad + \quad \alpha \leftrightarrow \beta \ .
\end{equation}
One final specialism is the case when $Q_{a \dot{a} b \dot{b} } = X_{a \dot{a}} Y_{b\dot{b}} $, in which case the answer can be recast again in differential form notation as
\begin{equation}
\label{ec:localisationformula2}
\frac{1}{2 \pi \rmi} \int_{\bPT} \bar{\pd} \Omega \wedge Q = \int_{\bR^{4}}
\bigg[
\omega_{\alpha,\beta} \wedge X\wedge Y \vert_{\alpha}
+ \frac{1 }{2} \mu_\alpha \pd_{0} \wedge (X\wedge Y ) \vert_{\alpha} \bigg]
\quad + \quad \alpha \leftrightarrow \beta \ .
\end{equation}

To apply these formulae we need the following $\mathbb{CP}_1$ derivatives :
\begin{align}
\pd_0 ( \dr \hat{g} \hat{g}^{-1} ) &= \hat{g} \dr \hat{u} \hat{g}^{-1} \, , \\
\pd_0 ( \hat{g}^{-1}\dr \hat{g} ) &= \dr \hat{u} + [ \hat{g}^{-1}\dr \hat{g}, \hat{u} ] \, , \\
\pd_0 ( A ) &= \pd_0 (B) = 0 \, , \\
\pd_0 (\hat{g}^{-1} A \hat{g} ) &= [\hat{g}^{-1} A \hat{g} ,\hat{u} ] \, , \\
\pd_0 \frac{1}{3} \Tr (\hat{g}^{-1} \dr g)^3 &= \dr ~\Tr( \hat{u} (\hat{g}^{-1} \dr g)^2) \, ,
\end{align}
in which used the definition $\hat{u} = \hat{g}^{-1}\pd_0 \hat{g}$.

\subsection{Fourth order Pole}
In section \ref{Gauged LMP action section}, we considered a different $(3,0)$-form given by
\begin{equation}\label{4th order}
\Omega = k \frac{{e}^{0} \wedge {e}^{\d{a}}\wedge {e}_{\d{a}}}{{\langle \pi \alpha \rangle}^{4}} = \frac{k'}{{\la \hat{\gamma} \alpha \ra}^{4}} \frac{d \zeta \wedge \theta^{\d{a}} \wedge \theta_{\d{a}}}{(\zeta - \alpha)^{4}} \, .
\end{equation}
Substituting this into the general integral expression above gives
\begin{equation}
I = - \frac{k}{{\la \hat{\gamma} \alpha \ra}^{4}} \frac{1}{2 \pi \rmi}
\int_{\bPT} \dr \zeta \wedge \dr \bar{\zeta} \, \pd_{\bar{\zeta}}
\bigg( \frac{1}{(\zeta - \alpha)^{4}} \bigg)
\wedge \theta^{\dot{a}} \wedge \theta_{\dot{a}} \wedge Q \ .
\end{equation}
Then, using the identity \eqref{eq:inhom_pole_der}, we find
\begin{equation}
I = - \frac{k}{6 {\la \hat{\gamma} \alpha \ra}^{4}}
\int_{\bPT} \dr \zeta \wedge \dr \bar{\zeta} \,
\Big( \pd_{\zeta}^{3} \delta (\zeta - \alpha) \Big)
\wedge \theta^{\dot{a}} \wedge \theta_{\dot{a}} \wedge Q \ .
\end{equation}
Applying integration by parts and completing the integral over $\bCP^{1}$ gives
\begin{equation}
I = \frac{k}{6 {\la \hat{\gamma} \alpha \ra}^{4}}
\int_{\bR^{4}} \pd_{\zeta}^{3} \Big( \theta^{\dot{a}} \wedge \theta_{\dot{a}} \wedge Q \Big) \Big\vert_{\alpha} \ .
\end{equation}
In order to distribute this over the argument, it is helpful to have the identities
\begin{equation}
\theta^{a} \big\vert_{\alpha} = \frac{\alpha_{a} \dr x^{a \dot{a}}}{\langle \hat{\gamma} \alpha \rangle} \ , \qquad
\pd_{\zeta} \theta^{a} \big\vert_{\alpha} = \hat{\gamma}_{a} \dr x^{a \dot{a}} \ , \qquad
\pd_{\zeta}^{2} \theta^{a} \big\vert_{\alpha} = 0 \ .
\end{equation}
Then, distributing the three $\pd_{\zeta}$-derivatives gives
\begin{equation}
I = \frac{k}{6 {\la \hat{\gamma} \alpha \ra}^{4}}
\int_{\bR^{4}} \bigg[
\frac{\alpha_{a} \alpha_{b} \Sigma^{ab}}{\langle \alpha \hat{\gamma} \rangle^{2}} \wedge \pd_{\zeta}^{3} Q \big\vert_{\alpha}
+ 6 \, \frac{\alpha_{a} \hat{\gamma}_{b} \Sigma^{ab}}{\langle \alpha \hat{\gamma} \rangle} \wedge \pd_{\zeta}^{2} Q \big\vert_{\alpha}
+ 6 \, \hat{\gamma}_{a} \hat{\gamma}_{b} \Sigma^{ab} \wedge \pd_{\zeta} Q \big\vert_{\alpha}
\bigg] \ .
\end{equation}
Converting this expression back into homogeneous coordinates (and making use of the fact that $Q$ was a $(0,2)$-form on twistor space meaning $\hat{\alpha}_{a} \dr x^{a \dot{a}} \wedge Q \vert_{\alpha} = 0$), this integral becomes
\begin{equation}
I = \frac{k}{6}
\int_{\bR^{4}} \alpha_{a} \alpha_{b} \Sigma^{ab}\wedge \pd_{0}^{3} Q \big\vert_{\alpha} \ .
\end{equation}

\section{Localisation Derivation with General Gaugings}\label{Sec:General gaugings}

In this appendix we describe in more detail the derivation of the gauged WZW$_4$ model from the gauged hCS$_6$ theory and the application of the localisation formulae above. We shall do this in a more general manner, allowing the gauging of an $H$ subgroup that acts as
\begin{equation}\label{eq:asymaction}
g\mapsto \rho_\beta(\ell) g \rho_\alpha(\ell^{-1}) \, , \quad B \mapsto \ell B \ell^{-1} - \dr \ell \ell^{-1} \, , \quad \ell \in H \subset{G} \,,
\end{equation}
in which $\rho_i: H \rightarrow G$ are group homomorphisms (algebra homomorphisms will be denoted with the same symbol). The covariant derivative is then given by \begin{equation}
\nabla g g^{-1} =\dr g g^{-1} + B_\beta - g B_\alpha g^{-1} \mapsto \rho_\beta(\ell) ( \nabla g g^{-1})\rho_\beta(\ell^{-1}) \, ,
\end{equation}
in which we ease notation by setting $B_i =\rho_i(B)$.

The starting point is the six-dimensional theory
\begin{equation}
S_{\mathrm{ghCS_6}}[\cA , \cB] = S_{\mathrm{hCS_6}}[\cA] - S_{\mathrm{hCS_6}}[\cB] + S_{\mathrm{bdy}}[\cA , \cB] \,,
\end{equation}
in which we specify a boundary interaction term
\begin{equation}
S_{\mathrm{bdy}}[\cA , \cB]= -\frac{q}{2\pi \rmi }\int_{\mathbb{PT} } \bar{\partial} \Omega \wedge \Tr_\mathfrak{g}\left( \cA \wedge \rho(\cB) \right) \, .
\end{equation}
Here we have introduced a parameter $q$, which will ultimately be set to one, to keep track of the contributions from this boundary term.
To specify this term we include an algebra homomorphism $\rho$ which only needs to be defined piecewise on the components of the support of $\pd \Omega$.
We could choose to dispense the higher-dimensional covariance and simply add different boundary terms specified only at the location of the poles but it is convenient to formally consider $\rho$ to be a defined as a piecewise map that takes values $\rho \vert_{\pi = \alpha , \beta } = \rho_{\alpha, \beta } $.

To define a six-dimensional theory requires imposing conditions that ensure the vanishing of the boundary term
\begin{equation}
\int_{\mathbb{PT} } \bar{\pd} \Omega \wedge \left( \Tr_{\mathfrak{g}}\left(\delta\cA \wedge (\cA - q \rho (\cB) )+q \rho(\delta \cB) \wedge \cA \right) - \Tr_{\mathfrak{h}}(\delta \cB \wedge \cB) \right) \, .
\end{equation}
We are required to cancel a term involving the inner product on the algebra $\mathfrak{h}$ with one on $\mathfrak{g}$, which can be achieved demanding
\begin{equation}
\Tr_\mathfrak{g} ( \rho(x) \rho(y) )\vert_{\alpha,\beta} = \Tr_\mathfrak{h} ( x y) \quad \forall x,y \in \mathfrak{h} \, .
\end{equation}
Note that as a consequence this implies
\begin{equation}\label{eq:anomv2}
\Tr_\mathfrak{g} ( \rho_\alpha (x) \rho_\alpha (y) ) = \Tr_\mathfrak{h} ( x y) = \Tr_\mathfrak{g} ( \rho_\beta (x) \rho_\beta (y) ) \, ,
\end{equation}
which is the anomaly-free condition allowing for the construction of a gauge-invariant extension to the WZW model for the gauge symmetry~\eqref{eq:asymaction}. With this condition the boundary term produced by variation is given by
\begin{equation}
\int_{\mathbb{PT} } \bar{\pd} \Omega \wedge \left( \Tr_{\mathfrak{g}}\left(\delta\cA \wedge (\cA - q \rho (\cB) )+q \rho(\delta \cB) \wedge (\cA - q^{-1} \wedge \rho(\cB))\right) \right) \, ,
\end{equation}
and is set to zero by the conditions
\begin{equation} \label{eq:ABboundarycond2}
\cA^{\fk} \big\vert_{\alpha , \beta} = 0 \ , \qquad
\cA^{\fh} \big\vert_{\alpha , \beta} =\rho(\cB) \big\vert_{\alpha , \beta} \ , \qquad
\pd_{0} \cA^{\fh} \big\vert_{\alpha , \beta} = \rho( \pd_{0}\cB) \big\vert_{\alpha , \beta} \ .
\end{equation}
It is noteworthy that if we impose all of these conditions from the outset, the contribution from the explicit boundary term $S_{\mathrm{bdy}}[\cA , \cB]$ would vanish.
However, from a four-dimensional perspective the boundary conditions on $\pd_0(\cA)$ lead to {\em differential} constraints on the fundamental fields and it is not clear that one should, or could, naively invoke them to produce a Lagrangian description.
Instead what we shall do is {\em only} impose the conditions $\cA^{\fk} \big\vert_{\alpha , \beta} = 0$ and $\cA^{\fh} \big\vert_{\alpha , \beta} =\rho(\cB) \big\vert_{\alpha , \beta}$, which can be solved algebraically and substituted into the Lagrangian without concern. Doing this one finds that $S_{\mathrm{bdy}}[\cA , \cB]$ does contribute, and when $q=1$ in particular, it provides a gauge invariant completion of the action. The boundary conditions that we have not imposed have not been forgotten, instead when $q=1$ they are recovered as on-shell equations in this four-dimensional theory. This provides an alternative view of the procedure; the explicit boundary term is serving to implement the constraints arising from $\pd_{0} \cA^{\fh} \big\vert_{\alpha , \beta} = \rho( \pd_{0}\cB) \big\vert_{\alpha , \beta}$ at the Lagrangian level.
We can see this explicitly by observing that if we just impose $\cA^{\fk} \big\vert_{\alpha , \beta} = 0$ and $\cA^{\fh} \big\vert_{\alpha , \beta} =\rho(\cB) \big\vert_{\alpha , \beta}$ then
\begin{equation}\begin{split}
    & \Big(\delta\cA \wedge (\cA - q \rho (\cB) )+q \rho(\delta \cB) \wedge \cA - \rho(\delta \cB) \wedge \rho(\cB)\Big)|_{\alpha,\beta} = 0 \,,
    \\
    & \partial_0 \Big(\delta\cA \wedge (\cA - q \rho (\cB) )+q \rho(\delta \cB) \wedge \cA - \rho(\delta \cB) \wedge \rho(\cB)\Big)|_{\alpha,\beta}
    \\ & \qquad =  (1-q)\delta(\partial_0\cA - \rho( \partial_0\cB) )\wedge \rho(\cB) |_{\alpha,\beta}  + (1+q)\rho(\delta\cB) \wedge (\partial_0 \cA - \rho(\partial_0 \cB))|_{\alpha,\beta} \,.
\end{split}\end{equation}
Therefore, for $q=1$ we see that the boundary equation of motion for $\mathcal{B}$ is precisely $\pd_{0} \cA^{\fh} \big\vert_{\alpha , \beta} = \rho( \pd_{0}\cB) \big\vert_{\alpha , \beta}$.

The localisation proceeds as follows. First, we change parametrisation $\cA = \cA'{}^{\hat{g}}$ and $\cB =\cB'{}^{\hat{h} } $ fixing some redundancy by demanding $\cA'$ and $\cB'$ have no $\mathbb{CP}_1$ legs. Second, we fix some of the residual symmetry preserved by the boundary conditions to set $\hat{g} \vert_{\beta} = \hat{h} \vert_{\alpha,\beta} = \id $ and $ \pd_0 \hat{h } \vert_{\alpha,\beta} = 0 $. The remaining fields are $\hat{g}\vert_{\alpha} =g $, $\hat{g}^{-1} \pd_0 \hat{g}\vert_{\alpha} = u $, $\hat{g}^{-1} \pd_0 \hat{g}\vert_{\beta} = \tilde{u} $ and the four-dimensional gauge fields $A$ and $B$ that arise from $\cA'$ and $\cB'$ once their holomorphicity is imposed.

We may now directly apply the localisation formulae~\eqref{ec:localisationformula2} to show that the hCS terms localise (without imposing any boundary conditions) to give
\begin{equation}
\begin{split}
S_{\mathrm{hCS}}[\mathcal{A}] \simeq & \int_{\mathbb{R}^4} \omega_{\alpha,\beta} \wedge \Tr_\fg ( A^g \wedge g^{-1} \dr g ) - \omega_{\alpha,\beta}\wedge \mathcal{L}_{\textrm{WZ}}[g] \\
& \qquad + \frac{1 }{2} \mu_\alpha\wedge \Tr_\fg ( A^g \wedge \dr u) + \frac{1 }{2} \mu_\beta \wedge \Tr_\fg ( A \wedge \dr \tilde{u}) \,,
\end{split}
\end{equation}
while $ S_{\mathrm{hCS}}[\cB]$ yields zero in this gauge. Let us first consider the terms involving $\omega_{\alpha,\beta}$. Since the gauge completion of the WZ term is
\begin{equation}
\label{eq:generalgwzw1}
\mathcal{L}_{\textrm{gWZ}}[g,B] = \mathcal{L}_{\textrm{WZ}}[g] + \Tr_\mathfrak{g} \left( g^{-1} \dr g \wedge
B_\alpha + \dr g g^{-1} \wedge B_\beta + g^{-1} B_\beta g B_\alpha \right) \,,
\end{equation}
we may express them (trace left implicit) as
\begin{equation}\begin{split}
& \omega_{\alpha,\beta} \wedge \left( A^g \wedge g^{-1} \dr g - \mathcal{L}_{\textrm{WZ}}[g] \right) \\
= \,& \omega_{\alpha,\beta} \wedge \left( A^g \wedge g^{-1} \dr g - \mathcal{L}_{\textrm{gWZ}}[g,B] + g^{-1} \dr g \wedge B_\alpha + \dr g g^{-1} B_\beta +g^{-1} B_\beta g B_\alpha \right) \\
= \, & \omega_{\alpha,\beta} \wedge \left( (A^g- B_\alpha) \wedge g^{-1} \nabla g - \mathcal{L}_{\textrm{gWZ}}[g,B] + A^g
\wedge B_\alpha - A \wedge B_\beta \right) \, .
\end{split}\end{equation}
To proceed we invoke the algebraic boundary conditions of eq.~\eqref{eq:ABboundarycond2}, which in differential form notation become
\begin{align}\label{eq:BCsol}
A = B_\beta - \bar{P}( \nabla g g^{-1} ) \Leftrightarrow A^g= P(g^{-1}\nabla g) + B_\alpha \, .
\end{align}
such that
\begin{equation}\begin{split}
& \omega_{\alpha,\beta} \wedge \left( A^g \wedge g^{-1} \dr g -\mathcal{L}_{\textrm{WZ}}[g] \right) \\
=\, & \omega_{\alpha,\beta} \wedge \left( P( g^{-1} \nabla g) \wedge g^{-1} \nabla g - \mathcal{L}_{\textrm{gWZ}}[g,B] + A^g
\wedge B_\alpha - A \wedge B_\beta \right) \\
=\, & -\frac{1}{2} g^{-1} \nabla g \wedge\star( g^{-1} \nabla g) -\omega_{\alpha,\beta} \wedge \left( \mathcal{L}_{\textrm{gWZ}}[g,B] - A^g
\wedge B_\alpha + A \wedge B_\beta \right)
\, .
\end{split}\end{equation}
Here in the last line we made use of the identity $
\omega \wedge P(\sigma)\wedge \sigma = - \frac{1}{2}\sigma\wedge \star \sigma $ for a 1-form $\sigma$. To treat the terms involving $\mu_{\alpha}$ and $\mu_\beta$ we may combine the algebraic boundary conditions with the properties $\mu_\alpha \wedge P(X) = \mu_\beta \wedge \bar{P}(X) = 0 $ such that $\mu_a\wedge A^g= \mu_\alpha B_\alpha$ and $\mu_\beta \wedge A= \mu_\beta B_\beta$. In summary we find
\begin{equation}
\begin{split}
S_{\mathrm{hCS}}[\mathcal{A}] &\simeq \int_{\mathbb{R}^4} -\frac{1}{2} \Tr_\fg\left(g^{-1}\nabla g \wedge \star g^{-1}\nabla g \right) - \omega_{\alpha,\beta}\wedge\left(\mathcal{L}_{\textrm{gWZ}}[g,B] + \Tr_\fg( A \wedge B_\beta - A^g B_\alpha ) \right) \\
& \qquad + \frac{1}{2} \mu_\alpha\wedge \Tr( B_\alpha \wedge \dr u) + \frac{1 }{2} \mu_\beta \wedge \Tr( B_\beta \wedge \dr \tilde{u}) \,.
\end{split}
\end{equation}
The localisation of the explicit boundary term yields, after using $\mu_a\wedge A^g= \mu_\alpha B_\alpha$,
\begin{equation}
\begin{split}
S_{\mathrm{bdy}}[\cA , \cB] \simeq & - q \int_{\mathbb{R}^4}\omega_{\alpha,\beta} \wedge \Tr_\fg ( A^g B_\alpha - A B_\beta )\\
& \qquad + \frac{1}{2} \mu_\alpha\wedge \Tr_\fg ( (\dr u + [B_\alpha , u]) B_\alpha ) + \frac{1 }{2} \mu_\beta\wedge \Tr_\fg ( (\dr \tilde{u} + [B_\beta , \tilde{u}]) B_\beta ) \, .
\end{split}
\end{equation}

The significance of the boundary term now becomes clear, as it serves to ensure manifest gauge invariance. When $q=1$ the terms $\omega_{\alpha,\beta} \wedge \Tr( A^g B_\alpha - A B_\beta )$ directly cancel. The contributions of the entire localised action that are wedged against $\mu_\alpha$ sum to
\begin{equation}
\mu_\alpha \wedge \Tr_\fg \left( (1- q) ~\dr u\wedge B_\alpha + 2 q ~ u F[B]_\alpha -2q ~\dr ( B_\alpha u) \right) \, .
\end{equation}
We can see that for $q=1$ we find a gauge-invariant field strength together with a total derivative term that we discard.
The terms wedged against $\mu_\beta$ give a similar contribution.
Hence the fully localised action becomes
\begin{equation}
\begin{split}
S \simeq & \int_{\mathbb{R}^4}
-\frac{1}{2} \Tr_\fg\left(g^{-1}\nabla g \wedge \star g^{-1}\nabla g \right) - \omega_{\alpha,\beta}\wedge \mathcal{L}_{\textrm{gWZ}}[g,B] \\
& \qquad + \mu_\alpha\wedge \Tr_\fg( u F[B]_\alpha ) + \mu_\beta \wedge \Tr_\fg( \tilde{u} F[B]_\beta) \, .
\end{split}
\end{equation}
Noting that the components of $u$ and $\tilde{u}$ in complement of $\fh$ decouple we can view $u$ and $\tilde{u}$ as $\fh$-valued and write
\begin{equation}
\begin{split}
S \simeq & \int_{\mathbb{R}^4}
- \frac{1}{2} \Tr_\fg\left(g^{-1}\nabla g \wedge \star g^{-1}\nabla g \right) - \omega_{\alpha,\beta}\wedge \mathcal{L}_{\textrm{gWZ}}[g,B] \\
& \qquad + \mu_\alpha\wedge \Tr_\fh( u F[B] ) + \mu_\beta \wedge \Tr_\fh( \tilde{u} F[B]) \, .
\end{split}
\end{equation}

\printbibliography

\end{document}